\documentclass[useAMS,usenatbib]{mn2e}
\usepackage{graphicx}
\usepackage{amssymb}
\usepackage{color}

\newcommand{\hide}[1]{}
\newcommand{\vc}[1]{\bmath{#1}} 
\newcommand{\lx}[1]{\rmn{#1}}
\newcommand{\did}{\lx{d}}

\newcommand{\secref}[1]{section \ref{#1}}
\newcommand{\secsand}[2]{sections \ref{#1} and \ref{#2}}
\newcommand{\secsdash}[2]{sections \ref{#1}--\ref{#2}}

\newcommand{\apref}[1]{Appendix \ref{#1}}
\newcommand{\Eqref}[1]{Equation (\ref{#1})}
\newcommand{\eqref}[1]{equation (\ref{#1})}
\newcommand{\eqsref}[1]{equations (\ref{#1})}
\newcommand{\Eqsand}[2]{Equations (\ref{#1}) and (\ref{#2})}
\newcommand{\eqsand}[2]{equations (\ref{#1}) and (\ref{#2})}
\newcommand{\Eqsdash}[2]{Equations (\ref{#1}--\ref{#2})}
\newcommand{\eqsdash}[2]{equations (\ref{#1}--\ref{#2})}
\newcommand{\exref}[1]{(\ref{#1})}
\newcommand{\figref}[1]{figure \ref{#1}}

\newcommand{\ffigref}[2]{figure \ref{#1} (#2 panel)}
\newcommand{\Ffigref}[2]{Figure \ref{#1} (#2 panel)}

\newcommand{\bea}{\begin{eqnarray}}
\newcommand{\eea}{\end{eqnarray}}
\newcommand{\beq}{\begin{equation}}
\newcommand{\eeq}{\end{equation}}
\newcommand{\lt}{\left}
\newcommand{\rt}{\right}
\newcommand{\dd}{\partial}
\newcommand{\const}{{\rm const}}
\newcommand{\eps}{\epsilon}

\newcommand{\nuii}{\nu_{ii}}
\newcommand{\mfp}{\lambda_{\lx{mfp}}} 
\newcommand{\vthi}{v_{\lx{th}i}}
\newcommand{\vthe}{v_{\lx{th}e}}
\renewcommand{\Re}{\lx{Re}}
\newcommand{\pperp}{p_\perp}
\newcommand{\piperp}{p_i^\perp}
\newcommand{\pperpi}{p_{2i}^\perp}
\newcommand{\dpperp}{\delta\pperp}
\newcommand{\ppar}{p_\parallel}
\newcommand{\pipar}{p_i^\parallel}
\newcommand{\ppari}{p_{2i}^\parallel}
\newcommand{\dppar}{\delta\ppar}
\newcommand{\vdel}{\vc{\nabla}}
\newcommand{\dpar}{\nabla_\parallel}
\newcommand{\vv}{\vc{v}}
\newcommand{\vperp}{v_\perp}
\newcommand{\vvperp}{\vv_\perp}
\newcommand{\vpar}{v_\parallel}
\newcommand{\vr}{\vc{r}}
\newcommand{\vk}{\vc{k}}
\newcommand{\vkperp}{\vk_\perp}
\newcommand{\kperp}{k_\perp}
\newcommand{\kpar}{k_\parallel}
\newcommand{\vu}{\vc{u}}
\newcommand{\dvu}{\delta\vu}
\newcommand{\vup}{\vu_{1i}^\perp}
\newcommand{\vE}{\vc{E}}
\newcommand{\Epar}{E_\parallel}
\newcommand{\vj}{\vc{j}}
\newcommand{\vB}{\vc{B}}
\newcommand{\dvB}{\delta\vc{B}}
\newcommand{\dvBperp}{\dvB_\perp}
\newcommand{\vBp}{\vB_1^\perp}
\newcommand{\Bbar}{\overline{B}}
\newcommand{\vBfrac}{\frac{\vBp}{B_0}}
\newcommand{\dBpar}{\delta B_\parallel}
\newcommand{\dBperp}{\delta B_\perp}
\newcommand{\vb}{\vc{\hat b}}
\newcommand{\dvb}{\delta\vb}
\newcommand{\gmax}{\gamma_{\rm max}}
\newcommand{\St}[1]{C[#1]}
\newcommand{\vP}{\textbfss{P}}
\newcommand{\vI}{\textbfss{I}}
\newcommand{\feo}{f_e^{(0)}}
\newcommand{\feone}{f_e^{(1)}}
\newcommand{\qperp}{q_{1i}^\perp}
\newcommand{\qpar}{q_{1i}^\parallel}
\newcommand{\gT}{\Gamma_T}
\newcommand{\opeak}{\omega_{\rm p}}
\newcommand{\kpeak}{k_{\rm p}}

\title[Theory of firehose \& gyrothermal instabilities]{
  A nonlinear theory of the parallel firehose and gyrothermal instabilities in a weakly collisional plasma}
\author[M. S. Rosin, A. A. Schekochihin, F. Rincon and S. C. Cowley]{
  M. S. Rosin$^{1}$\thanks{Current address: Department of 
Mathematics, University of California, 
520 Portola Plaza, Los Angeles, CA 90095, U.S.A.; 
Electronic address: msr35@math.ucla.edu}, 
  A. A. Schekochihin$^{2}$\thanks{Corresponding author; Electronic address: a.schekochihin1@physics.ox.ac.uk}, 
  F. Rincon$^{3}$,
  and S. C. Cowley$^{4,5}$\\
  $^{1}$ DAMTP, Centre for Mathematical Sciences, University of Cambridge, 
  Wilberforce Road, Cambridge, CB3 0WA, U.K.\\
  $^{2}$ Rudolf Peierls Centre for Theoretical Physics, University of Oxford, 
  1 Keble Road, Oxford, OX1 3NP, U.K.\\
  $^{3}$ Laboratoire d'Astrophysique de Toulouse-Tarbes, Universit\'e de Toulouse, CNRS, 
  14 avenue Edouard Belin, F-31400 Toulouse, France\\
  $^{4}$ EURATOM/CCFE Fusion Association, Culham Science Centre,
  Abingdon, OX14 3DB, U.K.\\
  $^{5}$ Blackett Laboratory, Imperial College, 
  Prince Consort Road, London, SW7 2AZ, U.K.
}
\begin{document}

\date{Submitted to MNRAS 21 February 2010; e-print {\tt arXiv:1002.4017}}

\pagerange{\pageref{firstpage}--\pageref{lastpage}} \pubyear{2009}

\maketitle

\label{firstpage}

\begin{abstract}
Weakly collisional magnetized cosmic plasmas have a dynamical tendency 
to develop pressure anisotropies with respect to the local direction of the 
magnetic field. These anisotropies trigger plasma instabilities at scales 
just above the ion Larmor radius $\rho_i$ and much below the mean free 
path $\mfp$. They have growth rates of a fraction of the ion cyclotron frequency, 
which is much faster than either the global dynamics or even local turbulence. 
Despite their microscopic nature, these instabilities dramatically modify the 
transport properties and, therefore, the macroscopic dynamics of the plasma. 
The nonlinear evolution of these 
instabilities is expected to drive pressure anisotropies towards marginal stability values, 
controlled by the plasma beta $\beta_i$. Here this nonlinear evolution is worked out in an 
{\em ab initio} kinetic calculation for the simplest analytically tractable 
example --- the parallel ($\kperp=0$) firehose instability in a high-beta plasma. 
An asymptotic theory is constructed, based on a particular physical ordering 
and leading to a closed nonlinear equation for the firehose turbulence. 
In the nonlinear regime, both analytical theory and the numerical solution 
predict secular ($\propto t$) growth of magnetic fluctuations. The fluctuations develop 
a $\kpar^{-3}$ spectrum, extending from scales somewhat larger than $\rho_i$ 
to the maximum scale that grows secularly with time ($\propto t^{1/2}$); 
the relative pressure anisotropy $(\pperp-\ppar)/\ppar$ tends 
to the marginal value $-2/\beta_i$. The marginal state is achieved via 
changes in the the magnetic field, not particle scattering. 
When a parallel ion heat flux is present, 
the parallel firehose mutates into the new {\em gyrothermal instability} 
(GTI), which continues to exist up to firehose-stable values of pressure anisotropy, 
which can be positive and are limited by the magnitude of the ion heat flux. 
The nonlinear evolution of the GTI also features secular growth of magnetic fluctuations, 
but the fluctuation spectrum is eventually dominated by modes around a maximal scale 
$\sim \rho_i l_T/\mfp$, where $l_T$ is the scale of the parallel temperature 
variation. Implications for momentum and heat transport are speculated about. 
This study is motivated by our interest in the dynamics of galaxy cluster 
plasmas (which are used as the main astrophysical example), 
but its relevance to solar wind and accretion flow plasmas is also briefly discussed. 
\end{abstract}


\section{Introduction} \label{sec:intro}

It has recently been realized in various astrophysics and space physics 
contexts that pressure anisotropies (with respect to the direction 
of the magnetic field) occur naturally and ubiquitously in magnetized 
weakly collisional plasmas.\footnote{As will be explained in detail in what follows, 
by weak collisionality we mean a state where Larmor motion is much faster than 
the collision rate, but large-scale dynamics occur on time scales slower 
than collisions, so collisions neither can be neglected nor are they sufficiently 
dominant to justify a fluid closure. \citet{Balbus2} calls this state a 
``dilute'' plasma.} 
They lead to very fast microscale instabilities, 
firehose, mirror, and others, whose presence is likely to fundamentally affect 
the transport properties and, therefore, both small- and large-scale dynamics of astrophysical 
plasmas --- most interestingly, the plasmas of galaxy clusters and accretion 
discs \citep{Hall1,Schek2,Schek3,Schek4,Sharma1,Sharma2,Lyutikov}. 
These instabilities occur even (and especially) 
in high-beta plasmas and even when the magnetic field is dynamically weak. 
The current state of theoretical understanding of this problem 
is such that we do not even have a set of well-posed macroscopic equations that govern 
the dynamics of a plasma in which the collisional mean free path exceeds the 
ion Larmor radius, $\mfp\gg\rho_i$ (equivalently, ion collision frequency is 
smaller than the ion cyclotron frequency, $\nuii\ll\Omega_i$). 
This is because calculating the dynamics at long spatial scales 
$l\gg\rho_i$ and slow time scales corresponding to frequencies $\omega\ll\Omega_i$ 
requires knowledge of the form of the pressure tensor and the heat 
fluxes, which depend on the 
nonlinear evolution and saturation of the instabilities triggered by 
the pressure anisotropies and temperature gradients. 
Since this is not currently understood, we do not have an effective mean-field theory 
for the large-scale dynamics. 

In the absence of a microphysical theory, it is probably sensible 
to assume that the instabilities will return the pressure anisotropies 
to the marginal level and to model large-scale dynamics on this basis, 
via a suitable closure scheme \citep{Sharma1,Sharma2,Schek2,Lyutikov,Kunz}.
This approach appears to be supported by the solar wind data 
\citep{Gary2,Kasper,Marsch,Hellinger1,Matteini2,Bale2}.
However, a first-principles calculation of the nonlinear evolution of 
the instabilities remains a theoretical imperative because, in order to 
construct the correct closure, we must 
understand the mechanism whereby the instabilities 
control the pressure anisotropy: do they scatter particles? do they modify 
the structure of the magnetic field? The calculation presented below 
will lead us to conclude that the latter mechanism is at work, at least 
in the simple case we are considering (see discussion in 
\secref{sec:scatt}), and indeed a sea of microscale 
magnetic fluctuations excited by the plasma instabilities will act to pin 
the plasma to marginal stability.  

In this paper, we present a theory of the nonlinear evolution of the 
simplest of the pressure-anisotropy-driven instabilities, the parallel 
($\kperp=0$) firehose instability and the gyrothermal instability \citep{GTI}. 
To be specific, we consider as our main application 
a plasma under physical conditions characteristic of galaxy clusters: 
weakly collisional, fully ionized, magnetized and approximately 
(locally) homogeneous. We will explain at the end the extent 
to which our results are likely to be useful in other contexts, 
e.g., accretion flows and the solar wind (\secref{sec:apps}). 

The plan of exposition is as follows. 
In \secref{sec:qualit}, we give an extended, qualitative, 
mostly low-analytical-intensity introduction to the problem, 
explain the relevant properties of the intracluster plasma (\secref{sec:clusters}), 
the origin of the pressure anisotropies (\secref{sec:aniso}),
sketch the linear theory of the firehose instability (\secref{sec:lin}),
the main principle of its nonlinear evolution (\secref{sec:nonlin}),
and show that a more complicated theory is necessary to work out 
the spatial structure of the resulting ``firehose turbulence'' (\secref{sec:FLR}).
In \secref{sec:main}, a systematic such theory is developed via 
asymptotic expansions of the electron and ion kinetics 
(the basic structure of the theory is outlined in the main 
part of the paper, while the detailed derivation is relegated 
to \apref{app:main}), culminating in a very simple one-dimensional equation  
for the nonlinear evolution of the firehose fluctuations (\secref{sec:eqn}), 
the study of which is undertaken in \secref{sec:firehose}. 
The results are a theoretical prediction for the nonlinear evolution 
and spectrum of the firehose turbulence (\secref{sec:results}) 
and some tentative conclusions about its effect on the momentum 
transport (\secref{sec:transp2}). 
In \secref{sec:GTI}, we extend this study to include the effect 
of parallel ion heat flux on the firehose turbulence: 
in the presence of a parallel ion temperature gradient, 
a new instability emerges (the gyrothermal instablity recently 
reported by \citealt{GTI} and recapitulated in \secref{sec:GTI_lin}) 
--- which, under some conditions, 
can take over from the firehose. For it as well, we develop a 
one-dimensional nonlinear equation (\secref{sec:GTI_eqn}), 
solve it to predict the nonlinear evolution and spatial 
structure of the gyrothermal turbulence (\secref{sec:GTI_nlin}) 
and discuss the implications for momentum transport (\secref{sec:GTI_transp}). 
A discussion of our results and of the ways in which they differ from previous work on 
firehose instability in collisionless plasmas 
is given in \secref{sec:diff}. A brief survey 
of astrophysical implications (both galaxy clusters 
and other contexts) follows in \secref{sec:apps}. 
Finally, \secref{sec:conc} contains a very concise 
summary of our findings and of the outlook for future work.
Note that while \secref{sec:qualit} is largely a pedagogical 
review of our earlier work \citep{Schek2,Schek3,Schek4}, 
most of the theory and results presented in 
\secsdash{sec:main}{sec:GTI} is new. 

A reader not interested in the technicalities of kinetic theory is advised 
to ignore \secref{sec:main} and \apref{app:main}. 
A reader only interested in the formal derivation may skip \secref{sec:qualit}, 
as \secref{sec:main} (supplemented by \apref{app:main}) and 
the sections that follow it can be read in a self-contained way. 

\section{Qualitative considerations}
\label{sec:qualit}

\subsection{Galaxy clusters: observations, questions, parameters}
\label{sec:clusters}

Galaxy clusters have long attracted the interest of both
theoreticians and observers both as dynamical systems in their own right
and as cosmological probes \citep{Bahcall,Peterson}. 
While gravitationally they are dominated by dark matter, 
most of their luminous matter is  
a hot, diffuse, fully ionized, X-ray emitting hydrogen plasma \citep{Sarazin} 
known as the intracluster medium, or ICM (the galaxies themselves 
are negligible both in terms of their mass and the volume they occupy). 
Crudely, we can think of an observable galaxy cluster as an 
amorphous blob of ICM about $1$~Mpc across, 
sitting in a gravitational well, with a density 
profile peaking at the center and decaying outwards. 
Observationally, on the crudest level, 
we know what the overall density and temperature 
profiles in clusters are \citep[e.g.,][]{Vikhlinin,Piffaretti,Leccardi,Cavagnolo}. 
Recent highly resolved X-ray observations 
reveal the ICM to be a rich, complicated,
multiscale structure displaying ripples, bubbles, filaments, waves, shocks, 
edges etc.\ \citep{Fabian1,Fabian_turb,Fabian2,Fabian4,Sanders1,Sanders2,Forman,Markevitch1}, 
temperature fluctuations \citep{Simionescu,Markevitch2,Fabian4,Million,Sanders3,Lagana}
and most probably also broad-band disordered turbulent motions 
\citep{Churazov3,Schuecker,Rebusco1,Rebusco2,Rebusco3,Graham,Sanders4,Sanders5,Ogrean}. 
Radio observations tell us that the ICM also hosts tangled 
magnetic fields, which are probably dynamically strong 
\citep{Carilli,Govoni1,Vogt,Kuchar,Clarke,Govoni2,Guidetti,Ferrari}. 

These and other observations motivate a number of questions about the ICM, 
which are representative of the problems  
generally posed for astrophysical plasma systems:\footnote{In \secref{sec:apps}, 
we will discuss some of the relevant questions for astrophysical contexts other 
than galaxy clusters. In \secref{sec:imp_clusters}, 
we will also give a brief survey of what in our view 
is the current state of play in answering the questions raised here in view 
of what we know about the plasma instabilities in the ICM and their likely 
saturation mechanisms.} 

\begin{itemize}

\item Can we explain the observed ICM temperature profiles,
in particular the apparent lack of a cooling catastrophe at the 
cluster core predicted by fluid models \citep{Fabian0,Binney,Peterson,Parrish2,Bogdanovic}? 
This requires modelling various heating processes involving 
conversion of the energy of plasma motions (turbulent or otherwise) 
into heat via some form of effective viscosity 
\citep[e.g.,][]{Omma1,Fabian3,Dennis,Chandran2,Guo,Brueggen,Kunz}, 
the dynamical effect of thermal instabilities arising in the magnetized ICM 
\citep{Balbus1,Parrish1,Quataert2,Sharma3,Sharma4,Parrish2,Parrish3,Bogdanovic,Ruszkowski2,GTI}, 
and the effective thermal conductivity 
of this medium with account taken of the tangled  
magnetic field \citep{Chandran1,Malyshkin,Narayan,Zakamska,Cho,Voigt}.

\item Can we construct theoretical and numerical models of 
the ICM dynamics that reproduce quantitatively the features we observe, 
e.g., the rise of radio bubbles \citep[e.g.,][]{Ruszkowski1,Dong},
the formation and propagation of shocks, fronts and sound waves, 
the structure of ICM velocity, density, temperature fluctuations? 

\item Can we explain the origin of the cluster magnetic 
fields (probably via some form of turbulent dynamo, see 
\citealt{Subramanian,Schek2,Ensslin,Xu}) 
and their observed spatial structure? 

\end{itemize}

Addressing these questions requires a theoretically sound 
mean-field theory for the ICM dynamics, i.e., a set of prescriptions 
for its effective transport properties (viscosity, thermal conductivity), 
which depend on the unresolved microphysics. 
Without such a theory, all we have is numerical simulations based 
on fluid models (see references above), which, while they can often 
be tuned to produce results that are visually similar to what is observed, 
are not entirely satisfactory because they lack a solid plasma-physical basis 
and because refining the numerical resolution often breaks the agreement 
with observations and requires retuning. 
A satisfactory transport theory is lacking because 
any plasma motions in the ICM that change the strength 
of the magnetic field trigger microscale plasma instabilities 
(see \secsand{sec:aniso}{sec:lin})
and we do not know what happens next. 

How some of these instabilities arise and evolve is discussed 
in greater detail below. In order to make this discussion more quantitative, 
we need to fix a few physical parameters that characterize the ICM.
In reality, these parameters vary considerably 
both between different clusters and within any individual cluster (as 
a function of radius: from the cooler, denser core to the hotter, more 
diffuse outer regions). However, for the purposes of this discussion, 
it is sufficient to adopt a set of fiducial values. 
Let us consider the plasma in the core of the Hydra A cluster
(also used as a representative example in our preceding papers, \citealt{Schek2,Schek4}), 
where the parameters are \citep{David,Ensslin}

\begin{itemize}

\item particle (ion and electron) number density 
\beq
n_i=n_e\sim6\times10^{-2}~\lx{cm}^{-3};
\eeq

\item measured electron temperature is 
\beq
T_e\sim 3\times10^7~\lx{K};
\eeq
the ion temperature is unknown, but assumed to be comparable, $T_i\sim T_e$;
then the ion thermal speed is 
\beq
\vthi=\lt(\frac{2T_i}{m_i}\rt)^{1/2}\sim7\times10^7~\lx{cm~s}^{-1}
\eeq
($m_i$ is the ion mass, $T_i$ is in erg); 
the ion Debye length is
\beq
\lambda_{Di} = \frac{\vthi}{\omega_{pi}} = \vthi\lt(\frac{4\pi e^2 n_i}{m_i}\rt)^{-1/2}
\sim 2\times10^5~\lx{cm};
\eeq

\item the ion-ion collision frequency (in seconds, assuming $n_i$ in cm$^{-3}$ and 
$T_i$ in K) is 
\beq
\nuii\sim 1.5 n_iT_i^{-3/2}\sim5\times10^{-13}~\lx{s}^{-1};
\eeq 
consequently the mean free path is 
\beq
\mfp=\frac{\vthi}{\nuii}\sim1.3\times10^{20}~\lx{cm};
\label{eq:mfp}
\eeq

\item the rms magnetic field strength is \citep{Vogt} 
\beq
B\sim7\times10^{-6}~\lx{G}; 
\eeq
consequently the plasma (ion) beta is 
\beq
\beta_i = \frac{8\pi n_iT_i}{B^2}\sim130,
\label{eq:beta}
\eeq 
the ion cyclotron frequency is 
\beq
\Omega_i = \frac{eB}{m_ic} \sim 0.07~\lx{s}^{-1} 
\eeq
($e$ is the elementary charge, $c$ the speed of light) 
and the ion Larmor radius is 
\beq
\rho_i = \frac{\vthi}{\Omega_i} \sim 10^9~\lx{cm};
\label{eq:rhoi_ICM}
\eeq
note that the magnetized-plasma condition $\rho_i\ll\mfp$ 
is satisfied extremely well;

\item the typical velocity of the plasma motions is 
\beq
U\sim2.5\times10^7~\lx{cm~s}^{-1}
\label{eq:U}
\eeq 
\citep[cf.][who consider a sample of clusters]{Sanders4,Sanders5},
while the typical length scale of these motions is 
\beq
L\sim2\times10^{22}~\lx{cm};
\label{eq:L}
\eeq 
consequently the Mach number is 
\beq
M=\frac{U}{\vthi}\sim0.3
\label{eq:M}
\eeq
(so the motions are subsonic, hence approximately incompressible 
on scales smaller than that of the mean density variation) 
and the Reynolds number based on collisional parallel viscosity is 
\beq
\Re=\frac{LU}{\mfp\vthi}\sim60, 
\label{eq:Re}
\eeq
assuming Kolmogorov scalings for turbulence, the viscous cutoff scale is
\beq
l\sim L\Re^{-3/4}\sim 10^{21}~\lx{cm}
\label{eq:l}
\eeq 
and the typical velocity at this scale is 
\beq
u \sim U\Re^{-1/4} \sim 10^7~\lx{cm~s}^{-1},
\label{eq:u}
\eeq
so the approximate rms rate of strain 
(assuming a viscous cutoff for the motions) is
\beq 
\gamma_0\sim \frac{u}{l}\sim\frac{U}{L}\,\Re^{1/2}\sim10^{-14}~\lx{s}^{-1}.
\label{eq:gamma0}
\eeq

\end{itemize}

\subsection{Origin of pressure anisotropy}
\label{sec:aniso}

If we consider length scales greater than $\rho_i$ and time scales longer 
than $\Omega_i$ 
(which is easily true for any large-scale dynamical processes in the ICM), 
the momentum equation for the plasma flow, characterized by the mean velocity
$\vu$, is \citep[e.g.,][]{Kulsrud} 
\bea
\nonumber
m_i n_i \frac{\did \vu}{\did t} &=& -\vdel \lt(\pperp + \frac{B^2}{8 \pi}\rt)\\ 
&&+\,\, \vdel \cdot \lt[\vb\vb \lt(\pperp - \ppar +\frac{B^2}{4\pi}\rt) \rt], 
\label{eq:gen_mom}
\eea
where ${\did \vu}/{\did t} = \dd/\dd t + \vu\cdot\vdel$ 
is the convective derivative, $\vb$ is the unit vector in the direction 
of the local magnetic field, $B$ is the field's strength, and 
$\pperp$ and $\ppar$ are the perpendicular and parallel plasma pressure, 
which are the only components of the plasma pressure tensor that survive 
at these long spatial and temporal scales:
\bea
\vP = \sum_s m_s\int\did^3\vv\,\vv\vv\,f_s &=& 
\pperp\lt(\textbfss{I} - \vb\vb\rt) + \ppar\vb\vb,\\
\pperp &=& \sum_s m_s\int\did^3\vv\,\frac{\vperp^2}{2}\, f_s,\\
\ppar &=& \sum_s m_s\int\did^3\vv\,\vpar^2 f_s,
\eea 
where $f_s$ is the distribution function for species $s$ ($s=i,e$), 
$\vv$ its velocity variable (particle's peculiar velocity), 
and $\vperp$ and $\vpar$ the projections of $\vv$
perpendicular and parallel to the magnetic field. 
The magnetic field is determined by the 
combination of Faraday's and Ohm's laws, which at these long scales 
takes the form of the ideal induction equation
\beq
\frac{\did \vB}{\did t} = \vB\cdot\vdel\vu - \vB\vdel\cdot\vu.
\label{eq:gen_ind}
\eeq

Without as yet going into the technicalities of kinetic theory, 
it is not hard to show that pressure anisotropies arise naturally 
in a weakly collisional plasma. Indeed, the first 
adiabatic invariant $\mu = \vperp^2/2B$ of a gyrating particle 
is conserved on time scales intermediate between the collision 
time and the cyclotron period (a nonempty interval when plasma is 
magnetized, $\nuii\ll\Omega_i$). Since $\pperp$ is proportional 
to the sum of the values of $\mu$ for all particles, $\pperp/B$ should 
be a conserved quantity, i.e., if the magnetic field changes (as a result 
of plasma motions into which the flux is frozen, see \eqref{eq:gen_ind}) 
then $\pperp$ should change accordingly. For the purposes of this qualitative 
discussion, we may momentarily ignore the fact that changing $B$ also causes 
$\ppar$ to change (in a different way from $\pperp$; see \apref{app:CGL}) 
and so conclude that changing $B$ will cause pressure anisotropies to develop. 

In the absence of collisions, the pressure anisotropies would track 
the field strength. If collisions do occur, even weakly, their 
effect will be to relax the system towards an isotropic pressure 
(and a Maxwellian distribution). Thus, there is a 
competition between changing $B$ inducing anisotropy 
and collisions causing isotropization. 
This can be modelled by the following heuristic equation:
\bea
\nonumber
\frac{1}{\pperp} \frac{\did \pperp}{\did t} &\sim& 
\frac{1}{B}\frac{\did B}{\did t} - \nuii\frac{\pperp-\ppar}{\pperp}\\
&=& \vb\vb:\vdel\vu - \nuii\frac{\pperp-\ppar}{\pperp},
\label{eq:Pressure} 
\eea
where we have used \eqref{eq:gen_ind} to express the change in the 
field strength in terms of the plasma flow velocity and assumed, 
for the purposes of this qualitative discussion, that plasma 
density is constant (i.e., the motions are incompressible). 
Considering what happens on time scales longer 
than the collision time, we conclude, after examining the right-hand 
side of \eqref{eq:Pressure}, that we should expect the typical 
(ion) pressure anisotropy in a moving plasma to be 
\beq
\Delta = \frac{\pperp-\ppar}{\pperp} \sim 
\frac{1}{\nuii}\frac{1}{B}\frac{\did B}{\did t} 
\sim \frac{\gamma_0}{\nuii},
\label{eq:Delta}
\eeq
where $\gamma_0$ is the typical rate of strain of the plasma 
motion.\footnote{A few tangential comments are appropriate here: 
\label{fn:comments} 
\begin{enumerate}
\item The electron pressure anisotropy is smaller by 
a factor of $\sim43$ because the electron collision frequency is 
$\sim(m_i/m_e)^{1/2}\nuii$. 
\item If we use \eqref{eq:Pressure} to write explicitly 
$\pperp-\ppar = (\pperp/\nuii)\vb\vb:\vdel\vu$ and substitute this 
into \eqref{eq:gen_mom}, we recover (to lowest order in $\nuii/\Omega_i$) 
the well known \citet{Braginskii} momentum equation with anisotropic viscosity, 
where $\pperp/\nuii\sim m_i n_i \vthi^2/\nuii$ is the Braginskii parallel 
viscosity coefficient.
\item If a Kolmogorov-style turbulence is assumed to exist in the ICM, 
the typical rate of strain $\gamma_0$ will be dominated by the motions 
at the viscous cutoff scale. However, as we saw in \secref{sec:clusters}, 
the Reynolds-number estimates for ICM do not give very large values 
and one might wonder whether calling these motions turbulence 
is justified \citep{Fabian_turb}. However, for our purposes, 
it is not important whether the rate of strain is provided 
by the viscous cutoff of a turbulent cascade or by a single-scale 
motion because either can change the magnetic field and thus cause pressure 
anisotropy \citep{Schek2}. 
\item For a purely compressive motion, $\Delta\sim-\vdel\cdot\vu/3\nuii$ 
(i.e., the anisotropy is still related to the change in the magnetic-field strength; 
see \eqref{eq:gen_ind}), but one has to work a little harder to show this. 
In the compressible case, one also discovers that heat fluxes contribute to 
the anisotropy alongside velocity gradients 
(this is done in \apref{app:aniso}; see \eqref{eq:D0}).
\end{enumerate}}
Thus, the pressure anisotropy is regulated by the ratio of the typical rate 
of change of the magnetic-field strength to the collision frequency.  

Substituting the numbers from \secref{sec:clusters}, we find 
that $|\Delta|\sim0.02$ in the core of Hydra A. Is this a large number? 
It turns out that it is a huge number because such anisotropies will 
make the plasma motion violently unstable. 

\subsection{Firehose instability}
\label{sec:lin}

While the full description of the plasma instabilities triggered by pressure 
anisotropies requires kinetic treatment, it is extremely straightforward 
to deduce the presence of the firehose instability directly from \eqref{eq:gen_mom}. 

Consider some ``fluid'' solution $(\vu_0,\vB_0,p_{0\perp},p_{0\parallel})$ of 
\eqsand{eq:gen_mom}{eq:gen_ind} that varies on long time and spatial scales --- that can be 
thought of as the turbulence and/or some regular magnetofluid motion caused by global 
dynamics. Let us now examine the linear stability of this solution with respect to 
high-frequency ($\omega\gg|\vdel\vu_0|$), short-scale ($k\gg|\vdel\vu_0|/u_0$)
perturbations $(\dvu,\dvB,\dpperp,\dppar)$. Mathematically, this is simply 
equivalent to perturbing a straight-magnetic-field equilibrium of \eqsand{eq:gen_mom}{eq:gen_ind}:
\bea
-m_i n_i\omega \dvu &=& -\vkperp\lt(\dpperp + \frac{B_0\dBpar}{4\pi}\rt) 
\nonumber \\
&& +\,\, \kpar \dvb \lt( p_{0\perp}-p_{0\parallel} + \frac{B_0^2}{4\pi}\rt) 
\nonumber \\  
&& -\,\,\kpar\vb_0 \lt[\dppar + \lt(p_{0\perp}-p_{0\parallel}\rt)\frac{\dBpar}{B_0}\rt],\\ 
\label{eq:lin_mom}
-\omega\frac{\dvB}{B_0} &=& \kpar\dvu - \vb_0\lt(\vk\cdot\dvu\rt),
\label{eq:lin_ind}
\eea
where $\dvb=\dvBperp/B_0$, we have used $\vk\cdot\dvb=-\kpar\dBpar/B_0$ (from $\vdel\cdot\vB=0$), 
and $\perp$ and $\parallel$ are with respect to the unperturbed magnetic field 
direction $\vb_0$. 
Pressure perturbations can only be calculated from the linearized kinetic 
equation \citep[see, e.g.,][]{Schek3}, but even without knowing them, we find 
that for the Alfv\'enically polarized modes, $\dvu\propto\vb_0\times\vk$, 
the dispersion relation is 
\beq
\omega = \pm\kpar\lt(\frac{p_{0\perp}-p_{0\parallel}}{m_i n_i} + v_A^2\rt)^{1/2}
\!\!\!= \pm\kpar c_s\lt(\Delta + \frac{2}{\beta}\rt)^{1/2},
\label{eq:disp_rel}
\eeq
where $v_A= B_0/\sqrt{4\pi m_in_i}$, $c_s = (p_{0\perp}/m_in_i)^{1/2}$, 
$\Delta = (p_{0\perp}-p_{0\parallel})/p_{0\perp}$ and 
$\beta = 8\pi p_{0\perp}/B_0^2$.  

\Eqref{eq:disp_rel} is simply the dispersion relation for Alfv\'en waves 
with a phase speed modified by the pressure anisotropy. If the pressure 
anisotropy is negative, $\Delta<0$, the associated stress opposes the Maxwell 
stress (the magnetic tension force), the magnetic-field lines become more easily 
deformable, the Alfv\'en wave slows down and, for $\Delta<-2/\beta$, turns into a 
nonpropagating unstable mode --- this is the firehose instability 
\citep{Rosenbluth,Chandra,Parker,Vedenov1,Vedenov2}. 
Its growth rate can, in general, be almost 
as large as the ion cyclotron frequency as $\kpar\rho_i$ approaches 
finite values (see \secref{sec:FLR}). 
For the ICM parameters given in \secref{sec:clusters},  
the instability is, therefore, many orders of magnitude faster than either 
the large-scale dynamics (typical turnover rate $\sim|\nabla\vu_0|\sim\gamma_0$) 
or collisions (typical rate $\nuii$). 

Thus, any large-scale motion that leads to a local decrease in the 
strength of the magnetic field\footnote{While turbulence on the average 
is expected to lead to the growth of the magnetic field (the dynamo 
effect; see, e.g., \citealt{Schek2} and references therein), 
locally there will always be regions where the field strength 
(temporarily) decreases. Decrease of the field and, consequently, negative 
pressure anisotropy can also result from expanding motion, which decreases 
the density of the plasma --- as, e.g., in the solar wind.} 
gives rise to a negative pressure anisotropy, 
which, in turn triggers the firehose instability, producing 
Alfv\'enically polarized fluctuations at small parallel scales 
--- unless the plasma beta is sufficiently low (magnetic field is 
sufficiently strong) for the magnetic tension to stabilize these 
fluctuations. Using the typical size of $\Delta$ estimated at the 
end of \secref{sec:aniso} for the Hydra A ICM parameters, 
we find that the typical beta below which the firehose is stable 
is $\beta\sim100$, which is quite close to the measured 
value (see \secref{sec:clusters}) --- perhaps not a coincidence? 

Positive pressure anisotropies also lead to instabilities (most 
importantly, mirror; see \citealt{Furth,Barnes,Tajiri,Hasegawa,Southwood,Hellinger2} 
and references therein), but they involve resonant particles and are 
mathematically harder to handle. 
We will not discuss them here \citep[see][]{Schek3,Schek4,Rincon}. 

\subsection{Nonlinear evolution of the firehose instability}
\label{sec:nonlin}

A nonlinear theory of the firehose instability can be constructed 
via a quasilinear approach, in which the unstable small-scale (perpendicular) 
fluctuations of the magnetic field on the average change the local 
magnetic-field strength and effectively cancel the pressure 
anisotropy \citep{Schek4}. In \eqref{eq:Delta}, let us 
treat the changing magnetic field as the sum of the large-scale 
field and the small-scale firehose fluctuations: $\vB=\vB_0 + \dvBperp$. 
Then the field strength averaged over small scales is 
\beq
\Bbar \approx B_0\lt(1 + \frac{1}{2}\overline\frac{|\dvBperp|^2}{B_0^2}\rt),
\label{eq:Bbar}
\eeq
where the overbar denotes the average (under which small-scale fluctuations 
vanish). The contribution from $\dvBperp$ is small, but 
for large enough $\kpar$, it is growing at a greater 
rate than the rate of change of the large-scale field, 
so its time derivative can be comparable to the time derivative of $B_0$. 
As $B_0$ is assumed to be decreasing, the growth of the fluctuations 
can then cancel this decrease and drive the total average pressure 
anisotropy to the marginal level, $\Delta = -2/\beta$. 
From \eqref{eq:Delta}, we get
\beq
\Delta \sim \frac{1}{\nuii}\lt(\frac{1}{B_0}\frac{\did B_0}{\did t} 
+ \frac{1}{2}\frac{\did}{\did t}\overline\frac{|\dvBperp|^2}{B_0^2}\rt)
= -\frac{2}{\beta}.
\label{eq:Delta_marg}
\eeq
The rate of change of $B_0$ is the typical rate of strain of the (large-scale) motion, 
$(1/B_0)\did B_0/\did t\sim-|\gamma_0|$.
The firehose growth rate $\gamma=-i\omega$ is given by \eqref{eq:disp_rel}. 
As long as the firehose fluctuations are smaller than the critical level
\beq 
\overline{\frac{|\dvBperp|^2}{B_0^2}}\sim\frac{|\gamma_0|}{\gamma}, 
\label{eq:crit}
\eeq
they cannot enforce the marginality condition expressed by \eqref{eq:Delta_marg}
and will continue growing until they reach the required strength 
(which is still small compared to the large-scale field because 
$|\gamma_0|/\gamma\ll1$ for sufficiently large $\kpar$). 
After that, their evolution becomes nonlinear and is determined by \eqref{eq:Delta_marg}, 
whence we find that their energy has to grow secularly:
\beq
\overline\frac{|\dvBperp|^2}{B_0^2} \sim \lt(|\gamma_0| - \frac{2\nuii}{\beta}\rt)t.
\label{eq:secular}
\eeq
As long as the large-scale field keeps decreasing, 
the small-scale fluctuation energy cannot saturate because if it did, 
its time derivative would vanish, the anisotropy would drop below marginal and 
the instability would come back. 

The secular growth given by \eqref{eq:secular} 
leads to $\delta B_\perp/B_0\sim1$ after roughly one turnover time ($\sim|\gamma_0|^{-1}$) 
of the large-scale background motion that produces the anisotropy in the first place --- thus, 
the magnetic field can develop order-unity fluctuations before this background motion decorrelates. 
What all this means for the large-scale dynamics on longer timescales, we do not know.   

In what follows, we will be guided by the simple ideas outlined above 
in constructing a more rigorous kinetic theory of the nonlinear firehose instability. 

\subsection{Effect of finite Larmor radius}
\label{sec:FLR}

We have so far carefully avoided discussing the magnitude of the 
wavenumber $\kpar$ of the firehose fluctuations, simply referring to them 
as ``small-scale,'' with the implication that their scale would be 
smaller than that of the background fluid dynamics that cause the instability. 
Examining the dispersion relation \exref{eq:disp_rel}, we see that 
the growth rate of the instability is proportional to $\kpar$, so 
the smaller the scale the faster the instability. This ultraviolet 
catastrophe cannot be resolved within the long-wavelength approximation, 
$k\rho_i\ll1$, in which \eqref{eq:gen_mom} is derived,\footnote{Which means 
that the equation is ill posed and cannot be solved without some 
kinetic prescription for the handling of small scales.} 
so finite-Larmor-radius (FLR) corrections must be brought in. 

Direct calculation of the linear firehose growth rate from the hot-plasma 
dispersion relation shows that the peak of the growth rate is 
at $\kpar\rho_i\sim|\Delta+2/\beta|^{1/2}$ for the parallel ($\kperp=0$) firehose 
\citep[][this result will emerge in \secref{sec:disp_rln}]{Kennel,Davidson} 
and, in general, at $k\rho_i\sim1$ for the oblique firehose with $\kperp\neq0$ 
\citep{Yoon,Hellinger3}. 
This means that the maximum growth rate of the instability is 
$\gmax \sim |\Delta + 2/\beta|\Omega_i \sim 10^{-3}$~s$^{-1}$ for $\kperp=0$ 
(see \secref{sec:disp_rln})
and $\gmax \sim |\Delta + 2/\beta|^{1/2}\Omega_i \sim 10^{-2}$~s$^{-1}$ for $\kperp\neq0$, 
where we have used the ICM parameters of \secref{sec:clusters} and 
the estimate of $\Delta$ from \secref{sec:aniso}. 

There are two conclusions to be drawn from this. 
First, the linear instability is enormously fast compared with the large-scale 
dynamics that cause it, so its nonlinear behaviour must be fundamentally 
important at all times. Second, in order to understand the spatial 
structure of the firehose fluctuations, we need a theory that takes the FLR 
effects explicitly into account because it is the FLR that sets the scale 
and the growth rate of the fastest-growing mode. 
We now proceed to construct such a theory for the simplest case --- 
the parallel ($\kperp=0$) firehose instability. 

\section{Kinetic theory}
\label{sec:main}


\subsection{Basic equations}
\label{sec:eqns}

The distribution function $f_s(t,\vr,\vv)$ satisfies the 
Vlasov-Landau kinetic equation
\beq
\frac{\dd f_s}{\dd t} + \vv\cdot\vdel f_s + 
\frac{q_s}{m_s}\lt(\vE + \frac{\vv\times\vB}{c}\rt)\cdot\frac{\dd f_s}{\dd\vv} 
= \St{f_s},
\label{eq:Vlasov}
\eeq
where $s=i,e$ is the particle species, $\vr$ its position, $\vv$ velocity, 
$q_s$ and $m_s$ are the charge and mass of the particle of species $s$ 
($q_e=-e$, $q_i=Ze$, $Z=1$ for hydrogen plasma), 
$\vE$ and $\vB$ are the electric and magnetic fields, 
and the term on the right-hand side is the collision operator. 
The electric and magnetic fields are determined from Maxwell's equations: 
quasineutrality 
\beq
\sum_s q_s n_s \equiv \sum_s q_s \int\did^3\vv\,f_s = 0
\label{eq:quasineut}
\eeq
($n_s$ is particle number density), 
Amp\`ere's law
\beq
\vj = \sum_s q_s n_s \vu_s \equiv \sum_s q_s\int\did^3\vv\,\vv\,f_s
= \frac{c}{4\pi}\vdel\times\vB
\label{eq:Ampere}
\eeq
($\vj$ is current density, $\vu_s$ is the mean velocity of the species $s$),
Faraday's law
\beq
\frac{\dd\vB}{\dd t} = -c\vdel\times\vE,
\label{eq:Faraday}
\eeq 
and $\vdel\cdot\vB=0$. 
Note that \eqsand{eq:quasineut}{eq:Ampere} are valid as long the 
particle motion is nonrelativistic and the scales we are interested 
in are larger than the Debye length. 

It is convenient for what follows to calculate the distribution 
function in terms of peculiar velocities $\vv'=\vv-\vu_s(t,\vr)$.
Transforming the variables $(t,\vr,\vv)\to(t,\vr,\vv')$, we find that 
\eqref{eq:Vlasov} takes the form
\bea
\nonumber
\frac{\dd f_s}{\dd t} &+& \vu_s\cdot\vdel f_s + \vv'\cdot\vdel f_s\\ 
&+& \lt[\frac{q_s}{m_s}\lt(\vE +  \frac{\vu_s\times\vB}{c} + \frac{\vv'\times\vB}{c}\rt)\rt.
\nonumber\\ 
&-&\lt.\frac{\dd\vu_s}{\dd t} - \vu_s\cdot\vdel\vu_s - \vv'\cdot\vdel\vu_s\rt]
\cdot\frac{\dd f_s}{\dd\vv'} = \St{f_s}.
\label{eq:Vlasovprime}
\eea
We will henceforth drop the primes, $\vv$ will be the peculiar velocity in all that follows. 
In this new formulation, the strategy for solving \eqsdash{eq:quasineut}{eq:Vlasovprime} 
is as follows. 

\subsection{Electron kinetics: Ohm's law and induction equation}
\label{sec:Ohm}

The electron kinetic equation can be expanded in the square root of the electron-ion mass ratio 
$(m_e/m_i)^{1/2}\approx0.02$, a natural small parameter for plasma. This expansion is carried 
out in \apref{app:els}, where we also explain what assumptions have to be made 
in order for it to be valid. The outcome of the mass-ratio expansion 
is that electrons are Maxwellian,\footnote{This means they do not contribute to the 
pressure anisotropy, which, to lowest order in the mass ratio, they indeed should not do, 
as pointed out already in footnote \ref{fn:comments}. Note that the validity of these 
statements depends on the ordering of the collision frequencies given by 
\eqref{eq:elcoll_order}.} 
isothermal ($T_e=\const$), and the electric field can be determined
in terms of $\vu_e$, $\vB$ and $n_e$ via a generalized Ohm's law: 
\beq
\vE + \frac{\vu_e\times\vB}{c} = -\frac{\vdel p_e}{en_e} = -\frac{T_e\vdel n_e}{e n_e}.
\label{eq:Ohm}
\eeq
This can now be recast in terms of moments of the ion distribution: 
from \eqref{eq:quasineut}, 
\beq
n_e=Zn_i
\eeq
and from \eqref{eq:Ampere}, 
\beq 
\vu_e = \vu_i - \frac{\vj}{en_e} = \vu_i - \frac{c}{4\pi e n_e}\,\vdel\times\vB,
\eeq
so \eqref{eq:Ohm} becomes
\beq
\vE + \frac{\vu_i\times\vB}{c} = - \frac{T_e\vdel n_i}{e n_i}
+ \frac{\lt(\vdel\times\vB\rt)\times\vB}{4\pi Ze n_i} 
\label{eq:Ohm_ions}
\eeq
and Faraday's law \exref{eq:Faraday} takes the form of the 
standard induction equation with a Hall term:
\beq
\frac{\dd\vB}{\dd t} = \vdel\times\lt[\lt(\vu_i 
- \frac{c}{4\pi Ze n_i}\,\vdel\times\vB\rt)\times\vB\rt].
\label{eq:ind_gen}
\eeq

\subsection{Ion kinetics: continuity and momentum equations}

To close this set of equations, we must determine $n_i$ 
and $\vu_i$. Integrating \eqref{eq:Vlasovprime}, 
we find that $n_i= \int\did^3\vv\, f_i$ 
satisfies the continuity equation
\beq
\frac{\dd n_i}{\dd t} + \vdel\cdot\lt(n_i\vu_i\rt) = 0.
\label{eq:cont}
\eeq 
The equation for $\vu_i$ (the ion momentum equation) follows 
from \eqref{eq:Vlasovprime} for $s=i$, by taking the $\vv$ moment and 
enforcing $\int\did^3\vv\,\vv\,f_i = 0$ (by definition of the peculiar 
velocity $\vv$), which gives
\beq
\frac{\dd\vu_i}{\dd t} + \vu_i\cdot\vdel\vu_i 
= - \frac{\vdel\cdot\vP_i}{m_i n_i} - \frac{ZT_e\vdel n_i}{m_i n_i} 
+ \frac{\lt(\vdel\times\vB\rt)\times\vB}{4\pi m_i n_i},
\label{eq:ionmom}
\eeq
where we have used \eqref{eq:Ohm_ions}, 
the second term on the right-hand side is the electron pressure gradient, 
and we have introduced the ion pressure tensor
\beq
\vP_i = m_i\int\did^3\vv\,\vv\vv f_i.
\eeq
It is in order to calculate $\vP_i$ in terms of $\vu_i$ and $\vB$ 
that we must solve the ion kinetic equation. We do this
by means of an asymptotic expansion in a physical small parameter. 

\subsection{Asymptotic ordering}
\label{sec:order}

The small parameter we will use is expressed in terms of the Mach 
and Reynolds numbers \citep{Schek3,Schek4}:\footnote{As already pointed 
out in footnote \ref{fn:comments}, our considerations do not depend 
on $\Re$ being large. If a single-scale flow is considered, 
our expansion is simply an expansion in Mach number.} 
\beq
\eps = \frac{M}{\Re^{1/4}} \sim 0.1,
\label{eq:eps}
\eeq
where we used the ICM parameters of \secref{sec:clusters}. 
This is the natural small parameter for the plasma 
motions because, using \eqsdash{eq:U}{eq:gamma0}, it is easy 
to see that
\beq
\frac{u}{\vthi}\sim\frac{\mfp}{l}\sim\eps,
\eeq
where $l$ is the viscous scale and $u$ the typical flow velocity 
at this scale. The typical rate of strain $\gamma_0\sim u/l$ 
is the relevant parameter for determining the size of the pressure anisotropy 
because, even though the viscous cutoff we are using is based on the parallel 
collisional viscosity and so motions can exist below this scale, 
these motions do not change the strength of the magnetic field 
\citep[see][]{Schek2}.\footnote{This statement applies to macroscopic 
motions: for example, Alfv\'enic turbulence below the parallel viscous 
scale that can occupy a wide range of scales all the way down to 
the ion Larmor scale \citep[e.g.,][]{Schek5}. The fast, microscale 
plasma fluctuations triggered by plasma instabilities, including 
the firehose fluctuations that will be considered in this paper, will, on 
the average, change the field strength (see \secref{sec:nonlin}). 
Accordingly, their ordering [\eqref{eq:B1order}] 
will be arranged in precisely such a way that they are able to have an effect 
comparable to the macroscale motions that produce $\gamma_0$.} 
Thus, the pressure anisotropy is (from \eqref{eq:Delta})
\beq
\Delta \sim \frac{\gamma_0}{\nuii} \sim \frac{u}{\vthi}\frac{\mfp}{l} \sim \eps^2.
\eeq

We solve the ion kinetic equation by asymptotic expansion in $\eps$. 
All ion quantities are expanded in $\eps$, so
\bea
\label{eq:fi_exp}
f_i &=& f_{0i} + f_{1i} + f_{2i} + f_{3i} + \cdots,\\
n_i &=& n_{0i} + n_{1i} + n_{2i} + n_{3i} + \cdots,\\
\vu_i &=& \vu_{0i} + \vu_{1i} + \cdots,\\
\vB &=& \vB_{0} + \vB_1 + \cdots.
\eea
The lowest-order quantities $n_{0i}$, $\vu_{0i}$, $\vB_0$ are 
associated with the motions that produce the pressure anisotropy and have 
the length scale $l$ and time scale $\gamma_0$, so we order 
\bea
\vu_{0i}\sim\eps\vthi,\quad
\vdel\vu_{0i}\sim\gamma_0\sim\eps^2\nuii.
\label{eq:order_u0}
\eea
Since the instability parameter is $\Delta + 2/\beta_i$, we must order 
$\vB_0$ so that 
\bea
\label{eq:Delta_order}
\frac{2}{\beta_i}\sim\Delta\sim\eps^2
\quad\Rightarrow\quad
\frac{B_0}{\sqrt{4\pi m_i n_{0i}}}\sim \eps\vthi.
\eea
The perturbations $n_{1i}$, $\vu_{1i}$, $\vB_1$ around this slow large-scale 
dynamics are assumed to be excited by the prallel ($\kperp=0$) firehose instability 
and have much shorter spatial and time scales. 
Their typical wavenumber is the one at which 
the instability's growth rate peaks 
and their time scale is set by this maximum growth rate
(see \secref{sec:FLR}): 
\beq
\kpar\rho_i \sim \lt|\Delta + \frac{2}{\beta_i}\rt|^{1/2}\!\!\! \sim \eps,
\quad
\gamma \sim \lt|\Delta + \frac{2}{\beta_i}\rt|\Omega_i \sim \eps^2\Omega_i.
\label{eq:order_k}
\eeq

In order to be able to proceed, we must order the time scales of 
the lowest-order (``equilibrium'') fields and of the fluctuations with respect 
to each other. 
Physically, they depend on different things and are not intrinsically related. 
However, our {\em a priori} consideration of the nonlinear evolution 
of the instability (\secref{sec:nonlin}) suggests that for the nonlinearity 
to become important, we must have (see \eqref{eq:crit})
\beq
\frac{\vB_1}{B_0} \sim \lt(\frac{\gamma_0}{\gamma}\rt)^{1/2}.
\label{eq:B1order}
\eeq
Since $\vB_1\sim\eps\vB_0$, this tells us that we must order
\beq
\gamma_0\sim\eps^2\gamma\sim\eps^4\Omega_i
\quad\Rightarrow\quad
\nuii\sim\eps^2\Omega_i,
\quad
\rho_i\sim\eps^2\mfp.
\label{eq:connection}
\eeq
These relations are, of course, not strictly right in the quantitative 
sense --- the Larmor radius is grossly overestimated here if we 
take the value of $\eps$ for the ICM given by \eqref{eq:eps} and then 
compare what \eqref{eq:connection} gives us as the value 
of $\rho_i$ with the ICM estimate in \secref{sec:clusters} 
(\eqref{eq:rhoi_ICM}). 
However, ordering $\rho_i$ this way allows us to capture 
all the important physics in our formal expansion. 
We will also argue in \secref{sec:nlin_qualit} that this 
ordering of the finite Larmor radius physics   
gets quantitatively better as the nonlinear regime proceeds
(see footnote \ref{fn:alt_order}). 
By the same token, the growth rate of the instability 
in the ICM is typically much larger than the collision rate, 
while we have ordered them similar --- but again, this ordering 
formally allows all the important physical effects 
to enter on a par with each other and also gets better in the 
nonlinear regime, where the firehose fluctuations grow slower.  

Let us summarize our ordering of the relevant time and spatial scales 
compared to $\kpar\vthi$ and $\kpar$, respectively: 
using \eqsand{eq:connection}{eq:order_k}, we have
\bea
\gamma_0\sim \eps^3\kpar\vthi,&\ &
\gamma\sim\nuii\sim\eps\kpar\vthi,\quad
\Omega_i\sim\eps^{-1}\kpar\vthi,\\
l^{-1}\sim\eps^2\kpar,&\ &
\mfp^{-1}\sim\eps\kpar,\quad
\rho_i^{-1}\sim \eps^{-1}\kpar. 
\eea

\subsection{Firehose fluctuations}
\label{sec:polarization}

The ordering we adopted, inasmuch as it concerns the properties of the 
firehose fluctuations, applies to the parallel firehose only, so we 
now explicitly restrict our consideration to the case 
of $\vdel_\perp=0$ for all first-order perturbations. 
Since $\vdel\cdot\vB=0$, this immediately implies 
\beq
B_1^\parallel=0, 
\label{eq:Bpar}
\eeq
so $\vB_1=\vBp$. 
Here and in what follows, $\parallel$ and $\perp$ refer to directions 
with respect to the unperturbed field $\vB_0$. 

The induction equation \exref{eq:ind_gen}, taken to the lowest order
in $\eps$, gives
\beq
\frac{\did}{\did t}\vBfrac = \dpar\vup
\label{eq:indperp}
\eeq
(all terms here are order $\eps^2\kpar\vthi$; see \secref{sec:order}; 
note that the Hall term in \eqref{eq:ind_gen} is subdominant by two 
orders of $\eps$). Here $\did/\did t = \dd/\dd t + \vu_0\cdot\vdel$ 
is the convective derivative, but, since $\vdel\vu_0\sim\eps^3\kpar\vthi$, 
the shearing of the perturbed field due to 
the variation of $\vu_0$ is negligible and we can replace 
$\did/\did t$ by $\dd/\dd t$ by transforming into the frame moving 
with velocity~$\vu_0$. 

In the continuity equation \exref{eq:cont} taken to the lowest order in $\eps$, 
setting $\vdel_\perp=0$ gives 
\beq
\frac{\did}{\did t}\frac{n_{1i}}{n_{0i}} = -\dpar u_{1i}^\parallel 
\eeq
(all terms are order $\eps^2\kpar\vthi$). 
Anticipating the form of the unstable perturbation, we will set 
\beq
n_{1i}=0,\quad u_{1i}^\parallel = 0
\label{eq:n1}
\eeq
without loss of generality. 
In \apref{app:f1}, we will explicitly prove that $n_{1i}=0$. 
In \apref{app:f2}, we will learn that $n_{2i}=0$ as well. 

Consider now  the ion momentum equation \exref{eq:ionmom}. 
In the lowest order of the $\eps$ expansion 
(terms of order $\eps\kpar\vthi^2$), it gives, upon using \eqref{eq:n1},  
\beq
\vdel\cdot\vP_{1i}=0. 
\eeq 
We will learn in \apref{app:grads} that this can be strengthened to set
\beq
\vP_{1i}=0. 
\label{eq:P1}
\eeq
In the next order ($\eps^2\kpar\vthi$), we get (using $n_{2i}=0$)
\beq 
\vdel\cdot\vP_{0i} + \vdel\cdot\vP_{2i} + ZT_e\vdel n_{0i} = 0.
\label{eq:mom2}
\eeq 
Averaging this over small scales eliminates the perturbed quantities, 
so we learn\footnote{This is simply the pressure balance for the large-scale 
dynamics, an expected outcome for a system with low Mach number. In \apref{app:Max}, 
we will show that the zeroth-order distribution is Maxwellian, so the pressure associated 
with it is a scalar, $p_{0i}=n_{0i}T_{0i}$, and \eqref{eq:P0} becomes 
$(T_{0i}+ZT_e)\vdel n_{0i} + n_{0i}\vdel T_{0i}=0$. 
Further discussion of the role played by the ion temperature gradient 
can be found in \secref{sec:GTI}.\label{fn:eqgrad}} 
\beq
\vdel\cdot\vP_{0i} + ZT_e\vdel n_{0i} = 0
\label{eq:P0}
\eeq
and, therefore, from \eqref{eq:mom2}, also 
\beq
\vdel\cdot\vP_{2i}=0
\label{eq:P2}
\eeq 
(confirmed in \apref{app:f2}). 
Finally, in the third order ($\eps^3\kpar\vthi^2$), 
the perpendicular part of \eqref{eq:ionmom} 
determines the perturbed velocity field:
\beq
\frac{\did \vup}{\did t} = -\frac{\lt(\vdel\cdot\vP_{3i}\rt)_\perp}{m_i n_{0i}} 
+ v_A^2\dpar\vBfrac,
\label{eq:momperp}
\eeq
where $v_A = B_0^2/4\pi m_in_{0i}$. 
There is no $ZT_e\vdel_\perp n_{3i}$ term in \eqref{eq:momperp} because 
we assume that the only small-scale spatial variations 
of all quantities are in the parallel direction. 
The ion pressure term $\lt(\vdel\cdot\vP_{3i}\rt)_\perp$ 
is to be calculated by solving the ion kinetic equation (see \secref{sec:exp}). 

To summarize, we are looking for perturbations 
such that $\vdel_\perp=0$, $n_{1i}=0$, $B_{1}^\parallel=0$, $\vu_{1i}^\parallel=0$, 
while $\vBp$ and $\vup$ satisfy \eqsand{eq:indperp}{eq:momperp}. 
Physically, this reflects the fact that the parallel ($\kperp=0$) 
firehose perturbations are Alfv\'enic in nature (have no compressive part). 
That it is legitimate to consider such perturbations separately from 
other types of perturbations is not {\em a priori} obvious, but will 
be verified by our ability to obtain a self-consistent solution of 
the ion kinetic equation, which will satisfy  \eqsref{eq:P1}, \exref{eq:P0}, 
and \exref{eq:P2} 
(see \apref{app:ions}). 

\subsection{Large-scale dynamics}
\label{sec:transp}

In \secref{sec:polarization}, equations for the first-order fields, 
$\vup$ and $\vBp$ emerged after expanding the induction equation \exref{eq:ind_gen}
and the continuity equation \exref{eq:cont} 
to lowest order in $\eps$ and the momentum equation \exref{eq:ionmom} up 
to the third order. If, using the ordering 
of \secref{sec:order}, we go to the next order 
and average over small scales to eliminate small-scale perturbations, 
we recover the equations for 
the large-scale (unperturbed) fields: the induction equation
\beq
\frac{\did\vB_0}{\did t} = \vB_0\cdot\vdel\vu_{0i} - \vB_0\vdel\cdot\vu_{0i}
\label{eq:indavg}
\eeq 
(all terms are order $\eps^3\kpar\vthi B_0$),
the continuity equation
\beq
\frac{\did n_{0i}}{\did t} = - n_{0i}\vdel\cdot\vu_{0i}
\label{eq:contavg}
\eeq
(all terms are order $\eps^3\kpar\vthi n_{0i}$), 
and the momentum equation
\beq
m_in_{0i}\frac{\did \vu_{0i}}{\did t} = -\vdel\cdot\vP_{2i} 
- \vdel\frac{B_0^2}{8\pi} + \frac{\vB_0\cdot\vdel\vB_0}{4\pi} 
\label{eq:momavg}
\eeq
(all terms are order $\eps^4m_i n_{0i}\kpar\vthi^2$). 
The divergence of the second-order ion pressure tensor here 
is with respect to the large-scale spatial variation 
(according to \eqref{eq:P2}, it has no small-scale dependence). 
Again, $\vP_{2i}$ is calculated from ion kinetics. 

\Eqsdash{eq:indavg}{eq:momavg} are precisely the kind of mean-field 
equations that are needed to calculate the large-scale dynamics of 
astrophysical plasmas. They look just like the usual fluid MHD 
equations, the only nontrivial element being the pressure term 
in the momentum equation \exref{eq:momavg}. The goal of kinetic 
theory is to calculate this pressure, which depends on the 
microphysical fluctuations at small scales. In this paper, we only do this 
for the parallel ($\kperp=0$) firehose fluctuations. For the mirror fluctuations, 
it is done in \citet{Rincon} (using a somewhat different, 
near-marginal-stability asymptotic expansion), 
while the oblique firehose fluctuations 
are a matter for future work. The implications of our results 
for the ion momentum transport will be discussed in \secref{sec:transp2}. 

\subsection{Solution of the ion kinetic equation}
\label{sec:exp}

We now proceed to use the ordering established in \secref{sec:order} 
to construct an asymptotic expansion of the ion kinetic equation. 
This procedure, while analytically straightforward, is fairly cumbersome 
and so its detailed exposition is exiled to \apref{app:ions}. 
The results are as follows. 

In the expansion of the ion distribution function (\eqref{eq:fi_exp}), 
$f_{0i}$ is found to be a Maxwellian (\apref{app:Max}), 
with density $n_{0i}$ and temperature $T_{0i}$ that have 
to satisfy the equilibrium pressure balance constraint 
(see \eqref{eq:P0} and \apref{app:grads}). 

The first-order perturbed distribution function, $f_{1i}$, 
is proportional to $\vb_0\cdot\vdel T_{0i}$ and is responsible for the 
large-scale collisional ion heat fluxes (\apref{app:grads}). 

The second-order perturbed distribution function $f_{2i}$ contains 
the pressure anisotropy. 
The corresponding second-order pressure tensor is diagonal: 
\beq
\vP_{2i} = p_{2i}\vI + (\pperpi-\ppari)\lt(\frac{1}{3}\,\vI - \vb_0\vb_0\rt),
\eeq
where $p_{2i}$ is the perturbed isotropic pressure
and $\pperpi-\ppari$ is the lowest-order pressure anisotropy. 
The isotropic part of the pressure is determined from the 
large-scale equations for density $n_{0i}$, temperature $T_{0i}$ 
and velocity $\vu_{0i}$ of the fluid --- this is explained in detail 
in \apref{app:transp}, but here let us just assume for simplicity that 
the zeroth-order 
density and temperature are constant ($\vdel n_{0i}=0$, $\vdel T_{0i}=0$), 
in which case the continuity equation \exref{eq:contavg} 
reduces to $\vdel\cdot\vu_{0i}=0$ and $p_{2i}$ then follows 
from enforcing this incompressibility constraint on the 
momentum equation \exref{eq:momavg}. 
The pressure anisotropy is calculated in \apref{app:aniso}:
\bea
\nonumber
\Delta(t) &\equiv& \frac{\pperpi-\ppari}{p_{0i}}\\
&=& \Delta_0 + 
\frac{3}{2}\int_0^t\did t' e^{-3\nuii(t-t')}
\frac{\dd}{\dd t'}\frac{\overline{|\vBp(t')|^2}}{B_0^2}, 
\label{eq:Delta_nlin}
\eea
where $p_{0i}=n_{0i}T_{0i}$ is the equilibrium pressure, 
the overbar denotes the averaging over small scales 
of the nonlinear feedback on the anisotropy from 
the firehose fluctuations, 
and $\Delta_0$ is the pressure anisotropy arising 
from the large-scale motions. In general, it contains 
contributions from changes in the magnetic field strength 
(because of the approximate conservation of the first adiabatic 
invariant, as discussed qualitatively in \secref{sec:aniso}), 
compression and heat fluxes (see \eqref{eq:D0}). 
When $n_{0i}$ and $T_{0i}$ are constant, only the 
anisotropy induced by the changes in field strength 
survives,\footnote{The effect of heat fluxes on the firehose 
turbulence is considered in \secref{sec:GTI}.} 
which is the case we will consider here:
\beq
\Delta_0 = \frac{\vb_0\vb_0:\vdel\vu_{0i}}{\nuii}
= \frac{1}{\nuii}\frac{1}{B_0}\frac{\did B_0}{\did t}
= \frac{\gamma_0}{\nuii}.
\label{eq:Delta0}
\eeq
This is exactly what was anticipated qualitatively --- 
see \eqref{eq:Delta}. 
Since we are interested in the firehose instability, we 
assume $\Delta_0<0$.\footnote{We have assumed the initial 
anisotropy $\Delta(0)=\Delta_0$. It is equally possible to 
start from any other value, including stable situations.   
In that case, the large-scale drivers of the pressure anisotropy 
will gradually build it up to the maximum (negative) level, 
$\Delta_0$, whereupon further evolution will proceed in the 
same way as discussed below. Mathematically, this amounts to 
mutiplying $\Delta_0$ in \eqref{eq:Delta_nlin} by 
$(1-e^{-3\nuii t})$ (solution of \eqref{eq:aniso}). 
An example of such a set up can be found in \citet{Schek4}. 
Note that if the initial fluctuation level is not infinitesimal, 
the nonlinear quenching of the anisotropy (discussed in the 
subsequent sections) can start before the maximum anisotropy 
is built up.} 

Finally, the third-order perturbed distribution function $f_{3i}$ 
is responsible for the third-order pressure tensor that 
appears in the perturbed ion momentum equation \exref{eq:momperp}. 
The relevant part of that tensor is calculated in \apref{app:fluct}. 
Assuming constant density and temperature (otherwise, there is again 
a contribution from the heat fluxes; see \apref{sec:GTI}), 
it may be written as follows
\beq
\frac{\lt(\vdel\cdot\vP_{3i}\rt)_\perp}{p_{0i}} = 
-\dpar\lt[\Delta(t)\,\frac{\vBp}{B_0}
+ \frac{\dpar\vup}{\Omega_i}\times\vb_0\rt],
\label{eq:P3_formula}
\eeq
where $\Delta(t)$ is given by \eqref{eq:Delta_nlin}. 

Let us now use these results to study the firehose turbulence 
(\secsdash{sec:eqn}{sec:results})
and its effect on the large-scale dynamics
(\secref{sec:transp2}). 

\section{Firehose turbulence}
\label{sec:firehose}

\subsection{Firehose turbulence equation}
\label{sec:eqn}

Using the results derived in \apref{app:ions} and summarized in \secref{sec:exp}, 
we find that the ion momentum equation \exref{eq:momperp}, which describes the evolution 
of the perturbed ion velocity, is, in the reference frame moving with $\vu_{0i}$, 
\beq
\frac{\dd\vup}{\dd t} = \frac{\vthi^2}{2}\,
\dpar\lt[\lt(\Delta(t) + \frac{2}{\beta_i}\rt)\frac{\vBp}{B_0}
+ \frac{\dpar\vup}{\Omega_i}\times\vb_0\rt], 
\label{eq:u1}
\eeq
where we have used \eqref{eq:P3_formula} for the pressure  
term in \eqref{eq:momperp}. 
Three forces appear on the right-hand side of this equation. 
First, there is the stress due to the anisotropy $\Delta$
of the ion distribution, given by \eqref{eq:Delta_nlin}. 
The latter equation is the quantitative form of the expression 
for $\Delta(t)$ that we guessed in \eqref{eq:Delta_marg}: the first term 
in \eqref{eq:Delta_nlin} 
is due to the slow decrease of the large-scale magnetic field, 
the second to the average effect of the growing small-scale fluctuations,
which strive to cancel that decrease. 
The second term in \eqref{eq:u1}, proportional to $1/\beta_i= v_A^2/\vthi^2$, 
is the magnetic tension force, which resists the perturbation of the 
magnetic-field lines and, therefore, acts against the pressure-anisotropy 
driven instability. The instability is marginal when $\Delta + 2/\beta_i\to-0$. 
Finally, the third term is the FLR effect, which, as was 
promised in \secref{sec:FLR} and as will shortly be demonstrated, 
sets the scale of the most unstable perturbations. 

Let us now combine \eqref{eq:u1} with the induction equation \exref{eq:indperp}
for the perturbed magnetic field, also taken in the reference frame 
moving with $\vu_0$. After differentiating \eqref{eq:indperp} once 
with respect to time, we get 
\beq
\frac{\dd^2\vBp}{\dd t^2} = \frac{\vthi^2}{2}\,
\dpar^2\lt[\lt(\Delta + \frac{2}{\beta_i}\rt)\vBp
+\frac{1}{\Omega_i}\frac{\dd\vBp}{\dd t}\times\vb_0\rt].
\label{eq:B1}
\eeq
In the second term on the right-hand side, we have used \eqref{eq:indperp} 
to express $\dpar\vup$ in terms of the time derivative of $\vBp$. 
\Eqref{eq:B1} with $\Delta(t)$ defined by \eqref{eq:Delta_nlin} is a closed
equation for the perturbed magnetic field with nonlinear feedback 
(last term in \eqref{eq:Delta_nlin}). 
This is the equation for the one-dimensional ($\kperp=0$) {\em firehose turbulence}. 
It represents the simplest nonlinear model for this kind of turbulence 
available to date.\footnote{The essential difference with the equation we 
derived in \citet{Schek4} is the FLR term, which removes the 
ultraviolet catastrophe of the long-wavelength firehose and thus 
allows \eqref{eq:B1} to handle non-monochromatic (multiscale) solutions. 
In \secref{sec:results}, we will see that this produces a much more 
complex behaviour than was seen in \citet{Schek4}, justifying the 
term ``firehose turbulence.''} 

\begin{figure}
\includegraphics[width=80mm]{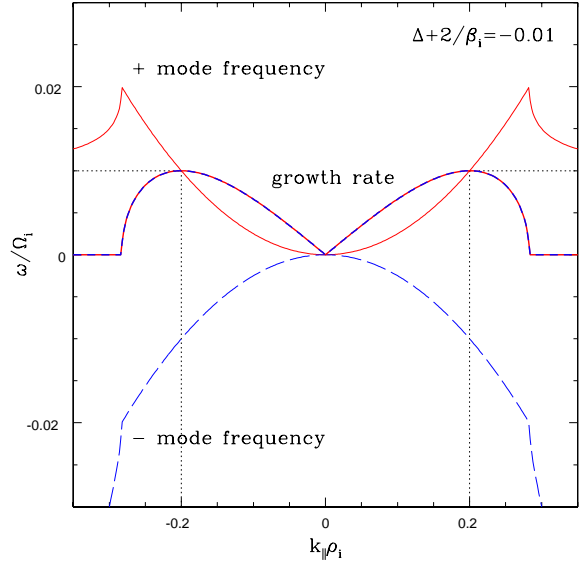}
\caption{Frequencies (thin lines) and growth rates (bold lines) of the unstable firehose 
modes (red/solid: the ``$+$'' mode; blue/dashed: the ``$-$'' mode) given by \eqref{eq:omega}. 
The instability parameter here is $\Delta+2/\beta_i=-0.01$. 
Dotted vertical lines indicate the wavenumber of fastest growth 
$\kpeak=0.2$ (\eqref{eq:kpeak}) and the dotted horizontal 
lines the corresponding maximum growth rate 
$\gmax = {\rm Im}\,\opeak = 0.01$ (\eqref{eq:opeak}).}
\label{fig:fh}
\end{figure}

\begin{figure*}
\includegraphics[width=85mm]{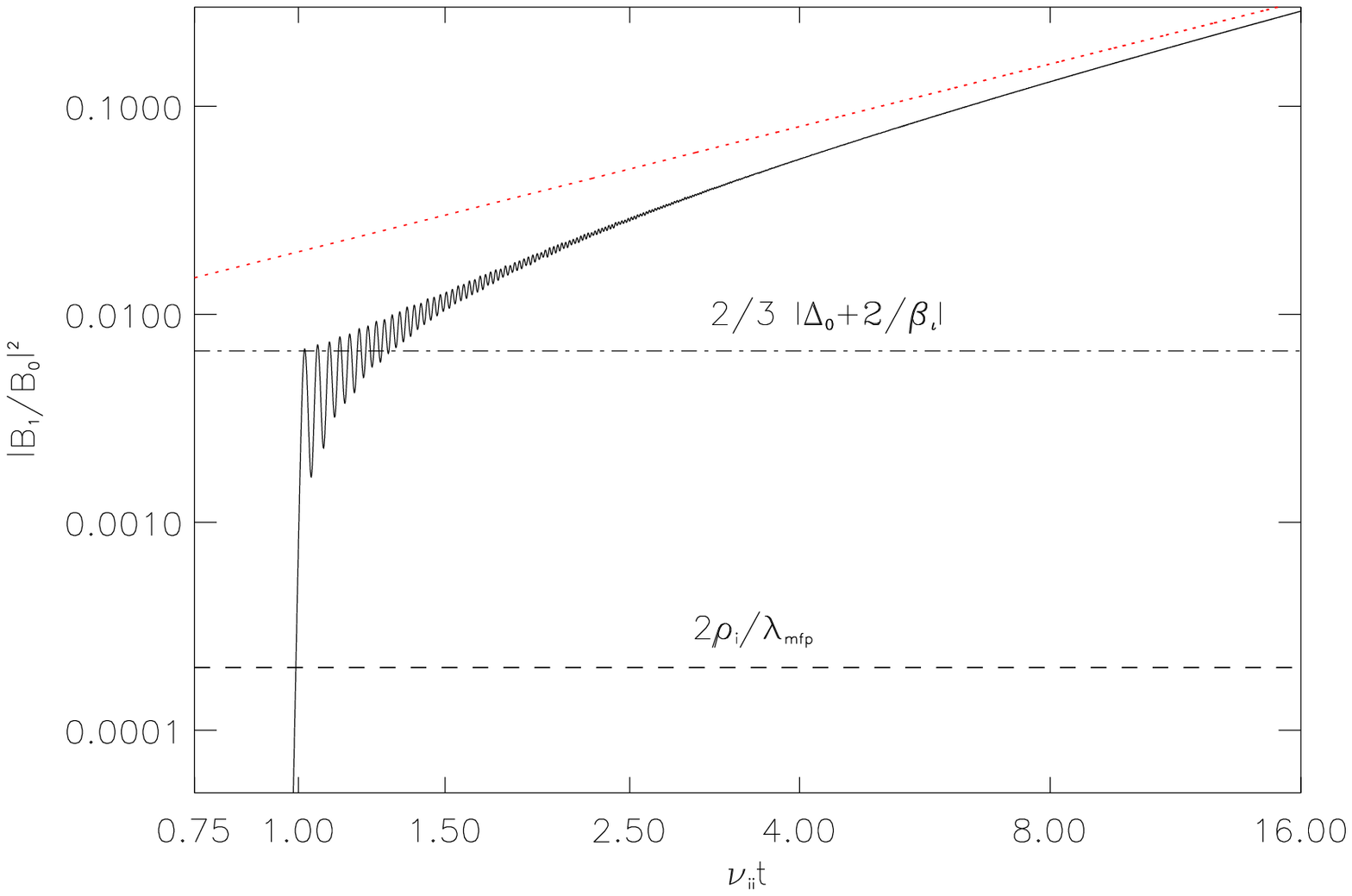}
\includegraphics[width=85mm]{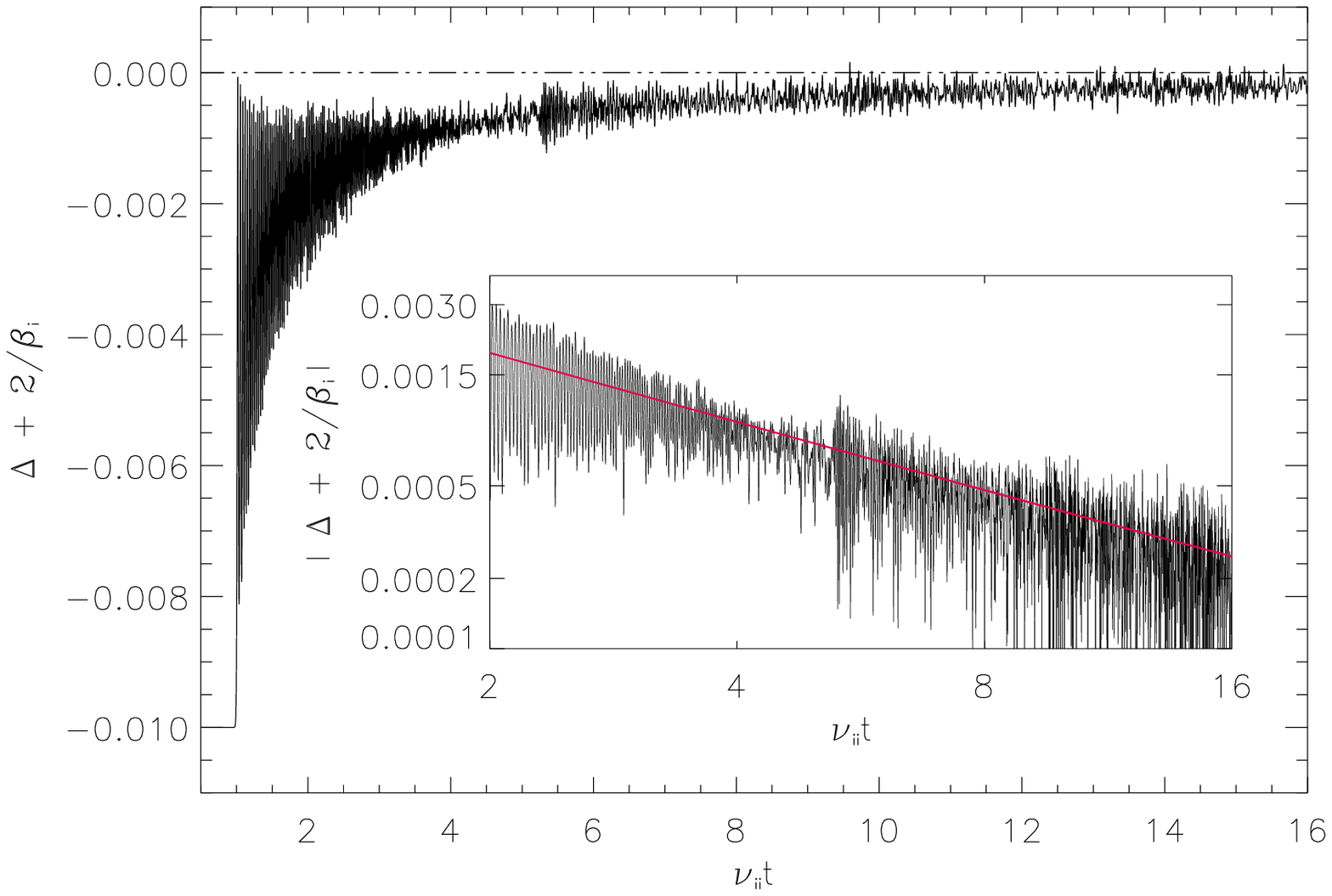}
\caption{{\it Left panel:} evolution of the magnetic energy 
$\overline{|\vBp|^2}/B_0^2 = \sum_k |A_k|^2$ with time 
in a numerical solution of \eqsand{eq:A}{eq:DeltaA} with parameters \exref{eq:num_params}; 
the time here is normalized using the collision frequency $\nuii$, 
not the cyclotron frequency $\Omega_i$; the two horizontal lines show the 
``collisional'' (lower line) and ``collisionless'' (upper line) estimates for the energy 
at which the nonlinear feedback turns on: \eqsand{eq:nlin_on}{eq:nlin_on_colless}, 
respectively; the red dotted line shows the nonlinear asymptotic 
given by \eqref{eq:Atot}. 
{\it Right panel:} evolution of the instability parameter (pressure 
anisotropy) $\Delta + 2/\beta_i$ in the same numerical solution. 
{\it Inset:} log-log plot of the evolution of $|\Delta+2/\beta_i|$; the red 
line shows the slope corresponding to $1/t$ (see \eqref{eq:kpt}).}
\label{fig:secular}
\end{figure*}

\subsection{Linear theory}
\label{sec:disp_rln}

In the linear regime, we may neglect the second term in 
\eqref{eq:Delta_nlin}, so $\Delta = \Delta_0$.
The linear dispersion relation for \eqref{eq:B1} is 
\beq
\lt[\omega^2 - \frac{\kpar^2\vthi^2}{2}\lt(\Delta+\frac{2}{\beta_i}\rt)\rt]^2 
\!\!= \frac{\kpar^4\vthi^4}{4}\frac{\omega^2}{\Omega_i^2}.
\label{eq:DR}
\eeq
This has four roots out of which two are unstable 
when $\Delta + 2/\beta_i<0$:
\beq
\frac{\omega}{\Omega_i} = \pm \frac{k^2}{4}
+ i\,\frac{|k|}{\sqrt{2}}\,\lt|\Delta+\frac{2}{\beta_i}\rt|^{1/2}
\!\!\!\sqrt{1-\frac{k^2}{k_0^2}}, 
\label{eq:omega}
\eeq
where $k=\kpar\rho_i$ and 
\beq
k_0=2\sqrt{2}\lt|\Delta+\frac{2}{\beta_i}\rt|^{1/2}
\label{eq:k0}
\eeq
\citep[this linear 
dispersion relation was first obtained by][]{Kennel,Davidson}. 
Unlike in the long-wavelength limit ($\kpar\rho_i\to0$), 
there is now a real frequency (so the firehose perturbation 
propagates while its amplitude grows exponentially and the vector 
$\vBp$ rotates; see \secref{sec:scalar}) 
and the growth rate has its peak at $\kpeak=k_0/\sqrt{2}$, 
so 
\bea
\label{eq:kpeak}
\kpeak &=& 2\lt|\Delta + \frac{2}{\beta_i}\rt|^{1/2},\\
\opeak &=& \lt(\pm 1 + i\rt)\lt|\Delta + \frac{2}{\beta_i}\rt|, 
\label{eq:opeak}
\eea 
where the complex peak frequency $\opeak$ is in units of $\Omega_i$. 
At $\kpar\rho_i>k_0$, there is no growth and 
the firehose perturbations turn into purely propagating 
Alfv\'en waves (modified by pressure anisotropy and dispersive FLR corrections). 

The dependence of the frequencies and growth rates of the two 
unstable modes on wavenumber given by \eqref{eq:omega} is plotted 
in \figref{fig:fh} for a representative value of the instability 
parameter $\Delta + 2/\beta_i=-0.01$ (this is the value used in 
the numerical solution of \secref{sec:num}). 

It should be pointed out here that in this theory, there is no dissipation of 
the magnetic fluctuations excited by the firehose. The most 
unstable wavenumber is set by dispersive effects; the stable  
modes are undamped. 

\subsection{Nonlinear evolution and spectrum}
\label{sec:results}

\subsubsection{Firehose turbulence equation in scalar form}
\label{sec:scalar}

Since the nonlinearity involves 
the spatially averaged perturbed magnetic energy, the firehose turbulence 
is compactly described 
in Fourier space not just in the linear but also in the nonlinear regime: 
this amounts to replacing $\dpar^2\to-\kpar^2$ in \eqref{eq:B1} and 
$\overline{|\vBp|^2} = \sum_{\kpar}|\vBp(\kpar)|^2$ in \eqref{eq:Delta_nlin}. 
A simple ansatz can now be used to convert \eqref{eq:B1} into scalar form. 
Let 
\bea
\label{eq:Bx}
\frac{B_{1x}}{B_0} &=& A_k(t)\cos\lt(\frac{k^2}{4}\,t + \phi_k\rt),\\
\frac{B_{1y}}{B_0} &=& A_k(t)\sin\lt(\frac{k^2}{4}\,t + \phi_k\rt),
\label{eq:By}
\eea
where the axes $(x,y)$ in the plane perpendicular to $\vb_0$ 
are chosen arbitrarily and we have non-dimensionalized wavenumbers and time: 
\beq
\kpar\rho_i\to k, \quad \Omega_i t\to t.
\label{eq:nondim}
\eeq  
This ansatz amounts to factoring out the rotation of the vector $\vBp(k)$
(the first term in \eqref{eq:omega}). 
The wavenumber-dependent but time-independent phase $\phi_k$ is determined 
by the initial condition. We assume $\phi_k=\phi_{-k}$, so 
$A_k^*=A_{-k}$ must be satisfied to respect the fact that $\vBp$ is a real field. 
The fluctuation amplitude $A_k(t)$ satisfies
\bea
\label{eq:A}
\frac{\dd^2 A_k}{\dd t^2} &=& \frac{k^2}{2}\lt[-\lt(\Delta + \frac{2}{\beta_i}\rt)
-\frac{k^2}{8}\rt]A_k,\\
\Delta(t) &=& \Delta_0 + \frac{3}{2}\int_0^t\did t' e^{-3\nu_*(t-t')}
\frac{\dd}{\dd t'}\sum_k|A_k(t')|^2, 
\label{eq:DeltaA}
\eea
where $\nu_*=\nuii/\Omega_i=\rho_i/\mfp$ and we remind the reader that $\Delta_0<0$. 
It is manifest in the form of \eqref{eq:A} how 
the dispersion relation \exref{eq:omega} (without the first term) 
is recovered. Note that there is no coupling between different 
wavenumbers modes in the sense that if a mode is not initially excited, 
it is never excited. The only effect that modes have on each other 
is via the sum over $k$ in \eqref{eq:DeltaA}, to which they all contribute. 

\begin{figure*}
\includegraphics[width=80mm]{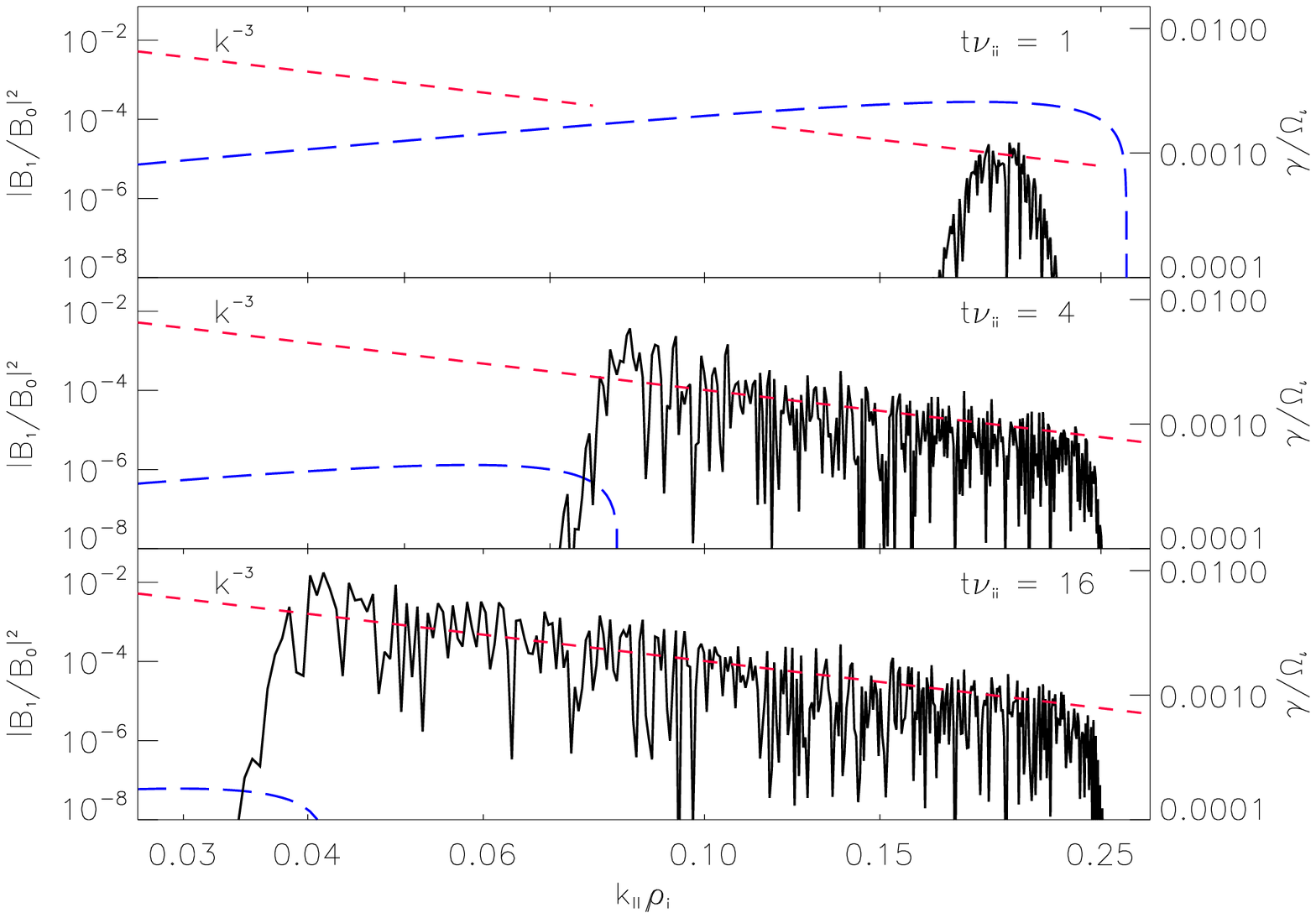}
\qquad
\includegraphics[width=80mm]{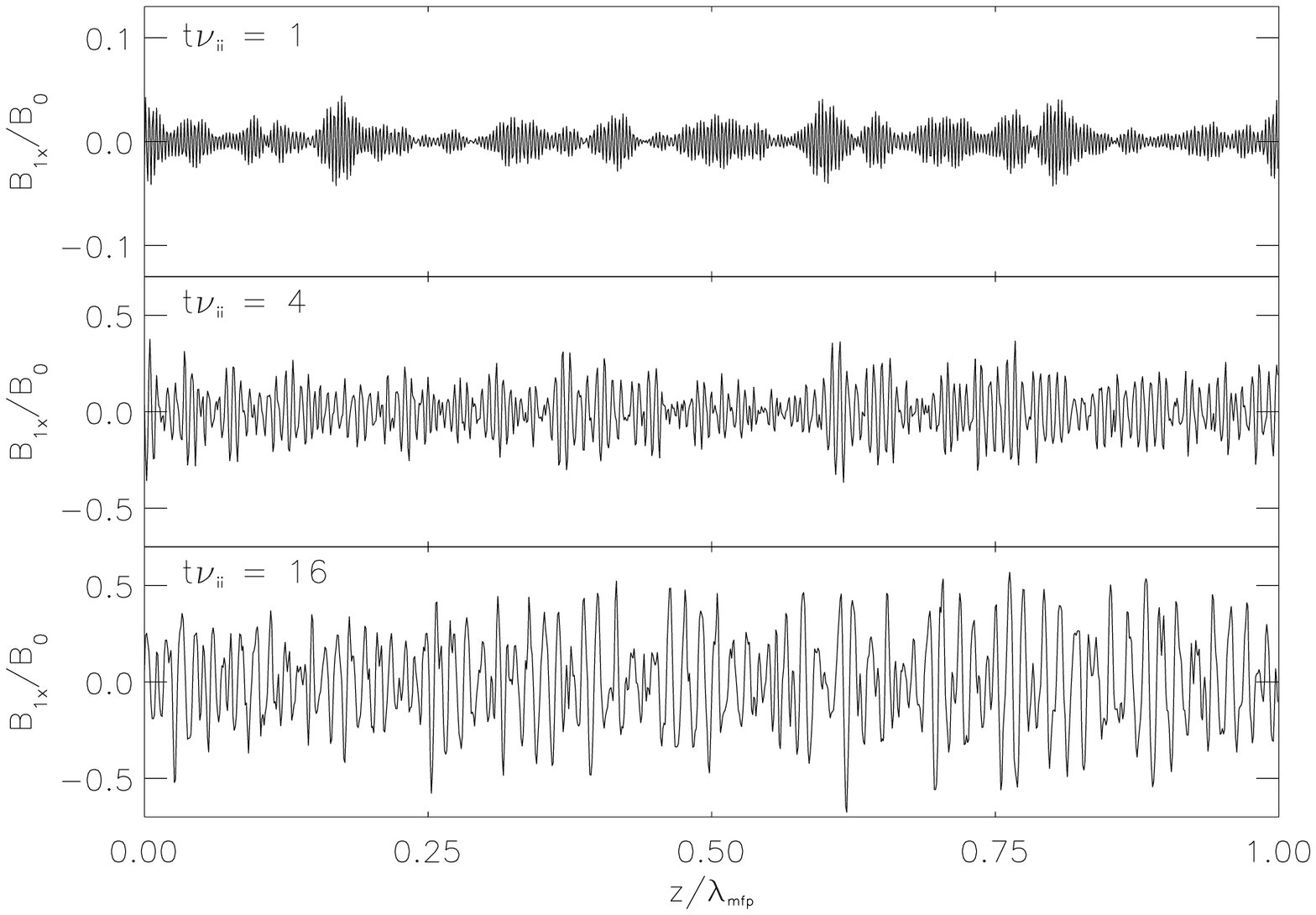}
\caption{{\it Left panel:} spectrum of the magnetic fluctuation energy 
at three specific times ($t\nuii = 1, 4, 16$) during the evolution 
shown in \figref{fig:secular}; red short-dashed lines show the $k^{-3}$ slope 
(\eqref{eq:spec}); blue long-dashed lines show the firehose growth rate (see \eqref{eq:omega}) 
for the instantaneous values of $\Delta+2/\beta_i$ at the times the spectra 
are plotted. {\it Right panel:} magnetic fluctuations in real space 
($B_{1x}/B_0$ vs.\ $z$) at the same times as the spectra in the left panel.}
\label{fig:spec}
\end{figure*}

\subsubsection{Qualitative picture}
\label{sec:nlin_qualit}

Already on the basis of linear theory and the qualitative considerations 
of \secref{sec:nonlin}, we can construct a fairly clear picture of the 
evolution of the firehose turbulence. 
Assuming a broad-band infinitesimal initial perturbation in $k$ space, 
at first, $\Delta=\Delta_0$ and all modes with $k<k_0$ (see \eqref{eq:k0})
will go unstable, with the fastest-growing one given by $\kpeak=k_0/\sqrt{2}$. 
Eventually the amplitude in this mode reaches the level at which the back-reaction 
becomes important: approximating \eqref{eq:DeltaA} by
\beq
\Delta(t)\simeq \Delta_0 + \frac{1}{2\nu_*}\frac{\dd}{\dd t}\sum_k |A_k|^2,
\label{eq:Delta_approx}
\eeq 
we find that the nonlinear contribution is comparable to $|\Delta_0+2/\beta_i|$ 
when (cf.\ \eqref{eq:crit}): 
\beq
\sum_k |A_k|^2\simeq 2\lt|\Delta_0+\frac{2}{\beta_i}\rt|\frac{\nu_*}{\gmax}
= \frac{2\nuii}{\Omega_i} = \frac{2\rho_i}{\mfp},
\label{eq:nlin_on}
\eeq
where $\gmax$ is the imaginary part of $\opeak$ given by \eqref{eq:opeak}. 
\Eqref{eq:nlin_on} only gives a good estimate of the critical 
amplitude if $3\nu_*$ is larger or not too much smaller than $\gmax$ (collisions are 
sufficiently strong). If $\nu_*\ll\gmax$ (as a subsidiary limit within our 
$\eps$ ordering), then a better approximation than 
\eqref{eq:Delta_approx} is to replace the collisional 
relaxation exponent in \eqref{eq:DeltaA} by unity, which gives 
\beq
\sum_k |A_k|^2\simeq\frac{2}{3}\lt|\Delta_0+\frac{2}{\beta_i}\rt|.
\label{eq:nlin_on_colless}
\eeq 

Once the nonlinear feedback becomes active, 
exponential growth must cease and secular growth starts because 
the anisotropy must be kept close to marginal: using \eqref{eq:Delta_approx}, 
we find, to dominant order,
\beq
\label{eq:Atot}
\Delta\simeq-\frac{2}{\beta_i}\quad\Rightarrow\quad
\sum_k|A_k|^2 \simeq 2 \lt|\Delta_0+\frac{2}{\beta_i}\rt|\nu_* t.
\eeq
This is valid regardless of which of the two estimates \exref{eq:nlin_on} 
or \exref{eq:nlin_on_colless} 
of the amplitude at the onset of nonlinearity was appropriate. 
This is because the effective 
growth rate associated with the secular growth decreases with time and so 
we will always eventually end up in the regime where the collisional relaxation 
exponent in \eqref{eq:DeltaA} is faster than the magnetic energy growth 
and \eqref{eq:Delta_approx} gives a good approximation of \eqref{eq:DeltaA}. 

The evolution of the fluctuation spectrum must be consistent with \eqref{eq:Atot}. 
As the magnitude of the total pressure anisotropy $\Delta$ approaches 
the marginal value, both the cutoff wavenumber $k_0(t)$ and 
the most unstable wavenumber $\kpeak(t)$ decrease, as they can still 
be estimated by \eqsand{eq:k0}{eq:kpeak} with $\Delta=\Delta(t)$. 
The modes whose growth has been thus switched off become oscillatory: 
from \eqref{eq:A}, it is obvious that for $k\gg k_0(t)$, 
\beq
A_k = c_1 e^{ik^2 t/4} + c_2 e^{-ik^2 t/4},
\eeq
where $c_1$ and $c_2$ are integration constants and  
$c_1^*=c_2$ because $A_k^*=A_{-k}$
(note that this oscillation of the amplitude 
is superimposed on the oscillation with the same frequency that was factored 
out in \eqsdash{eq:Bx}{eq:By}). Since these modes oscillate in time at a rate 
that is much larger than the rate of change of the anisotropy, they 
no longer contribute to the feedback term in \eqref{eq:DeltaA}.

Thus, as the range of growing modes, peaked at $\kpeak(t)$ and cut off 
at $k_0(t)$, sweeps from large to small wavenumbers, they leave 
behind a spectrum of effectively passive oscillations, whose amplitude 
no longer changes. 
Since there is no fixed special scale in the problem (except initial 
most unstable wavenumber), one expects the evolution to be self-similar
and the spectrum a power law. It is not hard to determine its exponent. 
Let $|A_k|^2\sim k^{-\alpha}$. 
Since the total energy must grow linearly (\eqref{eq:Atot}) 
\beq
\sum_k|A_k|^2 \sim \kpeak^{1-\alpha} \sim t \quad\Rightarrow\quad
\kpeak\sim t^{-1/(\alpha-1)}
\label{eq:total}
\eeq
(this is valid if $\alpha>1$; the extra power of $k$ comes from the 
integration over wavenumbers).
On the other hand, for the fastest-growing mode, 
we must have, assuming secular growth,  
\beq
\frac{1}{A_{\kpeak}}\frac{\dd A_{\kpeak}}{\dd t} \sim \frac{1}{t}
\sim \gmax \sim \lt|\Delta+\frac{2}{\beta_i}\rt|, 
\label{eq:mode}
\eeq
where the last relation follows from \eqref{eq:opeak}.
This gives us a prediction for the time evolution of the residual pressure anisotropy
and, via \eqref{eq:kpeak}, of the most unstable wavenumber (the infrared cutoff of the spectrum):
\beq
\lt|\Delta+\frac{2}{\beta_i}\rt| \sim \frac{1}{t},\quad 
\kpeak \sim \frac{1}{\sqrt{t}}.
\label{eq:kpt}
\eeq 
The only way to reconcile \eqsand{eq:total}{eq:kpt} is to set $\alpha=3$. 
Thus, we expect the one-dimensional firehose turbulence spectrum to scale as 
\beq
|A_k|^2\sim k^{-3}.
\label{eq:spec}
\eeq

The secular growth of the firehose fluctuations will continue until our asymptotic 
expansion becomes invalid, i.e., when the fluctuation amplitude 
is no longer small.\footnote{Note that while the amplitude grows and 
thus eventually breaks the ordering introduced in \secref{sec:order}, 
the stability parameter $|\Delta + 2/\beta_i|$ decreases, so 
the approximation of small Larmor radius gets quantitatively better 
with the growth of the firehose fluctuations moving to larger scales (\eqref{eq:kpeak})
--- equivalently, our ordering of $\rho_i$ introduced in \secref{sec:order}
(\eqref{eq:connection}) is quantitatively better satisfied. 
In fact, we could have chosen to construct our entire asymptotic 
theory by expanding close to marginal stability and so
ordering everything with respect to the small parameter 
defined as $\eps=|\Delta + 2/\beta_i|^{1/2}$ instead 
of \eqref{eq:eps} (this is the route followed in an analogous 
mirror instability calculation by \citealt{Rincon}).\label{fn:alt_order}} 
From \eqref{eq:Atot}, this happens at 
$t\sim \lt(\nuii|\Delta_0+2/\beta_i|\rt)^{-1}\sim |\gamma_0|^{-1}$, 
where dimensions have been restored. This is the time scale of the 
large-scale dynamics. Thus, as we have already explained in 
\secref{sec:nonlin}, there is no saturation of the firehose 
fluctuations on any faster time scale. 
Unsurprisingly, at the same time as the fluctuation 
amplitude becomes large enough to break our ordering, 
the scale of the fluctuations also breaks the ordering:  
substituting the above time scale into \eqref{eq:kpt},   
$\kpar\rho_i\sim (|\gamma_0|/\Omega_i)^{1/2} \sim \eps^2$,
or $\kpar\mfp\sim1$,  
while our original ordering assumption was $\kpar\rho_i\sim\eps$, 
or $\kpar\mfp\sim1/\eps$ (see \eqref{eq:order_k}). 


\begin{figure*}
\includegraphics[width=85mm]{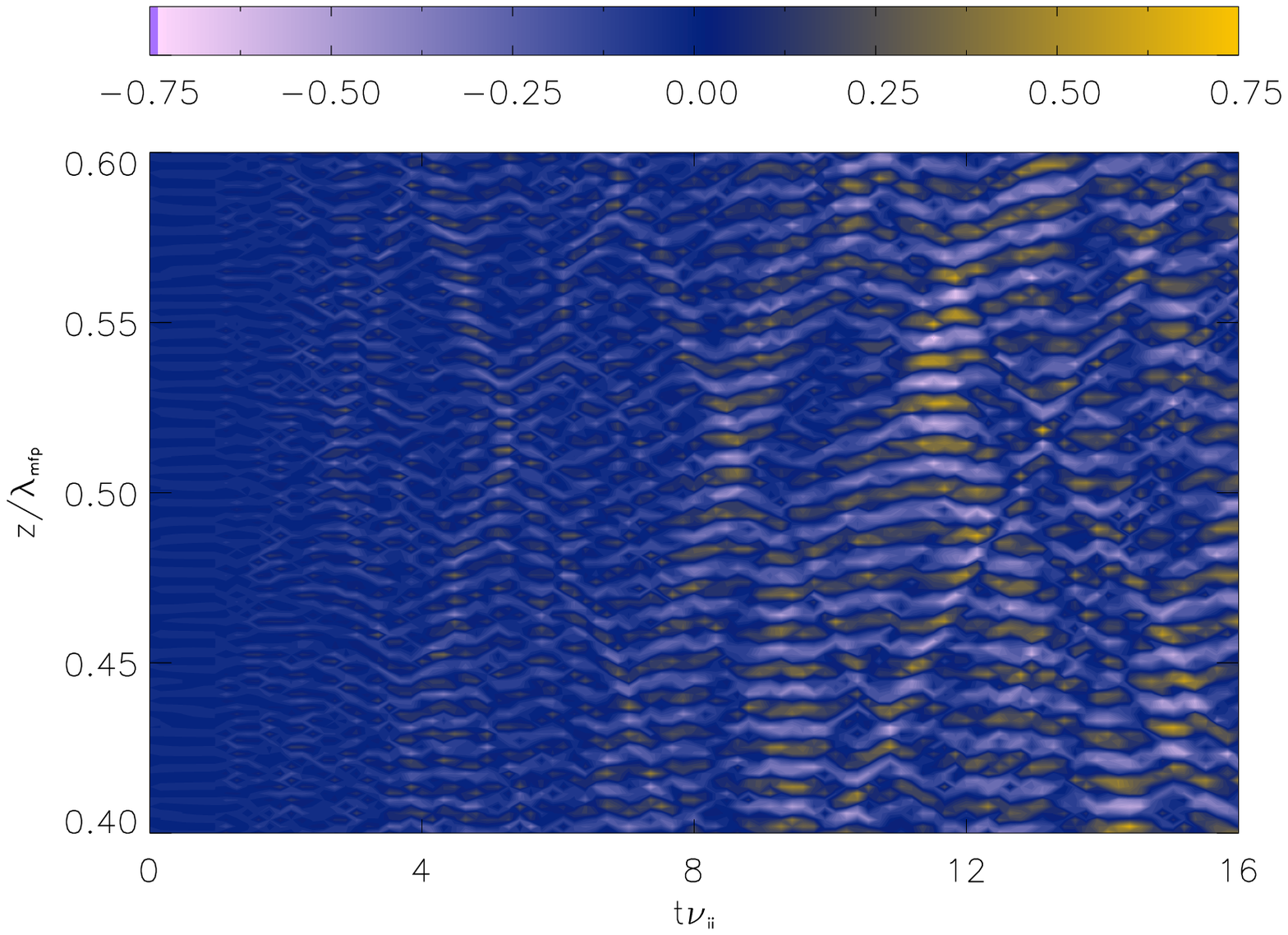}
\includegraphics[width=85mm]{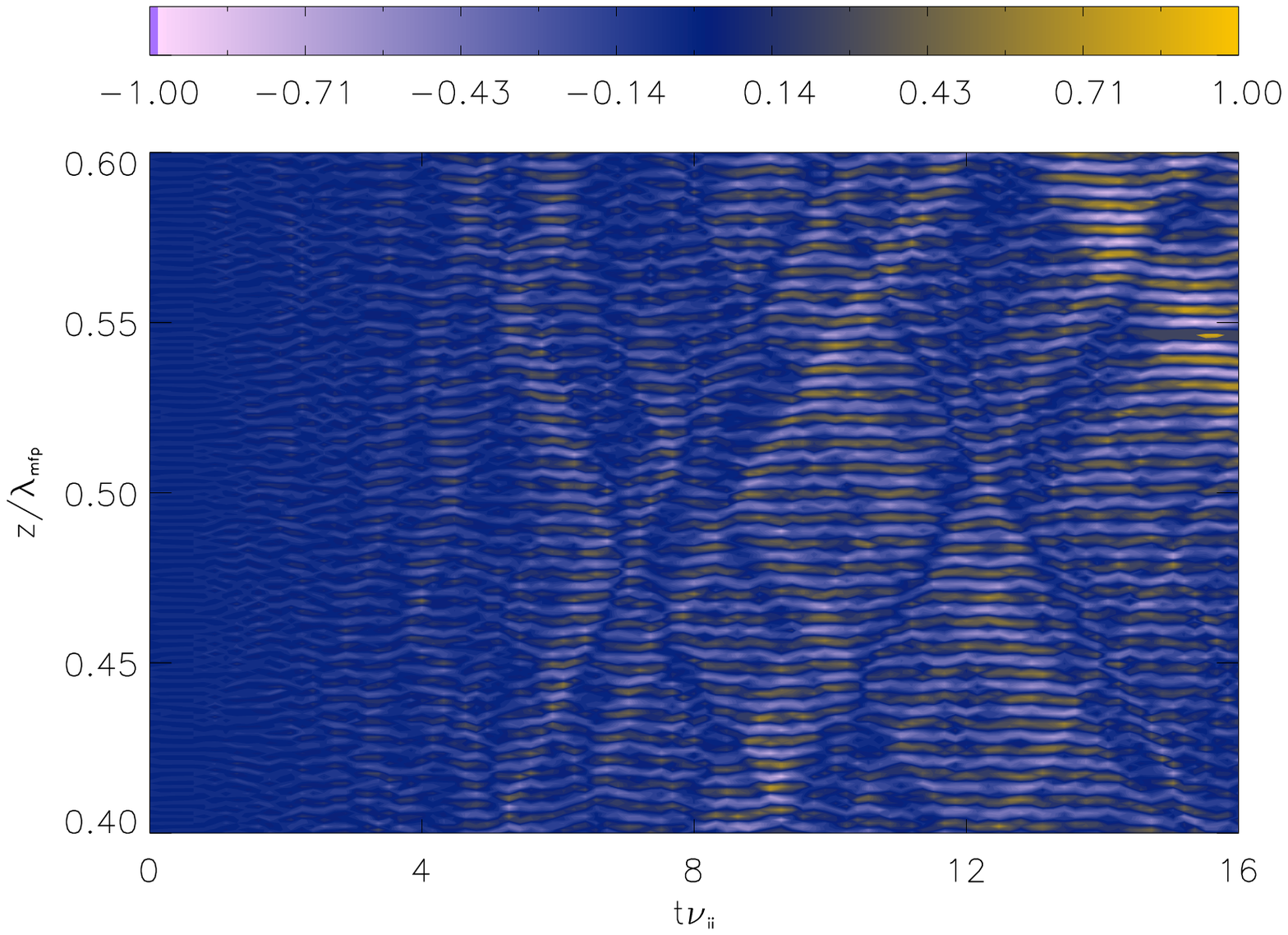}
\caption{{\it Left panel:} contour plot of the time evolution of the 
firehose turbulence for the numerical solution discussed in 
\secref{sec:num} --- the vertical axis is space (the middle fifth 
of our entire periodic domain), horizontal axis is time, the colours 
respresent the value of $B_{1x}/B_0$.
{\it Right panel:} the same plot for the gyrothermal turbulence
discussed in \secref{sec:GTI_num}.
Note that, as explained in the text, the firehose turbulence exhibits 
a gradual coarsening of the dominant structure with time, while the 
gyrothermal turbulence ends up dominated by a single scale.}
\label{fig:evolution}
\end{figure*}

\subsubsection{Numerical solution}
\label{sec:num}

The firehose turbulence equation \exref{eq:A} is one-dimensional, so it is 
very easy to solve numerically; \eqref{eq:DeltaA} is most conveniently 
solved in a differential form: 
\beq
\frac{\dd}{\dd t}\biggl(\Delta - \frac{3}{2}\sum_k |A_k|^2\biggr) 
= -3\nu_*(\Delta-\Delta_0) 
\eeq
with the initial condition $\Delta(0) = \Delta_0$.
Here we describe the results obtained 
from such a numerical calculation with the following parameters: 
\beq
\Delta_0=-0.02,\quad 
\frac{2}{\beta_i}=0.01,\quad 
\nu_*=\frac{\rho_i}{\mfp}=0.0001.
\label{eq:num_params}
\eeq 
This means that the maximum wavenumber at which firehose fluctuations 
can be excited is $k_0\simeq 0.28$ (\eqref{eq:k0}; see \figref{fig:fh}). 
We solve \eqref{eq:A} for 1024 wavenumbers in a periodic domain 
of size $\mfp$, so the smallest and the largest wavenumbers are 
(still normalized to $\rho_i$) $k_{\rm min}=2\pi\rho_i/\mfp\simeq0.00063$ 
and $k_{\rm max}\simeq 0.32$. The initial conditions are random 
amplitudes in each wavenumber (satisfying the reality condition $A_{-k}=A_k^*$). 
Note that with the parameters \exref{eq:num_params}, our ordering parameter 
is $\eps\sim0.1$, so we have chosen a spatial scale separation 
between collisions and the Larmor motion that substantially 
exceeds $1/\eps^2$ formally mandated by our ordering 
(\secref{sec:order}). This does not break anything and 
is in fact more realistic for the physical 
parameters in weakly collisional plasmas of interest 
(\secref{sec:clusters}). It also widens the scale 
interval available to the firehose 
turbulence spectrum and ensures that even deep 
in the nonlinear regime, when the wavenumber of the firehose 
fluctuations drops substantially, there is still a healthy 
scale separation between them and the collisional dynamics. 

The evolution of the total magnetic energy, 
$\overline{|\vBp|^2}/B_0^2 = \sum_k |A_k|^2$, 
is shown in \ffigref{fig:secular}{left}. 
Initially it grows exponentially at the (normalized) rate 
$\gmax = {\rm Im}\,\opeak$ 
(see \eqref{eq:omega}; this part of the evolution is trivial and so not shown). 
The exponential growth is followed by a secular, linear in time, 
growth of the energy in accordance with \eqref{eq:Atot}. 
The energy at which this nonlinear regime starts is closer to 
the estimate given by \eqref{eq:nlin_on_colless} than 
by \eqref{eq:nlin_on} because, as discussed above, 
we have taken a very small value of $\nu_*$. 
Note that in this and all subsequent figures, 
we have normalized time using the collision frequency $\nuii$, not the 
cyclotron frequency $\Omega_i$ --- this is indicated explicitly in the 
figures and should cause no confusion to an attentive reader. 

\Ffigref{fig:secular}{right} shows the time evolution of 
the instability parameter $\Delta + 2/\beta_i$. 
As expected, it is tending to the marginal stability value (zero). 
The inset shows that this approach to zero is consistent 
with the $1/t$ prediction (\eqref{eq:kpt}).\footnote{The oscillations
seen in the figure are not a numerical artefact. 
They are due to oscillatory transients --- \citet{Schek4} 
derived those analytically for a solution with only one Fourier mode.} 

The evolution of the spectrum of firehose fluctuations is 
illustrated by \ffigref{fig:spec}{left}. As anticipated in 
\secref{sec:nlin_qualit}, the spectral peak moves to smaller 
wavenumbers in the nonlinear regime. The spectrum extending 
from this moving peak to the original wavenumber of the 
fastest linear growth ($\kpeak=0.2$; see \eqref{eq:kpeak}) 
is statistically stationary and consistent with the $k^{-3}$ power law 
predicted by \eqref{eq:spec}. The instantaneous firehose 
growth rate is overplotted on the spectra in \ffigref{fig:spec}{left} 
and confirms that the position of the spectral peak closely 
follows the wavenumber of the fastest instantaneous growth 
of the firehose instability. 

\Ffigref{fig:spec}{right} shows 
snapshots of one of the components ($B_{1x}$) of 
the perturbed magnetic field corresponding to the spectra 
in \ffigref{fig:spec}{left}. The emergence of increasingly 
larger-scale fluctuations is manifest. Perhaps a better 
illustration of this real-space evolution of the firehose 
turbulence is \ffigref{fig:evolution}{left}, which is 
the space-time contour plot for the middle 
fifth of the domain. 

\subsection{Implications for momentum transport}
\label{sec:transp2}

Substituting the second-order pressure tensor calculated in 
\secref{sec:exp} into the large-scale momentum equation \exref{eq:momavg}, 
we get
\beq
m_in_{0i}\frac{\did \vu_{0i}}{\did t} = 
-\vdel\tilde p
+\vdel\cdot\lt[p_{0i}\vb_0\vb_0\lt(\Delta + \frac{2}{\beta_i}\rt)\rt],
\label{eq:momtransp}
\eeq
where, in the absence of density and temperature gradients, 
the total isotropic pressure $\tilde p = \pperpi + B_0^2/8\pi$ 
is set by the condition $\vdel\cdot\vu_{0i}=0$,\footnote{More generally, 
$\tilde p$ adjusts in such a way as to reconcile the pressure balance 
with the continuity and heat conduction equations; see \apref{app:transp}.} 
while the pressure anisotropy $\Delta$ is given by 
\eqref{eq:Delta_nlin}. A remarkable feature of \eqref{eq:momtransp} 
is that all of the effects of the magnetic field appear in the term 
proportional to $\Delta + 2/\beta_i$, which is precisely the 
instability parameter that the small-scale firehose turbulence 
described in \secref{sec:results} contrives to make vanish. 
In the marginal state that results, the tension force (the $2/\beta_i$ term)
is almost entirely cancelled by the combined pressure anisotropy due 
to large- and small-scale fields. 
This suggests that in regions of the plasma where the firehose 
is triggered (i.e., where the magnetic field is locally decreased 
by the plasma motion), the plasma motions become effectively 
hydrodynamic, with magnetic-field lines unable to resist  
bending by the flows.

Since the cancellation of the second term in \eqref{eq:momtransp} 
by the firehose turbulence also effectively removes the 
(parallel) viscosity of the plasma, these hydrodynamic motions 
are not dissipated. In a turbulent situation, this should 
enable a cascade to ever smaller scales. 
Obviously, once this happens, the original motion that caused 
the negative pressure anisotropy to develop is supplanted 
by other, faster motions on smaller scales. 
The theory developed above eventually breaks down  
because the scale separation that formed the basis 
of our asymptotic expansion is compromised: while the fluid 
motions penetrate to smaller scales, the firehose fluctuations 
move to larger scales (see \secref{sec:results}). 

Note also that the fluid motions produced by the turbulent cascade 
can give rise to both positive and negative 
pressure anisotropies --- and so, to have a full description 
of their further evolution, we must know the effect on  
momentum transport not just of the firehose but also of the 
mirror and other instabilities triggered by positive pressure 
anisotropies (locally increasing magnetic field strength). 
This is still work in progress (the mirror case is considered by \citealt{Rincon}). 
Another important adjustment to the viscous-stress reduction argument 
above has to do with the modification of the firehose instability 
by the parallel ion heat fluxes --- we now proceed 
to investigate this. 

\section{Gyrothermal turbulence}
\label{sec:GTI}

\begin{figure*}
\includegraphics[width=80mm]{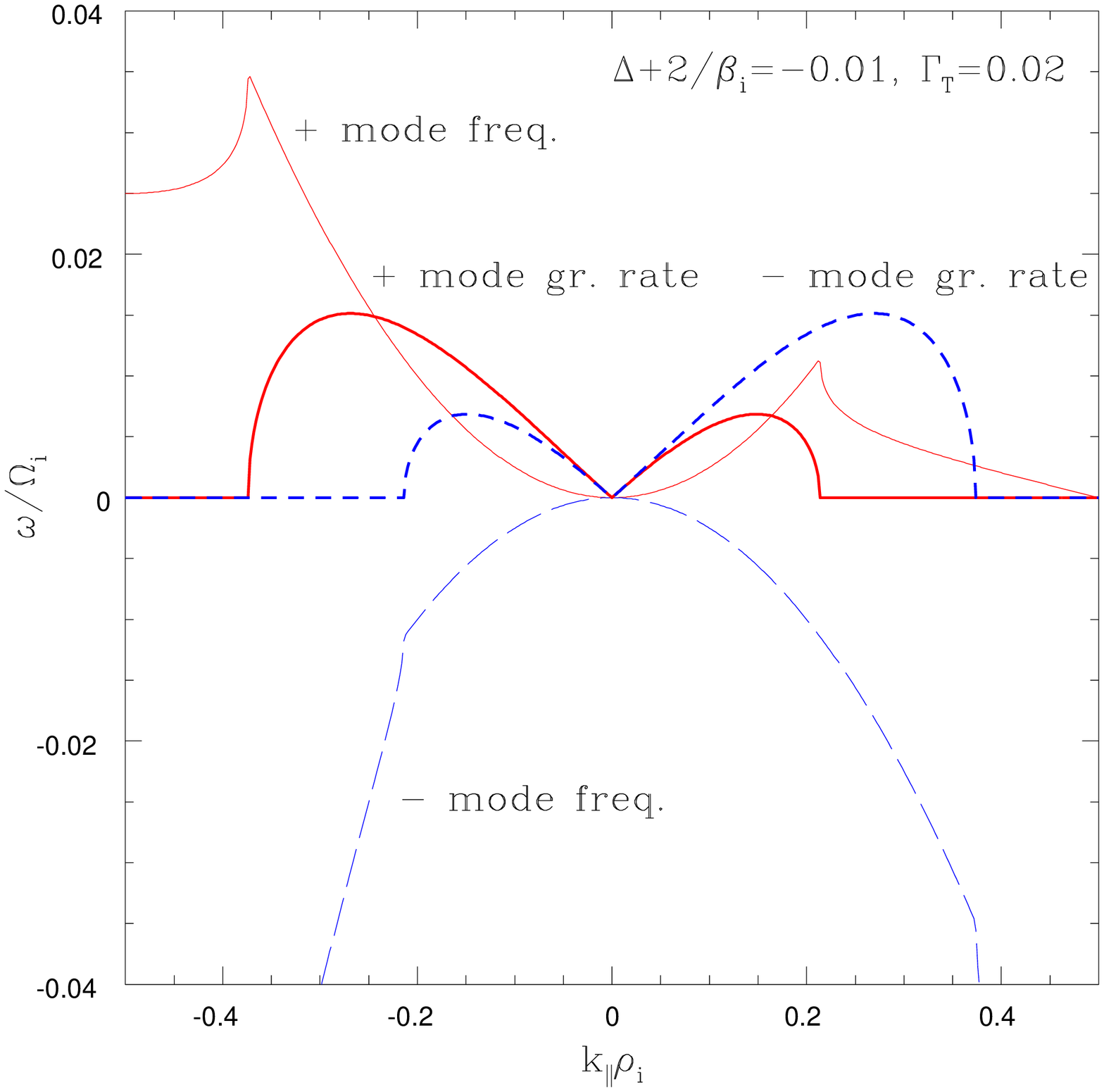}
\includegraphics[width=80mm]{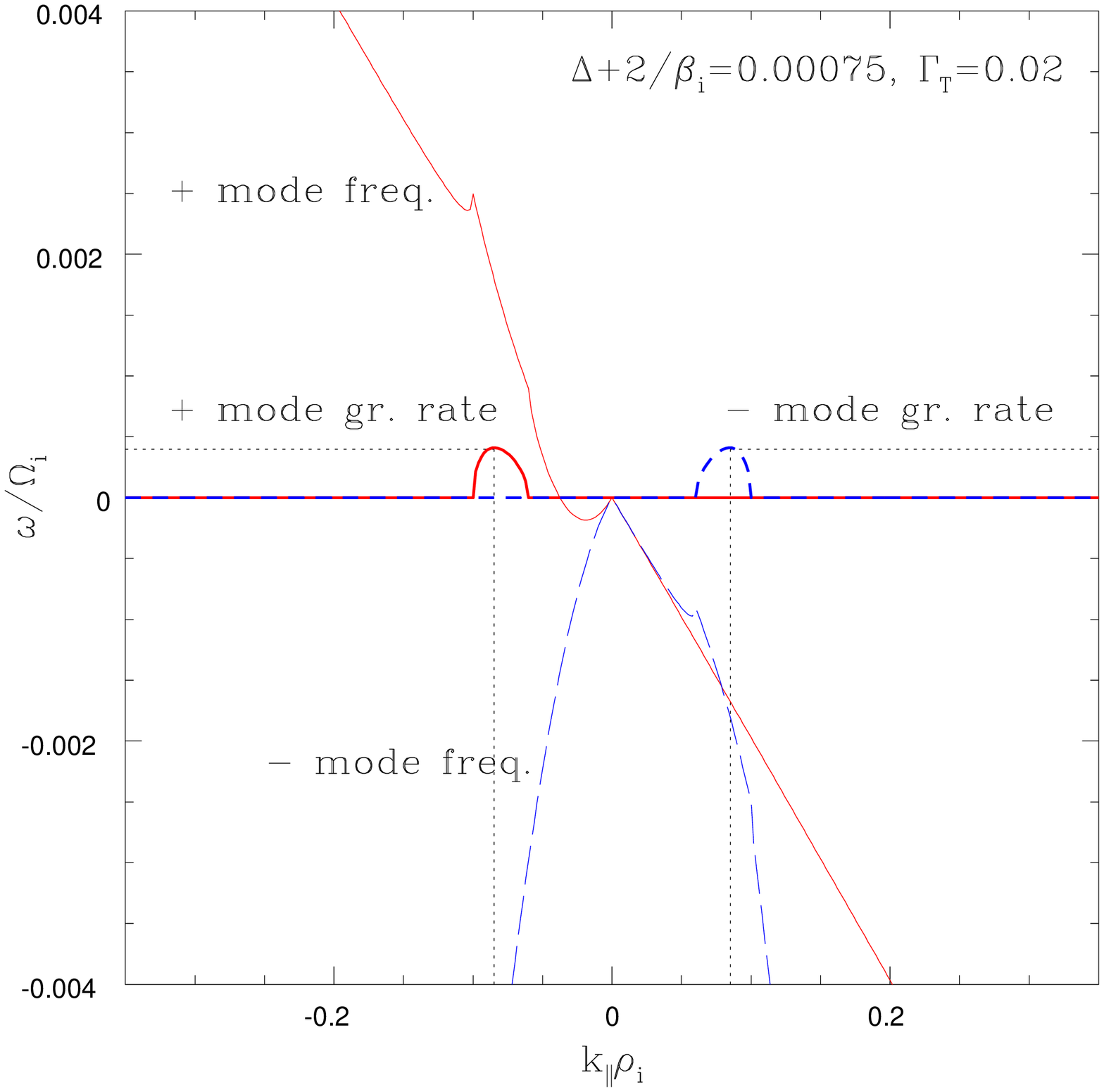}
\caption{{\it Left panel:} frequencies (thin lines) and growth rates (bold lines) 
of the unstable firehose modes (red/solid: the ``$+$'' mode; blue/dashed: the ``$-$'' mode) 
given by \eqref{eq:GTI_disp}; 
the parameters here are $\Delta+2/\beta_i=-0.01$ and $\gT=0.02$, 
so the instability parameter is $\Lambda=0.0054$ (\eqref{eq:Lambda}). 
{\it Right panel:} same, but for $\Delta + 2/\beta_i=0.00075$, 
so $\Lambda = 0.000025$ (close to marginal stability);  
dotted vertical lines indicate the wavenumber of fastest growth 
$\kpeak=0.085$ (\eqref{eq:GTI_lastk}) and the dotted horizontal 
lines the corresponding maximum growth rate 
$\gmax={\rm Im}\,\opeak = 0.0004$ (\eqref{eq:GTI_marg}).}
\label{fig:gti}
\end{figure*}

\subsection{Firehose turbulence equation with heat fluxes}
\label{sec:GTI_eqn}

As we briefly mentioned in \secref{sec:exp}, allowing 
a non-zero ion temperature gradient along the unperturbed 
magnetic field leads to substantial modifications. 
These are of two kinds. First, as shown in \apref{app:aniso}, 
the pressure anisotropy 
$\Delta_0$ caused by the large-scale dynamics contains 
contributions from the collisional parallel heat fluxes 
(proportional to $\vb_0\cdot\vdel T_{0i}$) and from 
compressive motions (as we pointed out in footnote \ref{fn:eqgrad}, 
the presence of a temperature gradient 
automatically implies a density gradient as well because 
of the requirement that pressure balance should be 
maintained; see \eqref{eq:pr_bal} and \apref{app:transp}). 
Instead of \eqref{eq:Delta0},  
valid in the incompressible case, we must use the more general 
\eqref{eq:D0}. This, however, does not change much: 
the unstable firehose fluctuations will grow in the manner 
described in \secref{sec:results}, first exponentially, 
then secularly, to compensate whatever pressure anisotropy 
is set up by the large-scale dynamics. The only change is the 
physical interpretation of the origin of the pressure anisotropy: 
as long as ion temperature gradients are present, 
the anisotropy is not tied exclusively to the change in 
the magnetic field. Physically, the heat-flux 
contributions to the anisotropy have to do with the fact 
that ``parallel'' and ``perpendicular'' heat flows along the 
magnetic-field lines somewhat 
differently and so imbalances between $\pperp$ and $\ppar$ 
can occur --- this can be seen already from the CGL 
equations (see \apref{app:CGL}). 

The second heat-flux-related modification of the theory developed thus far 
is more serious. It involves an additional contribution to the FLR 
term in the third-order pressure tensor (\eqref{eq:P3_formula}) 
and, therefore, to the firehose turbulence equation \exref{eq:B1}. 
This contribution was derived in \apref{app:fluct}, but suppressed 
in our previous discussion. It is given by \eqref{eq:P3} and 
consequently \eqref{eq:B1} now reads
\bea
\nonumber
\frac{\dd^2\vBp}{\dd t^2} &=& \frac{\vthi^2}{2}\,
\dpar^2\lt[\lt(\Delta + \frac{2}{\beta_i}\rt)\vBp\rt.\\
&&+\lt.\frac{1}{\Omega_i}\lt(\frac{\dd\vBp}{\dd t}
- \gT\frac{\vthi\dpar\vBp}{B_0}\rt)\times\vb_0\rt].
\label{eq:B1th}
\eea
We have introduced a dimensionless parameter 
measuring the magnitude of the parallel heat flux:\footnote{We stress that 
we are discussing the effect of the {\em ion} heat flux as the electrons 
are assumed isothermal at the scales we are considering (see \apref{app:els}).
We also stress that these heat-flux effects enter through the FLR 
terms in the plasma pressure tensor and are absent in, e.g., 
the lowest-order \citet{Braginskii} equations.}
\beq
\gT = \frac{1}{2}\frac{\vthi}{\nuii}\frac{\vb_0\cdot\vdel T_{0i}}{T_{0i}}
= \frac{1}{2}\frac{\mfp}{l_T},
\label{eq:gTdef}
\eeq
where $l_T$ is the parallel length scale of the ion temperature variation. 
We see that the functional form of the firehose turbulence equation 
is changed. We now proceed to study the effect of this change. 

\subsection{Linear theory: the gyrothermal instability}
\label{sec:GTI_lin}

The linear dispersion relation for \eqref{eq:B1th} is 
\beq
\lt[\omega^2 - \frac{\kpar^2\vthi^2}{2}\lt(\Delta_0+\frac{2}{\beta_i}\rt)\rt]^2 
\!\!= \frac{\kpar^4\vthi^4}{4}\frac{\lt(\omega+\kpar\vthi\gT\rt)^2}{\Omega_i^2}.
\label{eq:DRth}
\eeq
Like in the case of \eqref{eq:DR}, there are four roots of which two 
are potentially unstable:
\beq
\frac{\omega}{\Omega_i} = \pm\frac{k^2}{4} 
+ i\,\frac{|k|}{\sqrt{2}}
\sqrt{-\lt(\Delta+\frac{2}{\beta_i}\rt) \mp k\gT -\frac{k^2}{8}},
\label{eq:GTI_disp}
\eeq
where $k=\kpar\rho_i$. Instability occurs at wavenumbers 
for which the expression under the square root is positive. 
There is an interval of such unstable wavenumbers if and only if
\beq
\Lambda \equiv \gT^2-\frac{1}{2}\lt(\Delta + \frac{2}{\beta_i}\rt) >0.
\label{eq:Lambda}
\eeq
If this condition is satisfied, the ``$+$'' mode is unstable for 
\beq
\label{eq:plusint}
-4\lt(\gT+\sqrt{\Lambda}\rt) < k < -4\lt(\gT-\sqrt{\Lambda}\rt),
\eeq
and the ``$-$'' mode for 
\beq
\label{eq:minusint}
4\lt(\gT-\sqrt{\Lambda}\rt) < k < 4\lt(\gT+\sqrt{\Lambda}\rt).
\eeq
where we have assumed, without loss of generality, that $\gT>0$.
When $\Delta+2/\beta_i<0$, these two intervals intersect, so 
all modes with $|k|< k_0=4\lt(\gT + \sqrt{\Lambda}\rt)$ are unstable 
(others are pure propagating waves). 
When $\Delta+2/\beta_i>0$, the intervals are separated and there 
is an interval of stability 
at long wavelengths, viz., $|k|<4\lt(\gT - \sqrt{\Lambda}\rt)$. 

What is remarkable about all this is that not only the stability 
conditions and specific expressions for the firehose growth rate 
are modified by heat flux, but the presence of the heat flux 
allows for instability even when firehose is stable, $\Delta+2/\beta_i>0$
(but positive pressure anisotropy not too large and $\beta_i$ not too small, 
subject to \eqref{eq:Lambda}). 
This instability, called the {\em gyrothermal instability (GTI)}, 
leads to the growth of Alfv\'enically polarized fluctuations 
in the parameter regime in which they are otherwise stable 
\citep{GTI}.\footnote{Note that for $\Delta-1/\beta_i>0$, 
the mirror mode is unstable as well, but it involves growth of 
compressive fluctuations, $\dBpar\gg\dBperp$, at highly 
transverse wavenumbers $\kpar\ll\kperp$ \citep[see, e.g.,][]{Hellinger2},
while Alfv\'enic fluctuations are not affected by it to lowest order 
in the instability parameter $\Delta-1/\beta_i$ \citep{Rincon}.} 

\begin{figure*}
\includegraphics[width=85mm]{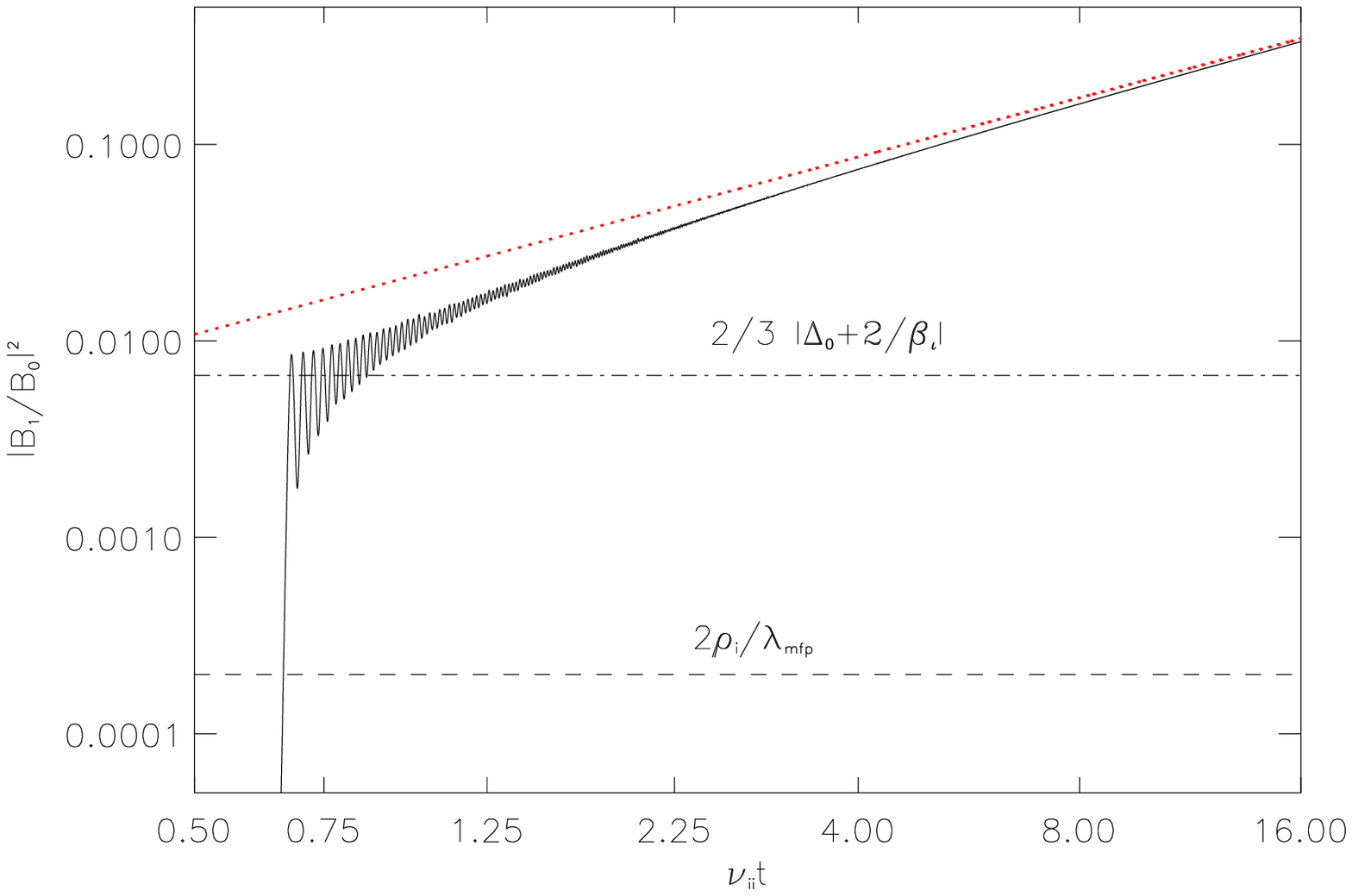}
\includegraphics[width=85mm]{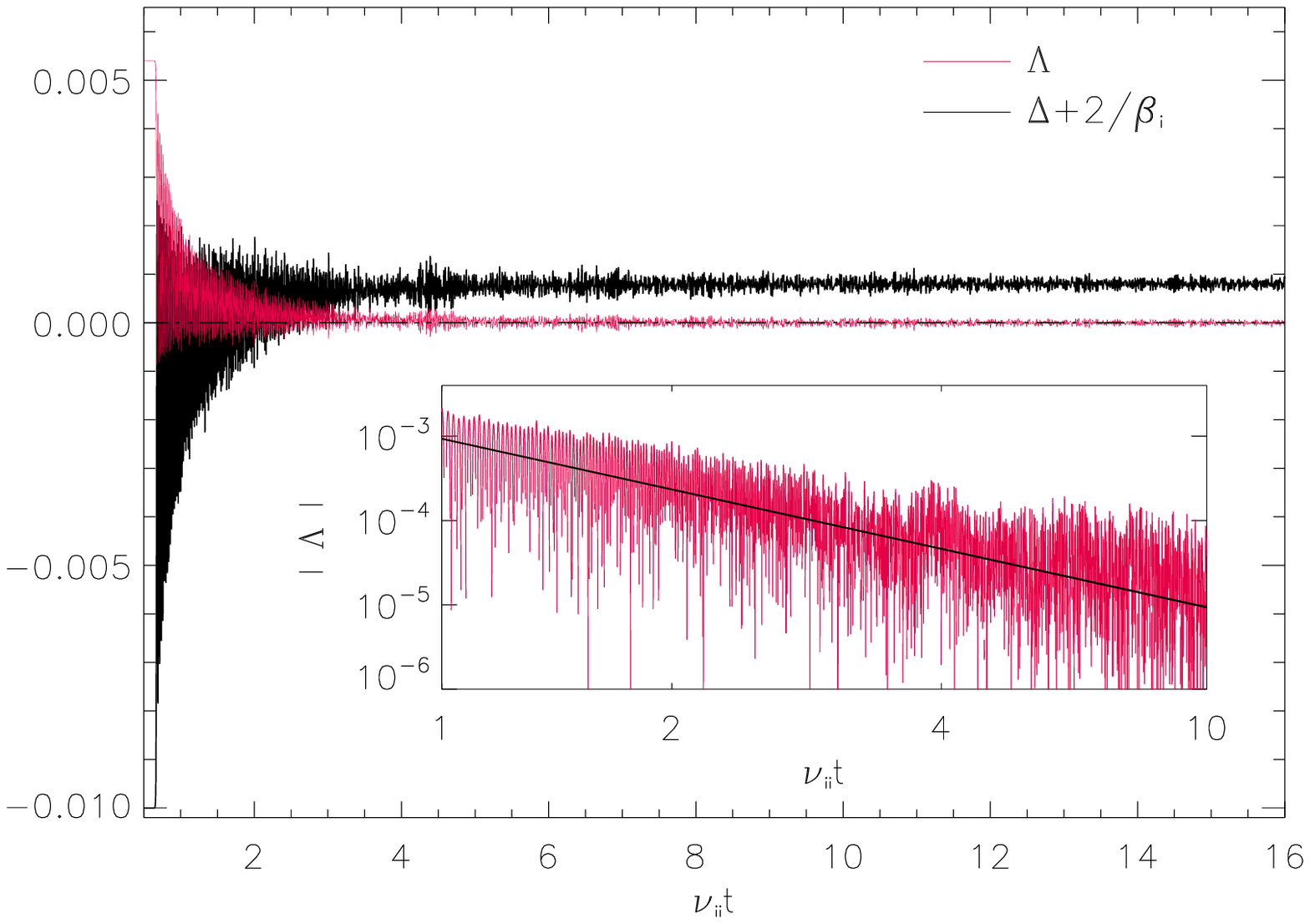}
\caption{{\it Left panel:} evolution of the magnetic energy 
$\overline{|\vBp|^2}/B_0^2 = \sum_k |A_k|^2$ with time 
in a numerical solution of \eqsand{eq:Ath}{eq:DeltaAth} with parameters \exref{eq:GTI_num_params}; 
the red dotted line shows the nonlinear asymptotic given by \eqref{eq:GTI_Atot};
this figure is the GTI analog of \ffigref{fig:secular}{left}. 
{\it Right panel:} evolution of the instability parameter $\Lambda$ (pink) 
and the pressure anisotropy parameter $\Delta + 2/\beta_i$ (black) in the same numerical 
solution. {\it Inset:} log-log plot of the evolution of $|\Lambda|$; the black 
line shows the slope corresponding to $1/t^2$ (see \eqref{eq:Lambda_to_zero}).}
\label{fig:GTI_secular}
\end{figure*}

The formulae for the wavenumber of the fastest-growing mode and the maximum growth 
rate for the combined firehose-GTI are straightforward to write down. 
As always with such formulae, they are not particularly 
illuminating in the general case, but are interesting in various
asymptotic limits. When the firehose instability parameter 
$\Delta+2/\beta_i<0$ and its magnitude is much larger than $\gT^2$, 
the effect of the heat flux is a small correction to the firehose 
instability already described in \secref{sec:disp_rln}. 
Conversely, when $|\Delta + 2/\beta_i|\ll\gT^2$, the GTI is 
dominant and, for the fastest growing mode, 
\bea
\kpeak &\simeq& \mp 6\gT,\\
\opeak &\simeq& 9\gT^2\lt(\pm1 + \frac{i}{\sqrt{3}}\rt),
\label{eq:GTI_omega}
\eea
where $\opeak$ is normalized to $\Omega_i$. 
Finally, close to the marginal state, $\Lambda\to+0$, we have
\bea
\label{eq:GTI_lastk}
\kpeak &\simeq& \mp 4\gT\lt(1+\frac{\Lambda}{\gT^2}\rt),\\
\opeak &\simeq& 4\gT^2\lt(\pm1 + \frac{i\sqrt{\Lambda}}{\gT}\rt).
\label{eq:GTI_marg}
\eea
Note that, unlike the firehose, the GTI has a definite preferred 
wavenumber that does not change as marginal stability is approached. 

\Ffigref{fig:gti}{left} shows the dependence of the frequencies and 
growth rates of the two unstable modes on wavenumber for a set 
of parameters for which the instability is a hybrid of firehose 
and GTI (these are the parameters used in the numerical solution 
of \secref{sec:GTI_num}). 
\Ffigref{fig:gti}{right} shows the same for the case 
in which the firehose is stable ($\Delta + 2/\beta_i>0$) and 
that is very close to marginal stability: we see that the instability 
only exists in the immediate neighbourhood of the last unstable wavenumber 
given by \eqref{eq:GTI_lastk}.

\subsection{Nonlinear evolution and spectrum}
\label{sec:GTI_nlin}

\subsubsection{Firehose-GTI turbulence equation in scalar form}
\label{sec:GTI_scalar}

As happened in \secref{sec:scalar}, \eqref{eq:B1th} can be reduced to 
one equation for a scalar field, although it is now a slightly more 
complicated transformation. Let us again non-dimensionalize time and 
space according to \eqref{eq:nondim} and introduce new fields $A^\pm_k(t)$ 
as follows: 
\beq
\frac{B_{1x}}{B_0} \pm i\,\frac{B_{1y}}{B_0} 
= A^\pm_k\exp\lt[\mp i\lt(\frac{k^2}{4}\,t + \phi_k\rt)\rt].
\label{eq:GTI_ansatz}
\eeq
With the ansatz \exref{eq:GTI_ansatz}, 
\eqref{eq:B1th} becomes
\bea
\label{eq:Ath}
\frac{\dd^2 A_k^\pm}{\dd t^2} &=& \frac{k^2}{2}\lt[-\lt(\Delta + \frac{2}{\beta_i}\rt)
\mp k\gT - \frac{k^2}{8}\rt]A_k^\pm,\\
\nonumber
\Delta(t) &=& \Delta_0 + \frac{3}{2}\int_0^t\did t' e^{-3\nu_*(t-t')}\times\\
&&\qquad\qquad\times\,\,\frac{\dd}{\dd t'}\sum_k\frac{|A_k^+(t')|^2+|A_k^-(t')|^2}{2}. 
\label{eq:DeltaAth}
\eea
It is now manifest how the dispersion relation \exref{eq:GTI_disp} emerges 
from \eqref{eq:Ath}. Unlike in the case of pure firehose turbulence 
($\gT=0$), the evolution of the mode now depends on the sign of its 
real frequency --- that is why we have two scalar equations. However, 
these equations have a symmetry: if we arrange initially that $A^+_k = A^-_{-k}$ 
(which we can always do by an appropriate choice of the phases $\phi_k$), 
then this relation will continue to be satisfied at later times. 
This also means that $A^\pm_k$ are real because, 
in order for $\vBp$ to be a real field, we must have (from \eqref{eq:GTI_ansatz})
$(A^+_k)^*=A^-_{-k}$ (we assume the phases satisfy $\phi_k=\phi_{-k}$).
The conclusion is that it is enough to solve just one of the two 
\eqsref{eq:Ath} --- either for the $+$ or the $-$ mode. The total energies of the 
two modes that enter \eqref{eq:DeltaAth} are equal.\footnote{The same approach 
could have been taken in \secref{sec:scalar}: instead of solving 
\eqref{eq:Ath} for a complex function $A_k$ subject to $A_k^*=A_{-k}$, we 
could have solved for one of two real functions $A_k^\pm$ subject 
to $A_k^+=A^-_{-k}$. The magnetic field is then recovered via 
\eqref{eq:GTI_ansatz}.} 

\begin{figure*}
\includegraphics[width=80mm]{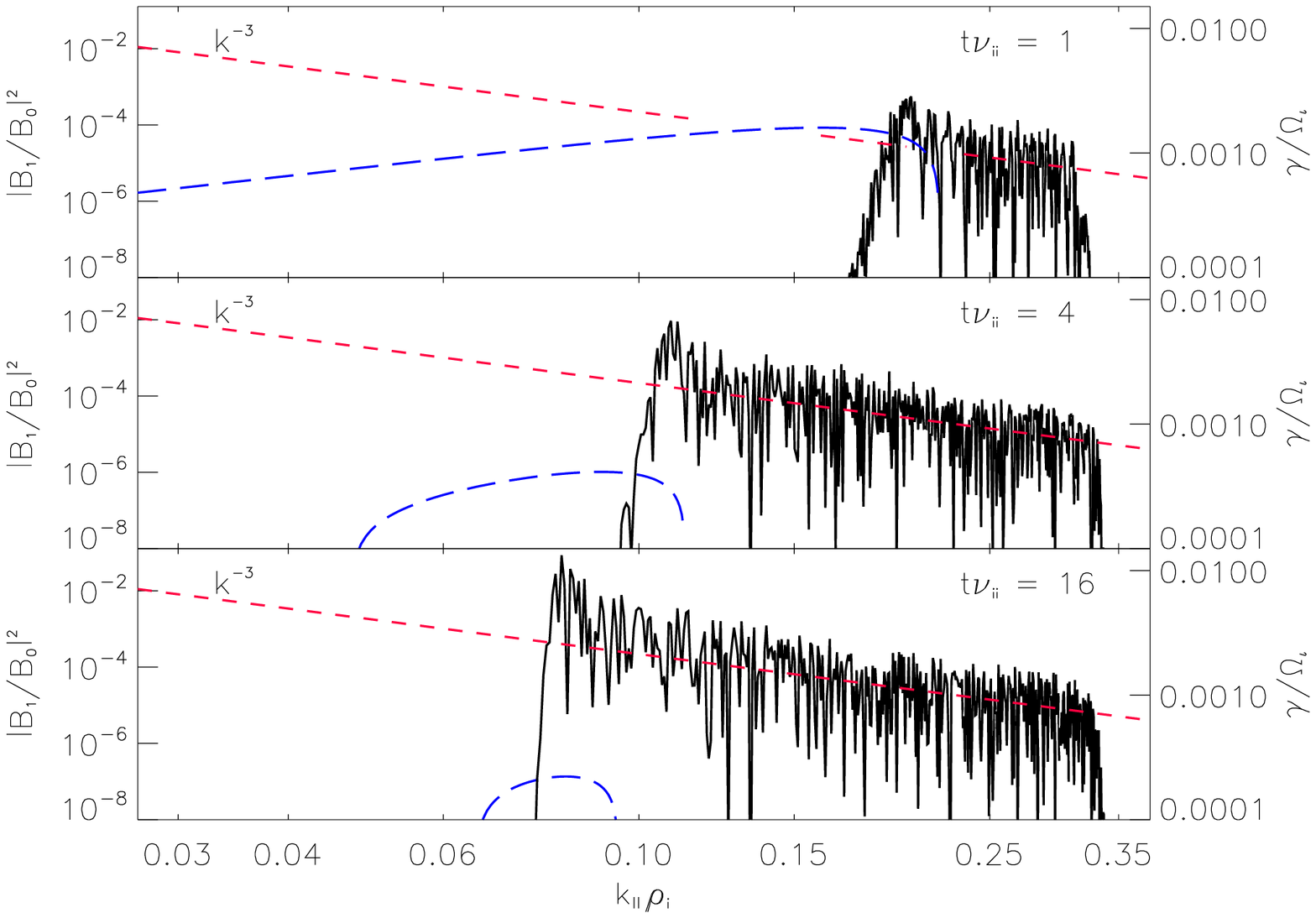}
\qquad
\includegraphics[width=80mm]{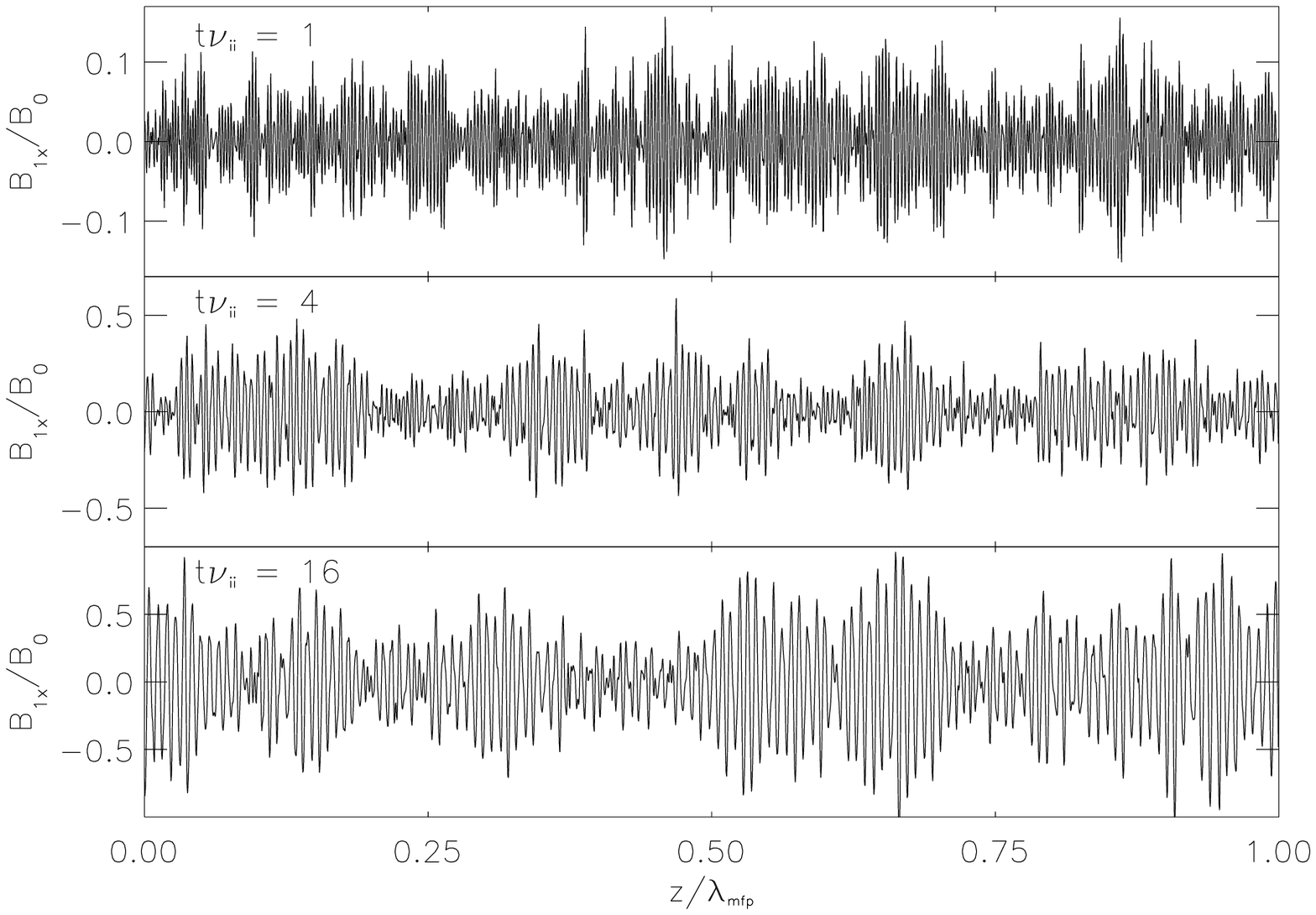}
\caption{{\it Left panel:} spectrum of the magnetic fluctuation energy 
at three specific times ($t\nuii = 1, 4, 16$) during the evolution 
shown in \figref{fig:GTI_secular}; red short-dashed lines show the $k^{-3}$ slope; 
blue long-dashed lines show the firehose/GTI growth rate (see \secref{sec:GTI_lin}) 
for the instantaneous values of $\Delta+2/\beta_i$ at the times the 
spectra are plotted. {\it Right panel:} magnetic fluctuations in real space 
($B_{1x}/B_0$ vs.\ $z$) at the same times as the spectra in the left panel.
This figure is the GTI analog of \figref{fig:spec}.}
\label{fig:spec_GTI}
\end{figure*}

\subsubsection{Qualitative picture}
\label{sec:GTI_qualit}

The evolution of the firehose-GTI turbulence is easy to predict 
arguing along the same lines as we did in \secref{sec:nlin_qualit}. 
Let us consider the case when initially the pressure anisotropy 
is negative and $-(\Delta_0 + 2/\beta_i) \gg 2\gT^2$,
i.e., the instability parameter $\Lambda_0>0$ (given by \eqref{eq:Lambda}
with $\Delta=\Delta_0$). In this regime, the heat flux does not matter 
and the evolution proceeds as in the case of the firehose turbulence: 
magnetic fluctuations grow and eventually the nonlinear feedback in 
\eqref{eq:DeltaAth} starts giving an appreciable positive 
contribution to the pressure anisotropy (estimates \exref{eq:nlin_on} 
and \exref{eq:nlin_on_colless} for the fluctuation amplitude at which 
this happens are still valid). A $k^{-3}$ spectrum will then form, with 
the infrared cutoff (wavenumber of maximum growth) moving to larger 
scales and $|\Delta + 2/\beta_i|$ decreasing (i.e., 
$\Delta + 2/\beta_i$ increasing and thus becoming less negative). 

The evolution of the gyrothermal fluctuations starts to differ from 
the pure firehose case after $|\Delta + 2/\beta_i|$ becomes comparable 
to $\gT^2$. The GTI is now the dominant instability mechanism. 
Since the fluctuations continue growing, $\Delta + 2/\beta_i$ continues 
to increase and will become positive, tending eventually to $2\gT^2$, 
so as to push the instability parameter $\Lambda$ (\eqref{eq:Lambda}) 
to zero and the GTI to its marginal state. As $\Lambda \to + 0$, 
the growth is concentrated in a shrinking neighbourhood of the 
wavenumber $\kpeak=4\gT$ (see \eqref{eq:GTI_lastk}). 
This means that the spectrum stops spreading towards lower 
wavenumbers and its infrared cutoff 
stabilizes at $\kpeak$. All the growth of magnetic energy is now 
provided by the growth of the one mode associated with $\kpeak$, 
which will soon tower over the rest of the spectrum. 

The growth is still secular: using \eqref{eq:DeltaAth} and the 
marginality condition $\Lambda=0$, we find to dominant order, 
analogously to \eqref{eq:Atot},
\beq
\label{eq:GTI_Atot}
\Delta\simeq 2\gT^2-\frac{2}{\beta_i}\quad\Rightarrow\quad
\sum_k|A_k|^2 \simeq 4\Lambda_0\nu_* t.
\eeq
Finally, we can calculate the evolution of the residual $\Lambda$. 
Analogously to \eqref{eq:mode}, the growing mode satisfies 
\beq
\frac{1}{A_{\kpeak}}\frac{\dd A_{\kpeak}}{\dd t} \sim \frac{1}{t}
\sim \gmax \sim 4\gT\sqrt{\Lambda}, 
\label{eq:GTI_mode}
\eeq
where we used \eqref{eq:GTI_marg} for $\gmax$. 
Therefore, 
\beq
\Lambda \sim \frac{1}{t^2}. 
\label{eq:Lambda_to_zero}
\eeq

As in the case of the firehose turbulence, the secular growth 
will continue until the fluctuation amplitude is no longer small: 
$t\sim (\nuii\Lambda_0)^{-1}\sim |\gamma_0|^{-1}$ 
(time scale of the large-scale dynamics). 
The key difference from the pure firehose case is that 
the fluctuations are now stuck at a microscopic spatial 
scale given by \eqref{eq:GTI_lastk}: restoring dimensions and using \eqref{eq:gTdef}, 
the corresponding wavenumber is 
\beq
\kpar\rho_i\sim \frac{\mfp}{l_T}
\label{eq:GTI_mink}
\eeq
(this scale is collisionless, $\kpar\mfp\gg1$, provided 
$l_T\ll\mfp^2/\rho_i$; for galaxy clusters, this is always 
true as is easy to ascertain by using the numbers 
from \secref{sec:clusters}). 
Thus, the gyrothermal turbulence is essentially {\em one-scale},
in the sense that fluctuations at this one scale become 
energetically dominant as marginal stability is approached 
at late stages of the nonlinear evolution.  

\subsubsection{Numerical solution}
\label{sec:GTI_num}

We have solved \eqsand{eq:Ath}{eq:DeltaAth} in a manner completely 
analogous to that described in \secref{sec:num}. 
The parameters we used are 
\bea
\nonumber
&&\Delta_0=-0.02,\quad 
\frac{2}{\beta_i}=0.01,\quad 
\gT=0.02,\\
&&\nu_*=\frac{\rho_i}{\mfp}=0.0001.
\label{eq:GTI_num_params}
\eea 
This implies that the instability parameter in the linear 
regime is $\Lambda=0.0054$ (\eqref{eq:Lambda}) 
and so the maximum unstable wavenumber is 
$k_0=4(\gT + \sqrt{\Lambda})\simeq 0.37$ 
(\eqsand{eq:plusint}{eq:minusint}; see \ffigref{fig:gti}{left}).
Our numerical solution now has 2048 wavenumbers, 
so $k_{\rm min}\simeq0.00063$ and $k_{\rm max}\simeq0.64$. 

As expected, the evolution of the total magnetic energy is 
similar to the case of pure firehose turbulence discussed 
in \secref{sec:num}: exponential, 
then secular growth (see \eqref{eq:GTI_Atot}) --- this is shown in 
\ffigref{fig:GTI_secular}{left}. The evolution of the instability 
parameter $\Lambda$ (\eqref{eq:Lambda}) towards its zero marginal 
value is given in \ffigref{fig:GTI_secular}{right}. The inset shows 
that this approach to zero is consistent with the $1/t^2$ 
prediction (\eqref{eq:Lambda_to_zero}). Also shown in 
\ffigref{fig:GTI_secular}{right} is 
the evolution of the pressure anisotropy parameter 
$\Delta+2/\beta_i$, which for the pure firehose 
used to be the instability parameter. Since $\Lambda\to0$, 
it should tend to $2\gT^2=0.0008$ and it indeed does. 

Finally, \ffigref{fig:spec_GTI}{left} illustrates the evolution of the 
spectrum of firehose/gyrothermal fluctuations. It follows the 
scenario outlined in \secref{sec:GTI_qualit}. At first it is 
similar to the firehose turbulence spectrum with the spectral 
peak moving towards larger scales leaving behind a $k^{-3}$ spectrum.
As the wavenumber of fastest growth $\kpeak$ approaches the 
value corresponding to the near-marginal GTI, 
$\kpeak=4\gT=0.08$ (see \eqref{eq:GTI_lastk}), 
the peak stays there and continues growing, eventually 
dominating all other modes. The emergence of a one-scale 
sea of gyrothermal fluctuations is further illustrated 
by \ffigref{fig:evolution}{right}, which shows what these 
fluctuations look like in real space as time progresses. 
The difference between them and the pure firehose fluctuations 
in \ffigref{fig:evolution}{left} is manifest: the gyrothermal 
ones stay at the same scale while the firehose ones become 
larger-scale as time progresses. 


\subsection{Implications for momentum and heat transport}
\label{sec:GTI_transp}

Let us now revisit the discussion of the effect of plasma 
instabilities on the momentum transport modification attempted 
for the pure firehose in \secref{sec:transp2}. 
As before, the combined large-scale viscous and Maxwell stress
is contained in the second term on the right-hand side 
of \eqref{eq:momtransp}. However, with parallel ion heat fluxes 
present, the nonlinear evolution of the GTI pushes 
the quantity $\Delta + 2/\beta_i$ not to zero but to 
a positive value $2\gT^2$, corresponding to the 
marginal state $\Lambda=0$ (\eqref{eq:GTI_Atot}). 
Since any smaller value of $\Delta + 2/\beta_i$ is 
GTI unstable, this leads to a curious conclusion that 
the momentum transport is now effectively 
determined by the ion heat flux:
\bea
\nonumber
m_in_{0i}\frac{\did \vu_{0i}}{\did t} &=& -\vdel\tilde p
+ \vdel\cdot\lt(p_{0i}\vb_0\vb_0 2\gT^2\rt)\\
&=& -\vdel\tilde p
+ \vdel\cdot\lt[\vb_0\vb_0 \frac{n_{0i}\lt(\vb_0\cdot\vdel T_{0i}\rt)^2}{m_i\nuii^2}\rt],
\label{eq:GTI_momtransp}
\eea
where we used \eqref{eq:gTdef} for $\gT$. This equation 
has to be supplemented with the transport and pressure-balance 
equations for $n_{0i}$, $T_{0i}$ and $\tilde p$ 
as explained in \apref{app:transp}. 

\Eqref{eq:GTI_momtransp} probably merits a careful study (which is outside the scope 
of this paper), but we would like to accompany it with a very important 
caveat. Since pressure anisotropy in the nonlinear state of the GTI 
can be positive, other plasma instabilities may be triggered. Thus, if 
$\Delta>1/\beta_i$, i.e., if $\gT^2>3/(2\beta_i)$, the plasma will be mirror 
unstable \citep[see][and references therein]{Hellinger2}. 
The magnetic fluctuations that the mirror instability produces 
are different from the GTI both in polarization ($\dBpar$, not $\dBperp$) 
and scale ($\kperp\gg\kpar$,  
$\kperp\rho_i\sim(\Delta-1/\beta_i)^{1/2}$, 
$\kpar\rho_i\sim\Delta - 1/\beta_i$ for the mirror, 
whereas for the GTI we had $\kperp=0$, $\kpar\rho_i\simeq 4\gT$). 
How they saturate and what they do to the effective 
pressure anisotropy is a matter under active current investigation 
\citep{Califano,Istomin,Rincon} --- and it is completely unknown 
how mirror and gyrothermal fluctuations might coexist.

The key question is whether the pressure anisotropy 
will be set by the GTI or the mirror marginal condition 
and if it is set by the latter ($\Delta = 1/\beta_i$, as, e.g., seems to be 
indicated by the solar wind data; see \citealt{Hellinger1,Bale2}), 
then whether a turbulent plasma has a way of suppressing 
the GTI by adjusting not the pressure anisotropy, but 
the heat flux to the marginal condition: $\gT^2=3/(2\beta_i)$. 
This raises the possibility that not only the pressure anisotropy 
but also the (ion) heat fluxes are determined by the marginal stability 
conditions of the firehose/GTI and mirror. Thus, plasma 
instabilities may be the crucial factor in 
setting both the momentum and heat transport properties 
of a weakly collisional plasma. 
We stress, however, that under the assumptions adopted in this 
paper, we have not produced a nonlinear mechanism for changing 
the ion heat flux and this remains a subject for future work. 


\section{Discussion and comparison with previous work}
\label{sec:diff}

\subsection{Marginal stability via particle scattering or via changing field structure?}
\label{sec:scatt}

It is not in itself particularly surprising that the nonlinear 
effect of an instability driven by pressure anisotropy is to 
produce fluctuations that effectively pin this pressure anisotropy 
at a value corresponding to marginal stability. 
Besides having direct observational support in the solar wind 
\citep{Gary2,Kasper,Marsch,Hellinger1,Matteini2,Bale2}, 
it makes sense as a fundamental theoretical expectation 
(Le Ch\^atelier's principle). One may be tempted to 
proceed to another, seemingly as reasonable, theoretical expectation 
that the mechanism for achieving this marginal state must be 
pitch-angle scattering of particles by the fluctuations leading 
to isotropization of pressure. 
While indeed physically reasonable, this is, however, not 
an inevitable conclusion. As we have shown above, particle scattering is, 
in fact, {\em not} the way the $\kperp=0$ firehose fluctuations make pressure 
anisotropy marginal (under the ordering assumptions we have adopted). 
Instead, the marginal state is achieved via a modification of the structure 
of the magnetic field: namely, secular growth of the microscale fluctuations 
cancels on average the decrease in the mean field that produced 
the pressure anisotropy thus pushing the latter to its marginal value. 
This was explained on an intuitive level in \secref{sec:nonlin} and the 
subsequent analytically rigorous developments showed that intuition 
to be correct. 

Considering this result, we must recognize it as physically 
reasonable on the following grounds. 
A particle travelling in a magnetic field will traverse a fluctuation 
with a given $\kpar$ over time $\sim1/\kpar\vthi$. 
This time is much longer than the ion cyclotron period 
if $\kpar\vthi\ll\Omega_i$, or, equivalanetly, if $\kpar\rho_i\ll 1$. 
If this condition is satisfied and if the frequency of the fluctuation 
$\omega\ll\Omega_i$, the fluctuation cannot change the first adiabatic 
invariant $\mu = \vperp^2/2B$ of the particle, so there cannot be 
very much pitch-angle scattering. 
In our calculation, as the pressure anisotropy (or, more precisely, 
the instability parameters $\Delta+2/\beta_i$ and $\gT$) were 
small, the parallel scale of the fluctuations generated by the $\kperp=0$ 
firehose or gyrothermal instabilities was substantially larger than 
the Larmor scale (see \secsand{sec:disp_rln}{sec:GTI_lin}) and, 
in the case of the firehose, it increased further in the nonlinear 
regime (see \secref{sec:nlin_qualit}). Thus, $\kpar\rho_i\ll1$ was 
satisfied at all times (as was $\omega\ll\Omega_i$), the plasma remained magnetized 
and pitch-angle scattering ineffective, so the rearrangement of the field structure 
was the only device available to the system to counteract the pressure 
anisotropy drive. It is possible that the oblique firehose 
(which is much harder to treat analytically than the parallel one) 
might produce fluctuations at the ion Larmor scale, so 
particle scattering by firehose fluctuations is not completely 
ruled out, but it certainly does not happen for the $\kperp=0$ 
case to which we have limited the scope of the present investigation. 

How important is it to know whether particle scattering is present? 
Recently, in the context of accretion-disc physics, \citet{Sharma1,Sharma2} 
proposed an {\em ad hoc} closure for numerical simulations, 
constraining the pressure anisotropy to lie within the marginal 
stability boundaries via artificial dissipation in the pressure equations
(the CGL equations given in \apref{app:CGL}). They argued that this was justified 
if it could be shown microphysically that plasma instabilities 
(in their case, ion cyclotron and firehose) produced fluctuations 
at the ion Larmor scale, where pitch-angle scattering of particles 
off the fluctuation ``foam'' isotropized pressure.\footnote{The same view 
was taken by \citet{Schek2} in their model 
of the dynamo action in a weakly collisional plasma and 
by \citet{Bale2} in interpreting their measurements of 
marginal pressure anisotropies in the solar wind.} 
As we have explained, our results for the parallel ($\kperp=0$) firehose 
do not support this picture. However, 
it is not obvious that the validity of a closure based on 
the {\em average} pressure anisotropy being maintained 
at the marginal level must be predicated on the presence of particle 
scattering. As we have shown above, a sea of secularly growing magnetic 
fluctuations far above the Larmor scale can produce the same effect. 
This, of course, does not excuse us from having to find the right 
microphysical theory for pressure isotropization if we are ever 
to have anything more than a plausible closure imposed by fiat.

One example of a context in which the presence or absence of scattering matters 
greatly is viscous heating of the plasma. The heating depends both 
on the pressure anisotropy and on the collision frequency 
(it is $\propto\nuii\Delta^2$; see \citealt{Kunz}), so, 
in order to calculate it correctly, we must know whether only 
the pressure anisotropy or also the (effective) collision frequency 
is modified by the firehose fluctuations. Assuming the Coulomb collision 
frequency unchanged, \citealt{Kunz} recently proposed a thermally 
stable heating mechanism for galaxy clusters (see a further 
short discussion in \secref{sec:cflows}). If microphysically 
justified in the most general case (i.e., not only for the 
parallel firehose, but also the oblique one, the mirror instability, etc.), 
this represents significant progress. 
Thus, having a detailed microphysical theory does make a difference 
not only for the analytical 
strength of the subject but also for explaining astronomically 
observed realities. 

\subsection{Quasilinear theories}
\label{sec:QL_prev}

The theory developed above is basically quasilinear in that 
the fluctuation amplitude is assumed small and it is found that 
such small fluctuations can drive the instability to 
a marginal state. There have been a number of 
quasilinear treatments of the firehose instability 
\citep{Shapiro,Kennel,Davidson,Gary0,Quest}, so it is perhaps 
useful to explain why they do not obtain similar results. 

The approach in such theories is to consider 
a {\em collisionless} plasma with some 
initial distribution that has a negative pressure anisotropy 
(let us call it $\Delta_0<0$) and work out how it relaxes.
The result is that a fluctuation level builds up, 
with\footnote{\citet{Hall2} argues qualitatively for a similar saturation 
level, but due to trapping of particles in firehose fluctuations. As the 
systematic kinetic calculation presented above shows, the trapping effect 
does not play a role at these amplitudes, at least not under the assumptions 
we adopted ($\kperp=0$ and relatively high collision frequency).} 
\beq
\frac{|\vBp|^2}{B_0^2}=\frac{2}{3}\lt|\Delta_0 + \frac{2}{\beta_i}\rt|,
\label{eq:QL_standard}
\eeq
which is small when the instability parameter $\Delta_0+2/\beta_i$ is small. 
This saturated fluctuation level suffices to 
marginalize the instability. 
This result is easily recovered in our theory if 
we formally set $\nuii=0$ in \eqref{eq:Delta_nlin}. 
This gives
\beq
\label{eq:Delta_QL}
\Delta(t) = \Delta_0 + \frac{3}{2}\frac{\overline{|\vBp(t)|^2}}{B_0^2}
\eeq 
and assuming saturation in the marginal state $\Delta=-2/\beta_i$, 
we recover \eqref{eq:QL_standard}. The classic 
work where this was first done is \citet{Shapiro} (a more 
detailed comparison is provided in \apref{app:ShSh}). 
We stress that in their calculation the saturation of the pressure 
anisotropy at the marginal level is not due to particle scattering 
any more than it was in ours because the fluctuations 
still have $\kpar\rho_i\ll1$, $\omega\ll\Omega_i$ and so 
conserve $\mu$ (see \secref{sec:scatt}). The internal energy stored 
in the pressure anisotropy is transferred into magnetic fluctuations 
until the pressure anisotropy is marginal. The magnetic fluctuations 
then persist because under the adopted appriximations there is no 
dissipation of the magnetic field. 

The difference in our approach is to include weak collisions 
and consider the case when the pressure anisotropy is constantly 
{\em driven} by the large-scale dynamics (which is physically 
where it comes from; see \secref{sec:aniso}). The steady level of 
the anisotropy is then set by the competition between collisions 
and the drive (\eqref{eq:Delta0}) and to offset this anisotropy and 
keep the instability marginal, the fluctuation level has to keep 
growing secularly rather than stay constant 
(the earlier quasilinear theories can then be 
interpreted to describe correctly what happens before one collision 
time has elapsed). 

\subsection{Driven anisotropy in a collisionless plasma}

It is interesting to inquire what would happen if the 
anisotropy were driven (rather than just initially imposed) 
but collisions not strong enough to balance the drive and 
impose a steady anisotropy. Formally speaking, our theory 
breaks down in this case because the equilibrium distribution 
cannot be proved Maxwellian. However, that is a technical issue 
and one could, in fact, reformulate our theory as a near-marginal 
expansion in the instability parameter $|\Delta+2/\beta_i|$. 
We expect that \eqref{eq:aniso} (or, equivalently, 
\eqref{eq:anisoB}), with the collisional relaxation term
removed, would still describe the evolution of the anisotropy: 
\beq
\frac{\dd\Delta}{\dd t} = 3\gamma_0 
+ \frac{3}{2}\frac{\dd}{\dd t}\frac{\overline{|\vBp(t)|^2}}{B_0^2},
\label{eq:Delta_colless}
\eeq  
where $\gamma_0$ (assumed negative) is the drive --- it contains all 
the terms in \eqref{eq:aniso} due the large-scale dynamics. 
Under these conditions, the driven part of anisotropy is constantly 
increasing and so again the fluctuations will have to grow 
secularly in order to keep it at the marginal level:
\beq
\frac{3}{2}\frac{\overline{|\vBp(t)|^2}}{B_0^2} = 
\lt|3\int_0^t\did t'\gamma_0(t') + \frac{2}{\beta_i}\rt|.
\label{eq:colless}
\eeq
This will, of course, break down once the fluctuation level 
is no longer small or collisions catch up. 

\subsection{Other nonlinear theories and simulations}

There exist a number of numerical studies of the nonlinear evolution 
of the firehose instability 
\citep{Berezin,Quest,Gary1,Hellinger4,Horton1,Horton2,Matteini1}. 
They mostly adopted the same relaxation-of-initial-anisotropy 
approach as the quasilinear theories discussed above 
and the results they report are broadly consistent in that the magnetic fluctuation 
energy saturates at a level scaling with the size of the initial anisotropy 
(see \eqref{eq:QL_standard}). A notable exception is the 
recent work of \citet{Matteini1} who consider the anisotropy driven 
by the expansion of the solar wind --- the fluctuation levels 
they see are probably well described by \eqref{eq:colless}. 

The spectrum of the firehose fluctuations has not previously been 
addressed analytically, but, perhaps vaguely in agreement with the 
results of \secref{sec:results}, some of the numerical evidence 
does point to the growing predominance of smaller wavenumbers 
in the nonlinear regime --- as the anisotropy approaches 
marginal level \citep{Quest,Matteini1}. 

Finally, to our knowledge, the effect of heat fluxes on the 
nonlinear behaviour of the firehose instability, studied 
in \secref{sec:GTI}, has not been specifically considered before. 
Note that although the heat fluxes are present 
in the numerical simulations of \citet{Sharma1,Sharma2}, 
their momentum equation does not have the gyroviscous 
and gyrothermal terms that regularize the firehose 
at small scales and give rise to the gyrothermal instability. 
The appropriate modification to the fluid equations suggested 
by \citet{GTI} should in principle enable one to study 
the spectrum of the firehose and GTI fluctuations numerically. 

\section{Astrophysical implications}
\label{sec:apps}

\subsection{Solar wind}

Much of the observational evidence about the firehose instability 
comes from the measurements of pressure anisotropies and 
fluctuation levels in the solar wind.\footnote{There are also 
some measurements indicating the presence of firehose 
fluctuations in the Earth's magnetotail 
\citep[see][and references therein]{Horton1,Horton2}.} 
Since the wind is expanding, 
both the local density and the local magnetic-field strength 
are dropping, so one expects a negative pressure anisotropy to develop: 
this can be described by \eqref{eq:Delta_colless}, where 
the drive is roughly $\gamma_0\sim -V_{\rm sw}/R$ (solar wind speed divided by the 
distance from the Sun) and the collisional 
relaxation is neglected.\footnote{Note, however, that although the mean free 
path in the solar wind is roughly comparable to 1 AU, one does find 
in the solar wind a strong correlation between pressure anisotropy and 
the estimated collisional age \citep{Bale2}, so modelling the solar-wind 
plasma as completely collisionless is possibly less valid than it 
might appear.} 
The evidence for this trend, negative pressure anisotropy developing 
with increasing distance from the Sun, is given by \citet{Matteini2}; 
a number of other papers also document the fact that 
the measured pressure anisotropies are bounded from below 
by the firehose marginal stability condition 
\citep{Kasper,Hellinger1,Bale2}.

\citet{Bale2} found increased levels of ion-Larmor-scale magnetic fluctuations 
close to this stability boundary --- presumably due to the firehose instability.  
There are also indications of an injection of energy into parallel 
wavenumbers just above the ion Larmor scale \citep{Podesta,Wicks} --- again, 
conceivably by the firehose instability. 
It is unclear how these firehose fluctuations coexist with the 
solar wind inertial- and dissipation-range turbulence --- as our 
theoretical understanding of this turbulence is still largely based 
on assuming isotropic equilibrium distributions 
\citep[see][and references therein]{Schek5}. 
This is one of the contexts in which the absence of a complete microphysical 
theory of the firehose turbulence and its effect on the 
plasma motions is particularly acutely felt. 

\subsection{Accretion discs}

Another such astrophysical context is hot accretion flows, 
in which the theoretical modelling of the 
long-standing problem of the angular momentum transport 
and radiative efficiency or inefficiency of the accretion 
has taken a new turn with the introduction of pressure anisotropies 
\citep{Quataert1,Sharma0,Balbus2,Islam,Sharma1,Sharma2}. 
The numerical model of \citet{Sharma1,Sharma2} consisted of a 
closure that pinned down the pressure anisotropies at 
marginal stability via artificial dissipation terms. 
As we explained in \secref{sec:scatt}, their 
assumption of microscale fluctuations 
scattering particles is not bourne out by the theory 
developed above for the parallel ($\kperp=0$) firehose 
fluctuations, although it is not excluded for other instabilities 
and, in any event, a closure based on marginal stability 
is probably a sensible choice. 

A key remaining unknown here is the fate of the microscale fluctuations 
over long (transport) time scales and the eventual structure 
of the tangled magnetic field that results --- a crucial question 
for accretion theories because they require knowledge of 
the Maxwell and Braginskii stresses in order to estimate the rate 
of the angular momentum transport \citep{Shakura}. The same problem 
of the magnetic-field structure arises in considerations of the 
ICM dynamics and magnetogenesis (see \secsand{sec:clusters}{sec:imp_clusters}). 

\subsection{Galaxy clusters}
\label{sec:imp_clusters}

In \secref{sec:qualit}, we discussed at length 
the basic properties of the galaxy cluster plasmas, the 
inevitability of pressure anisotropies and, therefore, plasma 
instabilities arising in a turbulent ICM, as well as the fundamental 
theoretical questions that this poses. These will not find their 
final resolution in this paper because it has only analyzed one 
of several plasma instabilities that must be understood. 
However, just like in the case of the solar wind and the accretion 
flows, an impatient astrophysicist can conceivably glimpse the 
contours of the eventual theory by constraining pressure anisotropies
and possibly also heat fluxes by the marginal stability conditions 
of the plasma instabilities --- with all the caveats and 
uncertainties already discussed above. 

Let us discuss how far this approach can take us in answering 
the three classes of physical problems that were described at the 
beginning of \secref{sec:clusters}.  

\subsubsection{Regulation of cooling flows} 
\label{sec:cflows}

The apparent refusal of the galaxy cluster cores to exhibit a cooling 
catastrophe \citep[e.g.,][]{Peterson} 
has long evaded a satisfactory theoretical explanation.
A comprehensive review of the relevant literature is outside 
the scope of this brief discussion. It is probably fair to summarize 
the two main physical mechanisms invoked to explain the relatively 
weak drop in the ICM temperature between the bulk and the core 
as thermal conduction and some form of viscous conversion into heat 
of the mechanical energy injected into the ICM 
by the central active galactic nuclei (probably in a 
self-regulating way; see, e.g., \citealt{Binney,Kaiser,Omma2,Ogrean,Teyssier}
and the references in \secref{sec:clusters}). 
It is clear that the latter mechanism cannot be ignored 
because the thermal conductivity of the ICM is unlikely 
to be sufficiently large \citep[e.g.,][]{Voigt} and  
at any rate, thermal conduction is a thermally unstable mechanism 
of balancing radiative cooling. \citet{Kunz} recently proposed that 
if a sufficient amount of turbulent power is assumed to be 
available, the viscous heating, regulated by the pressure 
anisotropy and, therefore, by the marginal stability of the 
mirror and/or firehose instabilities, can balance the cooling in 
a thermally stable way. They also found that assuming such a 
balance leads to reasonable predictions of the magnetic 
field strength, magnitude of the turbulent velocities and 
the outer scale of the turbulence in the ICM. 

\subsubsection{Temperature fluctuations and the GTI}

While detailed simulations of the turbulent ICM, 
bubble dynamics etc.\ similar to those of \citet{Sharma1,Sharma2} for accretion 
flows have not been attempted, 
the marginal stability condition for the GTI (\eqref{eq:Lambda}) 
could perhaps be used to impose a lower bound on the 
typical scale of temperature fluctuations in the ICM \citep{GTI}. 
Indeed, {\em if} the magnitude of the ion heat flux is 
limited so as to prevent the GTI from being unstable (see 
\secref{sec:GTI_transp}), then from \eqsand{eq:Lambda}{eq:gTdef}, 
we get
\beq
l_T \gtrsim \beta_i^{1/2}\mfp 
\sim 5\times10^{-4}\frac{T_i^{5/2}}{n_i^{1/2}B}
\sim 1.4\times10^{21}~{\rm cm},
\label{eq:lT}
\eeq
where $n_i$ is in cm$^{-3}$, $T_i$ is in K, $B$ is in G, 
and the numerical value has been computed for the plasma 
parameters in the core of Hydra A discussed in \secref{sec:clusters}
(see \eqsand{eq:mfp}{eq:beta}). Interestingly, kpc-scale 
temperature fluctuations are indeed observed in cool-core 
clusters \citep{Simionescu,Fabian4,Sanders3,Lagana}. 
Furthermore, if we use in \eqref{eq:lT} the physical parameters 
appropriate for the bulk of the cluster plasma, rather than 
the cores (say, $T_i\sim10^8$~K and $n_i\sim10^{-3}$~cm$^{-3}$) 
we would get much larger scales --- in the 100~kpc range, which 
is also consistent with reported observational values for the 
cluster bulk \citep{Markevitch2}. 

\subsubsection{Magnetogenesis}

In the presence of turbulence, the small-scale (fluctuation) dynamo 
mechanism generates a magnetic field --- this is certainly 
true in an MHD fluid \citep[e.g.,][]{Subramanian,Brandenburg}. 
How this mechanism works in a plasma susceptible to 
the microscale plasma instabilities remains a completely open problem. 
A rather speculative attempt by \citet{Schek2} 
to leapfrog the detailed microphysical derivations and model the 
large-scale dynamics based on the idea that the instabilities would always 
isotropize pressure towards marginal stability values 
led to a rather dramatic conclusion 
that the ICM might support self-accelerating, 
explosive dynamos. While this conclusion remains to be tested 
by more rigorous analytical approaches, it does illustrate the general 
conjecture that plasma instabilities are likely to result in radical changes 
of, rather than merely small corrections to, the large-scale dynamics of 
cosmic plasmas. A particular mystery in understanding the origin and 
structure of the magnetic field in the ICM is what determines the 
typical spatial scale of magnetic fluctuations, which observations 
suggest may be substantially smaller than the scale of the turbulent 
motions \citep[see further discussion and references in][]{Schek2}.\footnote{An example 
of such observations is \citet{Vogt}, although a more recent paper 
by the same group appears to revise this result \citep{Kuchar}.} 

\section{Conclusion}
\label{sec:conc}

Let us recapitulate what this paper has achieved and how it relates 
to what was known previously. It has been appreciated for some time 
that macroscale turbulence of magnetized weakly collisional plasma 
(exemplified by the ICM) will naturally produce pressure anisotropies, 
which will in turn trigger firehose 
and mirror instabilities at spatial and temporal microscales 
\citep[][see extended discussion in \secsand{sec:intro}{sec:qualit}]{Hall1,Schek3}. 
Since the pressure anisotropies are essentially due to local 
temporal change of the magnetic field strength, it is qualitatively 
intuitive that the nonlinear evolution of the instabilities is
governed by the tendency to cancel this change on average; hence 
it follows that in a driven system (see discussion in \secref{sec:diff})
the fluctuations must continue growing in the 
nonlinear regime, albeit secularly rather than exponentially 
\citep[][see \secref{sec:nonlin}]{Schek4}. 

In this paper, we have 
constructed a full {\em ab initio} (weakly) nonlinear kinetic theory 
of this process for the parallel ($\kperp=0$) firehose instability, 
which is the simplest analytically tractable case. 
The evolution not only of the fluctuation energy, but also of the 
full spectrum of the resulting firehose turbulence has been worked 
out, including the effect of gradual spreading of the fluctuations 
to ever larger scales as the nonlinearly compensated pressure anisotropy 
approaches its marginal-stability value (\secref{sec:results}). 
We have also extended our kinetic calculation to include the effect 
of ion temperature gradients parallel to the magnetic field 
(parallel heat fluxes). As was pointed out recently, they lead
to a new instability, the GTI, of parallel Alfv\'enic fluctutions
\citep[][see also \secref{sec:GTI_lin}]{GTI}. Here we have constructed 
a nonlinear theory of its evolution, featuring again a secular growth 
of magnetic fluctuations, but this time developing a spectrum 
heavily dominated by a particular scale (\secref{sec:GTI_nlin}). 

While a speculative discussion of the implications of
these results for transport in a general magnetized plasma 
(\secsdash{sec:transp2}{sec:GTI_transp}) 
and for particular astrophysical systems (\secref{sec:apps}) 
is possible, a full transport theory has to await, at the very least, 
the completion of similar {\em ab initio} kinetic investigations of 
the nonlinear evolution of the mirror instability 
\citep{Rincon} and of the oblique ($\kperp\neq0$) firehose.\footnote{There is 
a distinct possibility that constructing the most general theory will involve having to 
study how mirror and firehose/GTI fluctuations coexist (see \secref{sec:GTI_transp}).} 
Only then can one attempt to devise an effective mean field 
theory for the macroscale dynamics of cosmic plasmas 
based on solid microphysical foundations. 
A goal of this paper has been to establish a template 
for building these microphysical foundations. 

In the meanwhile, it appears sensible to rely on 
(or at least consider reasonable) the semiquantitative 
closure approach to the macroscale dynamics based on the assumption 
that average pressure anisotropies and, probably, also heat fluxes, 
are set by the marginal stability conditions of the microscale 
plasma instabilities --- an approach that has found strong 
observational support in the solar wind measurements 
\citep{Gary2,Kasper,Marsch,Hellinger1,Matteini2,Bale2} and has already 
yielded nontrivial and possibly sensible physical predictions for the 
evolution of cosmic magnetism \citep{Schek2}, 
accretion disk dynamics \citep{Sharma1,Sharma2}, 
and the turbulence and heating in the intracluster 
medium \citep{Lyutikov,Kunz} (see further discussion in \secref{sec:apps}).


\section*{Acknowledgments}
We thank S.~Balbus, S.~Bale, J.~Binney, D.~Burgess, W.~Dorland, 
G.~Hammett, T.~Heinemann, 
P.~Hellinger, R.~Kulsrud, M.~Kunz, M.~Markevitch, T.~Passot, E.~Quataert, 
J.~Stone and P.-L.~Sulem for useful discussions and suggestions 
at various stages of this project. 
This work was supported by an STFC studentship (MSR),
an STFC Advanced Fellowship (AAS), 
the STFC Astronomy Grant ST/F002505/2 (AAS and SCC),
the Leverhulme Trust International Network for Magnetized 
Plasma Turbulence (FR's travel), 
EPSRC and the European Communities under the contract of association 
between EURATOM and CCFE (SCC). 
The views and opinions expressed here do not necessarily reflect 
those of the European Commissioners.

\onecolumn
\appendix

\section{Kinetic theory: detailed derivation}
\label{app:main}

\subsection{Electrons}
\label{app:els}

\subsubsection{Mass-ratio ordering}

The kinetic equation \exref{eq:Vlasov} for electrons is (recall that $\vv$ is the peculiar velocity)
\beq
\label{eq:Vlasov_e}
\begin{array}{cccccccccccccccccccc} 
\displaystyle{\frac{\dd f_e}{\dd t}} & + &
\displaystyle{\vu_e\cdot\vdel f_e} & + &
\displaystyle{\vv\cdot\vdel f_e} & - &
\displaystyle{\Bigg[\frac{e}{m_e}} &
\displaystyle{\bigg(\vE + \frac{\vu_e\times\vB}{c}} & + &
\displaystyle{\frac{\vv\times\vB}{c}\bigg)} & + &
\displaystyle{\frac{\dd\vu_e}{\dd t}} & + &
\displaystyle{\vu_e\cdot\vdel\vu_e} & + &
\displaystyle{\vv\cdot\vdel\vu_e\Bigg]\cdot} &
\displaystyle{\frac{\dd f_e}{\dd \vv}} & = &
\displaystyle{\St{f_e},} \\
1&&
1&&
\lt(\frac{m_e}{m_i}\rt)^{-\frac{1}{2}}&&&
\lt(\frac{m_e}{m_i}\rt)^{-\frac{1}{2}}&&
\lt(\frac{m_e}{m_i}\rt)^{-1}&&
\lt(\frac{m_e}{m_i}\rt)^{\frac{1}{2}}&&
\lt(\frac{m_e}{m_i}\rt)^{\frac{1}{2}}&&
1 &&&
\lt(\frac{m_e}{m_i}\rt)^{-\frac{1}{2}}
\end{array}
\eeq
where we have labeled all terms according to their ordering 
in powers of $(m_e/m_i)^{1/2}$, while taking $T_e\sim T_i$  
The ordering has been done relative to $k\vthi f_e$ and we have assumed 
\bea
&&\frac{\dd}{\dd t}\sim\omega\sim k\vthi,\quad
\vdel\sim k\sim \rho_i^{-1} \sim \lt(\frac{m_e}{m_i}\rt)^{1/2} \frac{eB}{m_e c\,\vthe},\quad
\vu_e\sim\vthi,\quad
\vv\sim\vthe \sim \lt(\frac{m_e}{m_i}\rt)^{-1/2}\vthi,\quad
\vE\sim\frac{\vu_e\times\vB}{c},\\
&&\nu_{ei}\sim\nu_{ee}\sim \lt(\frac{m_e}{m_i}\rt)^{-1/2}\nu_{ii},\quad
\nuii \sim \omega,
\label{eq:elcoll_order}
\eea
where $\nu_{ei}$ and $\nu_{ee}$ are the electron-ion and electron-electron collision 
frequencies (they determine the ordering of the collision integral on the right-hand 
side of \eqref{eq:Vlasov_e}). 
We stress that these are formal orderings with respect to the mass-ratio expansion, 
not statements about the exact size of various quantities and their derivatives: 
thus, some of the quantities ordered as unity within the mass-ratio expansion 
(e.g., $k\rho_i$ or $\vu_e/\vthi$) 
will be ordered small in the subsidiary $\eps$ expansion to be used in 
solving the ion kinetics (see \secref{sec:order}). 

We now expand the electron distribution function in powers of $(m_e/m_i)^{1/2}$: 
$f_e = \feo + \feone + \cdots$. It turns out that we can learn all we 
need to know from just the two lowest orders in the expansion of \eqref{eq:Vlasov_e}. 
Note that we do not expand any of the fields --- exact $\vE$ and $\vB$ are kept. 

\subsubsection{Order $(m_e/m_i)^{-1}$: gyrotropic electrons}
\label{app:e_gyro}

To this order, \eqref{eq:Vlasov_e} is
\beq
-\frac{e}{m_e}\frac{\vv\times\vB}{c}\cdot\frac{\dd\feo}{\dd\vv} 
= -\Omega_e\,\frac{\dd\feo}{\dd\vartheta} = 0,
\eeq
where $\Omega_e=-eB/m_ec$ and $\vartheta$ is the gyroangle variable. 
Thus, in this order, we have learned that 
the lowest-order electron distribution function is gyrotropic 
(does not depend on $\vartheta$). 

\subsubsection{Order $(m_e/m_i)^{-1/2}$: Maxwellian electrons}
\label{app:e_Max}

To this order, \eqref{eq:Vlasov_e} is
\beq
\label{eq:e_order1}
\vv\cdot\vdel\feo 
- \frac{e}{m_e}\lt(\vE + \frac{\vu_e\times\vB}{c}\rt)\cdot\frac{\dd\feo}{\dd\vv}
- \Omega_e\frac{\dd\feone}{\dd\vartheta} = \St{\feo}.
\eeq
Let us multiply this equation by $1+\ln\feo$ and integrate over the entire 
phase space. This gives
\beq
\label{eq:Boltzmann_rhs}
\int\int\did^3\vr\did^3\vv\St{\feo}\ln\feo = 0
\eeq
because the left-hand side of \eqref{eq:e_order1} is an exact divergence 
in the phase space. 
Let us recall that, according to \citet{Boltzmann} $H$-theorem, 
\beq
\frac{\did}{\did t}\int\int\did^3\vr\did^3\vv f_e\ln f_e 
= \int\int\did^3\vr\did^3\vv\St{f_e}\ln f_e \le 0,
\eeq
where the inequality becomes equality only for a local Maxwellian 
distribution (the proof for plasmas can be found in, e.g., \citealt{Longmire}). 
Therefore, \eqref{eq:Boltzmann_rhs} implies that $\feo$ is a local Maxwellian:
\beq
\feo = \frac{n_e}{\lt(\pi\vthe^2\rt)^{3/2}}\,e^{-v^2/\vthe^2},\quad
\vthe = \sqrt{\frac{2T_e}{m_e}}.
\label{eq:feo}
\eeq 
Since $v$ is peculiar velocity, the mean flow $\vu_e$ has already been accounted for. 
Note that the perturbation expansion of $f_e$ can always be constructed in such a way 
that $n_e$ and $T_e$ in \eqref{eq:feo} are the exact density and temperature of the 
electron distribution. 

\subsubsection{Isothermal electrons}
\label{app:isoth}

More can be learned about the electrons without going to higher orders. 
Let us now substitute the expression \exref{eq:feo} for $\feo$ into \eqref{eq:e_order1} and 
gyroaverage this equation, $(1/2\pi)\int\did\vartheta$, to eliminate 
the term containing $\feone$:
\beq
\label{eq:e_order1_Max}
\vpar\vb\cdot\vdel\feo + \frac{e}{m_e}\vE\cdot\vb\,\frac{2\vpar}{\vthe^2}\,\feo 
= \lt[\frac{\vb\cdot\vdel n_e}{n_e} 
+ \lt(\frac{v^2}{\vthe^2} - \frac{3}{2}\rt)\frac{\vb\cdot\vdel T_e}{T_e}
+ \frac{e\Epar}{T_e}\rt]\vpar\feo = 0,
\eeq 
where $\Epar=\vE\cdot\vb$. 
Since \eqref{eq:e_order1_Max} must hold for all $v$, it follows from it that 
\bea
\label{eq:Epar}
\lefteqn{\Epar =  - \frac{T_e\vb\cdot\vdel n_e}{en_e},}&&\\
\lefteqn{\vb\cdot\vdel T_e = 0.}&&
\eea
The second equation means that electrons (to lowest order) are isothermal along 
the magnetic-field lines, a standard outcome of the mass-ratio expansion \citep{Snyder2,Schek5}, 
valid up to parallel scales $\sim\mfp(m_i/m_e)^{1/2}$ (the electron thermal 
conduction scale; see, e.g., \citealt{Lithwick,Schek5}). For our fiducial ICM parameters, 
we have $\mfp(m_i/m_e)^{1/2}\sim6\times10^{21}$~cm, which is larger 
than the scale $l$ of the motions that have the highest rate of strain 
(see \secref{sec:clusters}). For turbulent plasmas, this implies 
globally isothermal electrons ($T_e=\const$) because the field lines are stochastic. 
We will adopt this assumption of globally isothermal electrons in all our calculations. 

\subsubsection{Generalized Ohm's law}

Let us again go back to \eqref{eq:e_order1}, multiply it by $m_e\vv$ and integrate over 
the velocity space. The result is the electron momentum equation to lowest order in 
the mass-ratio expansion: 
\beq
en_e\lt(\vE + \frac{\vu_e\times\vB}{c}\rt) = - \vdel\cdot\int\did^3\vv m_e\vv\vv\feo = -\vdel p_e 
= - T_e\vdel n_e,
\label{eq:elmom}
\eeq
where the electron pressure is isotropic because the distribution is Maxwellian, $p_e=n_eT_e$, 
and the gradient only affects $n_e$ because $T_e=const$ (\secref{app:isoth}). 
Note that \eqref{eq:Epar} is simply the parallel part of \eqref{eq:elmom}. 
\Eqref{eq:elmom} is the generalized Ohm's law, \eqref{eq:Ohm}. We have thus arrived 
at the starting point of the derivation in \secref{sec:Ohm}. 

\subsection{Ions}
\label{app:ions}

\subsubsection{Ordering}

The kinetic equation \exref{eq:Vlasov} for ions is 
\beq
\label{eq:Vlasov_i}
\begin{array}{lcccccccccccccccccccc} 
&
\displaystyle{\frac{\dd f_i}{\dd t}} & + &
\displaystyle{\vu_i\cdot\vdel f_i} & + &
\displaystyle{\vv\cdot\vdel f_i} & + &
\displaystyle{\Bigg[\frac{Ze}{m_i}} &
\displaystyle{\bigg(\vE + \frac{\vu_i\times\vB}{c}} & + &
\displaystyle{\frac{\vv\times\vB}{c}\bigg)} & - &
\displaystyle{\frac{\dd\vu_i}{\dd t}} & - &
\displaystyle{\vu_i\cdot\vdel\vu_i} & - &
\displaystyle{\vv\cdot\vdel\vu_i\Bigg]\cdot} &
\displaystyle{\frac{\dd f_i}{\dd \vv}} & = &
\displaystyle{\St{f_i},} \\\\
\lx{Equil.} & \eps^3 && \eps^3 && \eps^2 &&& \eps^2 && \eps^{-1} && \eps^4 && \eps^4 && \eps^3 &&& \eps\\
\lx{Pert.} & \eps^2 && \eps^2 && \eps &&& \eps && 1 && \eps^3 && \eps^3 && \eps^2 &&& \eps^2\\
\end{array}
\eeq
where we have labeled all terms according to their ordering in 
powers of $\eps$. The ordering has been done relative to $\kpar\vthi f_i$ 
using the assumptions explained in \secref{sec:order}. 
The first row of orderings in \eqref{eq:Vlasov_i} applies to 
the equilibrium (lowest-order) quantities and their gradients. 
The second row gives the lowest order in which perturbed 
quantities appear in each term of 
the kinetic equation. 

\subsubsection{Expansion of the Lorentz force}
\label{app:Lforce}

A particular explanation is in order regarding the ordering and the expansion 
of the Lorentz force. The Lorentz force is given in terms of 
$n_i$ and $\vB$ by \eqref{eq:Ohm_ions}. Expanding this equation in $\eps$, 
we have to three lowest orders 
\beq
\label{eq:Lforce}
\begin{array}{cccccccccccc}
\displaystyle{\frac{Ze}{m_i}\lt(\vE + \frac{\vu_i\times\vB}{c}\rt)} &=&-&
\displaystyle{\frac{ZT_e\vdel n_{1i}}{m_i \lt(n_{0i} + n_{1i} + n_{2i}\rt)}} &-&
\displaystyle{\frac{ZT_e\vdel \lt(n_{0i} + n_{2i}\rt)}{m_i \lt(n_{0i} + n_{1i}\rt)}} &-&
\displaystyle{\frac{ZT_e\vdel n_{3i}}{m_i n_{0i}}} &+&
\displaystyle{v_A^2\dpar\frac{\vBp}{B_0}} &+&\cdots, \\\\
&&& \eps && \eps^2 && \eps^3 && \eps^3 &&
\end{array}
\eeq
where we have used $B_1^\parallel=0$ (see \eqref{eq:Bpar}). 
The ordering of the Lorentz force in \eqref{eq:Vlasov_i} follows from \eqref{eq:Lforce}. 
Note that, in order to keep $\eps^3$ precision, 
we have to keep perturbed densities in the denominators of the first two terms 
on the right-hand side. However, as promised in \secref{sec:polarization}, 
we will see in \secref{app:f1} that $n_{1i}=0$, so 
the contributions to the Lorentz force will start at order $\eps^2$ and 
\eqref{eq:Lforce} will simplify to read  
\beq
\label{eq:Lforce2}
\frac{Ze}{m_i}\lt(\vE + \frac{\vu_i\times\vB}{c}\rt) = 
- \frac{ZT_e\vdel \lt(n_{0i} + n_{2i}\rt)}{m_i n_{0i}}
- \frac{ZT_e\vdel n_{3i}}{m_i n_{0i}}
+ v_A^2\dpar\frac{\vBp}{B_0} + \cdots
\eeq
In \secref{app:f2}, we will find that $n_{2i}=0$ as well. 

\subsubsection{Order $\eps^{-1}$: gyrotropic equilibrium}

We now proceed to expand the ion kinetic equation \exref{eq:Vlasov_i}. 
To lowest order, $\eps^{-1}$, we get (cf.\ \secref{app:e_gyro})
\beq
\frac{Ze}{m_i}\frac{\vv\times\vB_0}{c}\cdot\frac{\dd f_{0i}}{\dd\vv} 
= - \Omega_i\,\frac{\dd f_{0i}}{\dd\vartheta} = 0,
\eeq
where $\Omega_i=ZeB_0/m_i c$. Thus, the ion equilibrium distribution is gyrotropic.
We will express the fact that $f_{0i}$ is independent of the gyroangle $\vartheta$ 
by writing $f_{0i}$ as a function of two velocity variables, $v=|\vv|$ 
and $\vpar = \vv\cdot\vb_0$. In the derivation that follows these variables are 
more convenient than the perhaps more intuitive pair $(\vperp,\vpar)$. 
Thus, 
\beq
f_{0i} = f_{0i}(t,\vr,v,\vpar).
\eeq
Hence follows an identity that will be useful shortly both for $f_{0i}$ and 
other gyrotropic functions:
\beq
\label{eq:df0dv}
\frac{\dd f_{0i}}{\dd\vv} = 
\frac{\vv}{v}\lt(\frac{\dd f_{0i}}{\dd v}\rt)_{\vpar} +
\vb_0\lt(\frac{\dd f_{0i}}{\dd\vpar}\rt)_{v}.
\eeq

\subsubsection{Order $\eps^{0}$}

In the next order, \eqref{eq:Vlasov_i} is
\beq
\frac{Ze}{m_i}\frac{\vv\times\vBp}{c}\cdot\frac{\dd f_{0i}}{\dd\vv} 
- \Omega_i\,\frac{\dd f_{1i}}{\dd\vartheta} = 0, 
\eeq
where we have again used $(Ze/m_ic)(\vv\times\vB_0)\cdot\dd/\dd\vv = - \Omega_i\dd/\dd\vartheta$. 
Using \eqref{eq:df0dv}, we get
\beq
\frac{\dd f_{1i}}{\dd\vartheta} = 
\frac{1}{\Omega_i}\frac{Ze}{m_i}\frac{\vv\times\vBp}{c}\cdot\vb_0\lt(\frac{\dd f_{0i}}{\dd\vpar}\rt)_{v}
= \lt(\vb_0\times\vvperp\rt)\cdot\frac{\vBp}{B_0}\lt(\frac{\dd f_{0i}}{\dd\vpar}\rt)_{v}. 
\eeq
Noticing that $\vb_0\times\vvperp = \dd\vvperp/\dd\vartheta$, we integrate this equation:
\beq
f_{1i} = \vvperp\cdot\frac{\vBp}{B_0}\lt(\frac{\dd f_{0i}}{\dd\vpar}\rt)_{v} 
+ g_{1i}(t,\vr,v,\vpar),
\label{eq:f1}
\eeq
where $g_{1i}$ is an arbitrary function (the gyrotropic part of the first-order perturbed 
distribution). 

Thus, all we have learned at this order is the gyroangle dependence of $f_{1i}$. This will 
be a general feature of our expansion: since the gyroangle derivative in \eqref{eq:Vlasov_i}
is the lowest-order term, what we learn about each perturbed distribution function
$f_{1i}$, $f_{2i}$, $f_{3i}$, \dots, 
at the lowest order in which it first appears will always be its dependence on $\vartheta$. 

\subsubsection{Order $\eps^{1}$: Maxwellian equilibrium}
\label{app:Max}

At this order, \eqref{eq:Vlasov_i} is
\beq
\label{eq:eps1}
\vv\cdot\vdel f_{1i} + \frac{Ze}{m_i}\lt(\vE + \frac{\vu_i\times\vB}{c}\rt)_1\cdot\frac{\dd f_{0i}}{\dd\vv}
- \Omega_i\,\frac{\dd f_{2i}}{\dd\vartheta} 
+ \underbrace{\frac{Ze}{m_i}\frac{\vv\times\vBp}{c}\cdot\frac{\dd f_{1i}}{\dd\vv}}_{I_1}
= \St{f_{0i}},
\eeq
where, using \eqref{eq:f1} and other tricks already employed in the two previous sections, 
we can express the last term on the left-hand side of \eqref{eq:eps1} as follows
\beq
I_1 = \Omega_i\lt(\vb_0\times\vvperp\rt)\cdot\frac{\vBp}{B_0}
\lt[\vvperp\cdot\frac{\vBp}{B_0}\lt(\frac{\dd^2 f_{0i}}{\dd\vpar^2}\rt)_{v} 
+ \lt(\frac{\dd g_{1i}}{\dd\vpar}\rt)_v\rt] 
= \Omega_i\,\frac{\dd}{\dd\vartheta}
\lt[\frac{1}{2}\lt(\vvperp\cdot\frac{\vBp}{B_0}\rt)^2\lt(\frac{\dd^2 f_{0i}}{\dd\vpar^2}\rt)_{v} 
+ \vvperp\cdot\frac{\vBp}{B_0}\lt(\frac{\dd g_{1i}}{\dd\vpar}\rt)_v\rt].
\label{eq:term1}
\eeq

Collisions have made their first appearance at this order and we can now prove that 
$f_{0i}$ is a Maxwellian. The proof is similar to the one for electrons in \secref{app:e_Max}: 
we multiply \eqref{eq:eps1} by $1+\ln f_{0i}$ and integrate over the entire phase space. 
All terms on the left-hand side vanish because, to the order at which we are computing them, 
they are all full derivatives with respect to the phase-space variables. Thus, 
\beq
\int\int\did^3\vr\did^3\vv\St{f_{0i}}\ln f_{0i} = 0 
\quad\Rightarrow\quad
f_{0i} = \frac{n_{0i}}{\lt(\pi\vthi^2\rt)^{3/2}}\,e^{-v^2/\vthi^2},
\quad \vthi = \sqrt{\frac{2T_{0i}}{m_i}}.
\label{eq:f0i}
\eeq

\subsubsection{Order $\eps^{1}$ continued: more information about $f_{1i}$}
\label{app:f1} 

The fact that $f_{0i}$ is a Maxwellian 
allows us to uncover three important additional pieces of information. 
First, from \eqref{eq:f1}, we learn that $f_{1i}$ is gyrotropic:
\beq
f_{1i} = g_{1i}(t,\vr,v,\vpar). 
\label{eq:f1_gyro}
\eeq 
Second, we can now prove that $n_{1i}=0$, as promised in \secsand{sec:polarization}{app:Lforce}. 
Using \eqsand{eq:term1}{eq:f0i} in \eqref{eq:eps1}, gyroaveraging this equation, 
$(1/2\pi)\int\did\vartheta$, and substituting for the Lorentz force the 
lowest-order expression from \eqref{eq:Lforce}, we get 
\beq
\vpar\lt(\dpar f_{1i} 
+ \frac{ZT_e}{T_{0i}}\frac{\dpar n_{1i}}{n_{0i}}\,f_{0i}\rt) = 0
\quad\Rightarrow\quad
\lt(1+\frac{ZT_e}{T_{0i}}\rt)\dpar n_{1i} = 0,
\label{eq:dparn1}
\eeq
where the second equation has been obtained by cancelling $\vpar$ in the first equation 
and integrating it over velocities. 
We have used the shorthand $\dpar=\vb_0\cdot\vdel$, which henceforth 
will be employed wherever fast parallel variation of the perturbed 
quantities is involved (for slow parallel gradients, we will continue writing 
$\vb_0\cdot\vdel$ explicitly to emphasize that $\vb_0$ is curved on the 
large scales). \Eqref{eq:dparn1} implies that we may set 
\beq
n_{1i}=0
\label{eq:n1zero}
\eeq
(q.e.d.; see \eqref{eq:n1}) 
and absorb whatever slow-varying density perturbation may arise into $n_{0i}$. 
After eliminating $n_{1i}$ from \eqref{eq:dparn1}, we get
\beq
\dpar f_{1i}=0,
\label{eq:dparf1}
\eeq 
so $f_{1i}$ has no small-scale spatial variation at all. 
Note that this confirms \eqref{eq:P1}, which was derived from the 
ion momentum equation in the $\eps^1$ order (i.e., it is the velocity 
moment of \eqref{eq:eps1}) and restricted the fast spatial 
variation of the first-order pressure tensor. 
\Eqref{eq:Lforce2} is also now confirmed. 

\subsubsection{Order $\eps^{1}$ continued: gyroangle dependence of $f_{2i}$}
\label{app:f2_gyro} 

Finally, we go back to \eqref{eq:eps1} to 
determine the gyroangle dependence of $f_{2i}$. Since $f_{1i}$ does not 
have a small-scale part, the first two terms drop out. Using the fact that 
$f_{0i}$ is a Maxwellian and \eqref{eq:term1}, we integrate \eqref{eq:eps1} 
with respect to the gyroangle and get
\beq
f_{2i} = \vvperp\cdot\frac{\vBp}{B_0}\lt(\frac{\dd f_{1i}}{\dd\vpar}\rt)_{v} 
+ g_{2i}(t,\vr,v,\vpar),
\label{eq:f2}
\eeq
where $g_{2i}$ is the gyrotropic part of $f_{2i}$ (so far arbitrary). 

\subsubsection{Order $\eps^{2}$: role of equilibrium density and temperature gradients} 
\label{app:grads}

At this order, \eqref{eq:Vlasov_i} becomes, upon substitution of the Maxwellian $f_{0i}$ 
and the lowest-order ($\eps^2$) expression for the Lorentz force from \eqref{eq:Lforce2}
\bea
\nonumber
\frac{\dd f_{1i}}{\dd t} &+& \vv\cdot\vdel f_{0i} + \vpar\dpar f_{2i} 
+ \frac{ZT_e}{T_{0i}}\frac{\vv\cdot\vdel n_{0i}+ \vpar\dpar n_{2i}}{n_{0i}}\,f_{0i}\\
&-& \Omega_i\,\frac{\dd f_{3i}}{\dd\vartheta} 
+ \underbrace{\frac{Ze}{m_i}\lt(\frac{\vv\times\vB_2}{c}\cdot\frac{\dd f_{1i}}{\dd\vv}
+ \frac{\vv\times\vBp}{c}\cdot\frac{\dd f_{2i}}{\dd\vv}\rt)
+ \frac{2\vpar\lt(\dpar\vup\rt)\cdot\vvperp}{\vthi^2}\, f_{0i}}_{I_2}
= \St{f_{1i}},
\label{eq:eps2}
\eea 
where we have explicitly enforced the assumption that perturbed quantities 
have no fast perpendicular spatial dependence. 
Analogously to \eqref{eq:term1}, upon using \eqsand{eq:f1_gyro}{eq:f2} 
and noticing that $\vvperp = -\dd(\vb_0\times\vv)/\dd\vartheta$, 
we find that the last two terms on the left-hand side of \eqref{eq:eps2} are 
a full gyroangle derivative:
\beq
I_2 = \Omega_i\frac{\dd}{\dd\vartheta}
\lt[\vvperp\cdot\frac{\vB_2^\perp}{B_0}\lt(\frac{\dd f_{1i}}{\dd\vpar}\rt)_v
+ \frac{1}{2}\lt(\vvperp\cdot\frac{\vBp}{B_0}\rt)^2\lt(\frac{\dd^2 f_{1i}}{\dd\vpar^2}\rt)_{v} 
+ \vvperp\cdot\frac{\vBp}{B_0}\lt(\frac{\dd g_{2i}}{\dd\vpar}\rt)_v
- \frac{2\vpar\dpar\vup}{\vthi^2}\cdot\frac{\vb_0\times\vvperp}{\Omega_i}\,f_{0i}\rt].
\label{eq:term2}
\eeq
In view of \eqref{eq:term2} and of the gyroangle independence 
of $f_{1i}$ (\eqref{eq:f1_gyro}), the gyroaverage of \eqref{eq:eps2} is
\beq
\frac{\dd f_{1i}}{\dd t} + \vpar\lt[\vb_0\cdot\vdel f_{0i} + \dpar g_{2i} 
+ \frac{ZT_e}{T_{0i}}\frac{\vb_0\cdot\vdel n_{0i}+\dpar n_{2i}}{n_{0i}}\,f_{0i}\rt]
= \St{f_{1i}}.
\label{eq:eps2_gavg}
\eeq
Since $f_{1i}$ has no fast spatial gradients (\eqref{eq:dparf1}),  
averaging \eqref{eq:eps2_gavg} over small scales gives
\beq
\frac{\dd f_{1i}}{\dd t} + \vpar\lt[\vb_0\cdot\vdel f_{0i} + 
\frac{ZT_e}{T_{0i}}\frac{\vb_0\cdot\vdel n_{0i}}{n_{0i}}\,f_{0i}\rt] 
= \St{f_{1i}}.
\label{eq:eps2_avg}
\eeq
This equation determines $f_{1i}$ purely in terms of the equilibrium 
density and temperature gradients. 
Since the time variation of the equilibrium is slow, it is clear that $f_{1i}$ will 
converge to a steady solution after a few collision times. Then $\dd f_{1i}/\dd t$ 
in \eqref{eq:eps2_avg} can be neglected and the solution obtained by inverting the 
linearized collision operator. Since we are not interested in exact collisional 
transport coefficients here, instead of the full Landau collision operator, 
we will use a very simple model one --- the Lorentz pitch-angle scattering 
operator \citep[see, e.g.,][]{Helander}, 
so \eqref{eq:eps2_avg} becomes in steady state 
\beq
\St{f_{1i}} = \nuii\,\frac{\dd}{\dd\xi}\frac{1-\xi^2}{2}\frac{\dd f_{1i}}{\dd\xi} 
= \xi v\lt[\lt(1+\frac{ZT_e}{T_{0i}}\rt)\frac{\vb_0\cdot\vdel n_{0i}}{n_{0i}} 
+ \lt(\frac{v^2}{\vthi^2} - \frac{3}{2}\rt)\frac{\vb_0\cdot\vdel T_{0i}}{T_{0i}}\rt] f_{0i},
\label{eq:f1_Lorentz}
\eeq 
where $\xi=\vpar/v$ and $\nuii$ is the collision frequency, whose dependence on 
$v$ is not important here and is suppressed for simplicity. 
The solution of \eqref{eq:f1_Lorentz} that satisfies \eqref{eq:n1zero} is 
\beq
f_{1i} = -\frac{\vpar}{\nuii}\lt[\lt(1+\frac{ZT_e}{T_{0i}}\rt)\frac{\vb_0\cdot\vdel n_{0i}}{n_{0i}} 
+ \lt(\frac{v^2}{\vthi^2} - \frac{3}{2}\rt)\frac{\vb_0\cdot\vdel T_{0i}}{T_{0i}}\rt] f_{0i}
= -\frac{\vpar}{\nuii}\lt(\frac{v^2}{\vthi^2} - \frac{5}{2}\rt)\frac{\vb_0\cdot\vdel T_{0i}}{T_{0i}}f_{0i}.
\label{eq:f1_sln}
\eeq
The second, simplified, expression above is obtained by noticing that 
the temperature and density gradients are, in fact, related by the 
equilibrium pressure balance, \eqref{eq:P0}, which was obtained 
in \secref{sec:polarization} from the ion momentum equation in 
the $\eps^2$ order. It is easily 
recovered by taking the velocity moment of \eqref{eq:eps2} and averaging out the 
small scales. Since $f_{0i}$ is a Maxwellian, the pressure balance takes the form 
(previewed in footnote \ref{fn:eqgrad})
\beq
\lt(1 + \frac{ZT_e}{T_{0i}}\rt)\frac{\vdel n_{0i}}{n_{0i}} 
+ \frac{\vdel T_{0i}}{T_{0i}} = 0, 
\label{eq:pr_bal}
\eeq
whence immediately follows the final expression for $f_{1i}$ in \eqref{eq:f1_sln}.

Let us note two useful properties of the solution \exref{eq:f1_sln}. 
First, $f_{1i}$ makes no contribution to the pressure tensor:
\beq
\vP_{1i} = m_i\int\did^3\vv\,\vv\vv f_{1i} = 0,
\eeq
a result we promised in \secref{sec:polarization} (\eqref{eq:P1}). 
Second, the derivative of $f_{1i}$ with respect to $\vpar$ is isotropic:
\beq
\lt(\frac{\dd f_{1i}}{\dd\vpar}\rt)_v = 
-\frac{1}{\nuii}\lt(\frac{v^2}{\vthi^2} - \frac{5}{2}\rt)\frac{\vb_0\cdot\vdel T_{0i}}{T_{0i}}f_{0i},
\quad
\lt(\frac{\dd^2 f_{1i}}{\dd\vpar^2}\rt)_v = 0,
\label{eq:df1dvpar}
\eeq
which leads to vanishing of one of the terms in \eqref{eq:term2}. 

We will see in \apref{app:transp} that $f_{1i}$ encodes the ion collisional heat flux
(\eqref{eq:f1_sln} is the standard form of the appropriate 
contribution to the perturbed distribution function; 
see, e.g., equation (D16) of \citealt{Schek5}). 
We will carry the ion-temperature-gradient effect contained in $f_{1i}$ 
through to the end of this calculation because it will interesting and 
instructive to see how contributions from the ion heat flux arise in the 
problem. However, this is not the main effect we are after and 
an impatient reader attempting to follow this derivation may find 
the following simplification useful. 
If one assumes by fiat that 
$\vdel T_{0i}=0$, then $\vdel n_{0i}=0$ as well (from \eqref{eq:pr_bal}) 
and in all the calculations that follow one may set $f_{1i}=0$ 
and $f_{0i}=\const$, which substantially reduces the amount of algebra. 

\subsubsection{Order $\eps^{2}$ continued: more information about $f_{2i}$}
\label{app:f2}

Staying at this order, we can learn more about $f_{2i}$ and $f_{3i}$. 
Subtracting \eqref{eq:eps2_avg} from \eqref{eq:eps2_gavg}, we get
\beq
\vpar\lt(\dpar g_{2i} + \frac{ZT_e}{T_{0i}}\frac{\dpar n_{2i}}{n_{0i}}\,f_{0i}\rt) = 0
\quad\Rightarrow\quad
\lt(1 + \frac{ZT_e}{T_{0i}}\rt)\dpar n_{2i} = 0,
\label{eq:dparn2}
\eeq
analogously to \eqref{eq:dparn1}. 
We have used the fact that, as follows from \eqref{eq:f2}, 
$n_{2i} = \int\did^3\vv f_{2i} = \int\did^3\vv g_{2i}$. 
Similarly to the argument in \secref{app:f1}, this implies 
that $n_{2i}$ has no fast spatial variation and so we can set 
\beq
n_{2i}=0.
\eeq 
\Eqref{eq:dparn2} then implies
\beq
\dpar g_{2i} = 0,
\label{eq:dparg2}
\eeq
i.e., $g_{2i}$ has no small-scale spatial dependence. 
Since the first term in 
\eqref{eq:f2} does not contribute to the second-order pressure tensor (because 
the derivative of $f_{1i}$ is a function of $v$ only), we have 
\beq
\vP_{2i} = m_i\int\did^3\vv\,\vv\vv\, g_{2i},
\label{eq:P2def}
\eeq 
and so, in view of \eqref{eq:dparg2}, $\vP_{2i}$ has no small-scale dependence. 
This confirms \eqref{eq:P2}, derived in \secref{sec:polarization} 
from the ion momentum equation. Note that the tensor $\vP_{2i}$ 
will be needed in the large-scale momentum equation \exref{eq:momavg}. 
It will contain the lowest-order pressure anisotropy. 

\subsubsection{Order $\eps^{2}$ continued: gyroangle dependence of $f_{3i}$}
\label{app:f3_gyro} 

Subtracting \eqref{eq:eps2_gavg} from \eqref{eq:eps2} and using \eqref{eq:f2}, we get
\beq
\Omega_i\frac{\dd f_{3i}}{\dd\vartheta} = I_2
+ \vvperp\cdot\lt[\vdel f_{0i} + \frac{ZT_e}{T_{0i}}\frac{\vdel n_{0i}}{n_{0i}}\,f_{0i}\rt]
+ \frac{\vpar\lt(\dpar\vBp\rt)\cdot\vvperp}{B_0}\lt(\frac{\dd f_{1i}}{\dd\vpar}\rt)_{v},
\label{eq:f3_angle}
\eeq
where $I_2$ is given by \eqref{eq:term2} (note that the second derivative of $f_{1i}$ 
vanishes there; see \eqref{eq:df1dvpar}). Using again the fact that 
$\vvperp = -\dd(\vb_0\times\vvperp)/\dd\vartheta$, 
we integrate \eqref{eq:f3_angle} and get
\bea
\nonumber
f_{3i} &=& \vvperp\cdot\lt[\frac{\vB_2^\perp}{B_0}\lt(\frac{\dd f_{1i}}{\dd\vpar}\rt)_v
+ \frac{\vBp}{B_0}\lt(\frac{\dd g_{2i}}{\dd\vpar}\rt)_v\rt]\\
&&-\,\, \frac{\vb_0\times\vvperp}{\Omega_i}\cdot\lt[\frac{2\vpar\dpar\vup}{\vthi^2}\,f_{0i}
+ \frac{\vpar\dpar\vBp}{B_0}\lt(\frac{\dd f_{1i}}{\dd\vpar}\rt)_{v}
+ \lt(\frac{v^2}{\vthi^2}-\frac{5}{2}\rt)\frac{\vdel T_{0i}}{T_{0i}}\,f_{0i}\rt]
+ g_{3i}(t,\vr,v,\vpar),
\label{eq:f3}
\eea
where $g_{3i}$ is the gyrotropic part of $f_{3i}$ (so far arbitrary)
and we have used \eqref{eq:pr_bal} to simplify the terms that contain 
equilibrium gradients. 

We will see in \secref{app:fluct} that we do not need to know either $g_{3i}$ 
or $\vB_2$ in order to calculate the third-order ion pressure tensor $\vP_{3i}$ 
and close the ion momentum equation \exref{eq:momperp} for the firehose 
perturbations. The only remaining quantity we do need is $g_{2i}$ --- we will 
now derive the equation for it by going to next order in the $\eps$ expansion.

\subsubsection{Order $\eps^{3}$}
\label{app:eps3}

At this order, \eqref{eq:Vlasov_i} is, upon substitution of the Maxwellian $f_{0i}$, 
gyrotropic $f_{1i}$, and \eqref{eq:Lforce2} for the Lorentz force, 
\bea
\nonumber
\frac{\did f_{0i}}{\did t} + \frac{\did f_{2i}}{\did t} 
&+& \vv\cdot\vdel f_{1i} + \vpar\dpar f_{3i}
+ \lt(\frac{ZT_e}{T_{0i}}\frac{\vpar\dpar n_{3i}}{n_{0i}} 
- \frac{2}{\beta_i}\frac{\lt(\dpar\vBp\rt)\cdot\vvperp}{B_0}\rt)f_{0i} 
- \frac{ZT_e}{m_i}\frac{\vdel n_{0i}}{n_{0i}}\cdot
\lt[\frac{\vv}{v}\lt(\frac{\dd f_{1i}}{\dd v}\rt)_{\vpar} 
+ \vb_0\lt(\frac{\dd f_{1i}}{\dd\vpar}\rt)_v\rt]\\
\nonumber
&-& \Omega_i\,\frac{\dd f_{4i}}{\dd\vartheta}
+ \underbrace{\frac{Ze}{m_i}
\Biggl[\frac{\vv\times\vBp}{c}\cdot\frac{\dd f_{3i}}{\dd\vv}
+ \frac{\vv\times\vB_2}{c}\cdot\frac{\dd f_{2i}}{\dd\vv}}_{I_3} 
+ \frac{\vv\times\vB_3}{c}\cdot\vb_0\lt(\frac{\dd f_{1i}}{\dd\vpar}\rt)_v\Biggr]\\
&+& \frac{\did\vup}{\did t}\cdot\frac{2\vvperp}{\vthi^2}\,f_{0i}
+ \frac{2\vv\vv:\vdel\lt(\vu_{0i} + \vu_{2i}\rt)}{\vthi^2}\,f_{0i} 
- \vpar\lt(\dpar\vup\rt)\cdot\frac{\vvperp}{v}\lt(\frac{\dd f_{1i}}{\dd v}\rt)_{\vpar}
= \St{f_{2i}} + C[f_{1i},f_{1i}],
\label{eq:eps3}
\eea
where $\did/\did t = \dd/\dd t + \vu_{0i}\cdot\vdel$ is the convective 
derivative (with respect to the large-scale flow), $\beta_i=\vthi^2/v_A^2$. 
Note that $\vdel f_{1i}$ in the above equation is with respect to slow spatial variation. 
Note also that the collision operator at this order has two parts: the linearized operator 
describing interaction of $f_{2i}$ with the Maxwellian equilibrium $f_{0i}$ 
and the nonlinear operator, denoted $C[f_{1i},f_{1i}]$, 
describing interaction of $f_{1i}$ with itself. 

We will only ever need the gyroaverage of \eqref{eq:eps3}. Many terms then vanish or 
simplify. What happens is mostly straightforward: the gyrovaerages of $\dd/\dd\vartheta$ 
are zero, the gyroaverages of the velocities are done using the identities 
\beq
\lt<\vv\rt> = \vpar\vb_0,\quad 
\lt<\vv\vv\rt> = \frac{\vperp^2}{2}\lt(\vI - \vb_0\vb_0\rt) + \vpar^2\vb_0\vb_0
\eeq
(henceforth angle brackets denote $(1/2\pi)\int\did\vartheta$).
There are a few terms that are perhaps not obvious and so require explanation. 

First consider the term $\vv\cdot\vdel f_{1i}$. We showed above that $f_{1i}$ is 
a function of $v$, $\vpar$, and $\vr$. However, the spatial gradient here is still 
taken at constant $\vv$. Since one of the new velocity variables 
$\vpar=\vv\cdot\vb_0$ is a function of $\vv$ and $\vr$, we have
\beq
\vv\cdot\lt(\vdel f_{1i}\rt)_{\vv} = \vv\cdot\lt(\vdel f_{1i}\rt)_{v,\vpar} 
+ \lt(\vv\vv:\vdel\vb_0\rt)\lt(\frac{\dd f_{1i}}{\dd\vpar}\rt)_v
\quad\Rightarrow\quad
\lt<\vv\cdot\lt(\vdel f_{1i}\rt)_{\vv}\rt> 
= \vpar\vb_0\cdot\vdel f_{1i} + \lt(\vdel\cdot\vb_0\rt)\frac{\vperp^2}{2}
\lt(\frac{\dd f_{1i}}{\dd\vpar}\rt)_v.
\eeq
In the final expression, $\vdel f_{1i}$ is now understood to be 
at constant $v$ and $\vpar$. Since $\vdel\cdot\vb_0=-\lt(\vb_0\cdot\vdel B_0\rt)/B_0$, 
the additional term that has emerged is readily interpreted as the mirror force
associated with the large-scale variation of the magnetic field. 

Now let us turn to the two terms in \eqref{eq:eps3} denoted by $I_3$:
the second of these terms gives 
\beq
\lt<\frac{Ze}{m_i}\frac{\vv\times\vB_2}{c}\cdot\frac{\dd f_{2i}}{\dd\vv} \rt> 
= \lt<\frac{Ze}{m_i}\frac{\vv\times\vB_2}{c}\cdot
\lt[\frac{\vBp}{B_0}\lt(\frac{\dd f_{1i}}{\dd\vpar}\rt)_v
+ \vb_0\lt(\frac{\dd g_{2i}}{\dd\vpar}\rt)_v\rt]\rt>
= \Omega_i\vpar\lt(\vb_0\times\frac{\vB_2^\perp}{B_0}\rt)\cdot
\frac{\vBp}{B_0}\lt(\frac{\dd f_{1i}}{\dd\vpar}\rt)_v,
\label{eq:B2f2}
\eeq 
where we have used \eqref{eq:f2} and the fact that 
$\lt(\dd f_{1i}/\dd\vpar\rt)_v$ only depends on $v$ (\eqref{eq:df1dvpar}); 
the first term, upon substitution of \eqref{eq:f3}, gives 
\bea
\nonumber
\lt<\frac{Ze}{m_i}\frac{\vv\times\vBp}{c}\cdot\frac{\dd f_{3i}}{\dd\vv} \rt>
&=& \lt<\frac{Ze}{m_i}\frac{\vv\times\vBp}{c}\cdot\Biggl\{
\frac{\vB_2^\perp}{B_0}\lt(\frac{\dd f_{1i}}{\dd\vpar}\rt)_v
+ \vb_0\,\vvperp\cdot\frac{\vBp}{B_0}\lt(\frac{\dd^2 g_{2i}}{\dd\vpar^2}\rt)_v
\Biggr.\rt.\\
\nonumber
&& +\,\, \frac{\vb_0}{\Omega_i}\times\lt[\frac{2\vpar\dpar\vup}{\vthi^2}\,f_{0i}
+ \frac{\vpar\dpar\vBp}{B_0}\lt(\frac{\dd f_{1i}}{\dd\vpar}\rt)_{v}
+ \lt(\frac{v^2}{\vthi^2}-\frac{5}{2}\rt)\frac{\vdel T_{0i}}{T_{0i}}\,f_{0i}\rt]\\
\nonumber
&&\lt.\Biggl.-\,\,\vb_0\,\frac{\vb_0\times\vvperp}{\Omega_i}\cdot\lt[\frac{2\dpar\vup}{\vthi^2}\,f_{0i}
+ \frac{\dpar\vBp}{B_0}\lt(\frac{\dd f_{1i}}{\dd\vpar}\rt)_{v}\rt]
+ \vb_0\lt(\frac{\dd g_{3i}}{\dd\vpar}\rt)_{v}\Biggr\}\rt>\\
\nonumber
&=& \Omega_i\vpar\lt(\vb_0\times\frac{\vBp}{B_0}\rt)\cdot
\frac{\vB_2^\perp}{B_0}\lt(\frac{\dd f_{1i}}{\dd\vpar}\rt)_v\\
\nonumber
&& +\,\, \vpar\lt(\vb_0\times\frac{\vBp}{B_0}\rt)\cdot
\Biggl\{\vb_0\times\lt[\frac{2\vpar\dpar\vup}{\vthi^2}\,f_{0i}
+ \frac{\vpar\dpar\vBp}{B_0}\lt(\frac{\dd f_{1i}}{\dd\vpar}\rt)_{v}
+ \lt(\frac{v^2}{\vthi^2}-\frac{5}{2}\rt)\frac{\vdel T_{0i}}{T_{0i}}\,f_{0i}\rt]\Biggr\}\\
&& -\,\,\frac{\vBp}{B_0}\cdot\lt<\lt(\vb_0\times\vvperp\rt)
\lt(\vb_0\times\vvperp\rt)\rt>\cdot\lt[\frac{2\dpar\vup}{\vthi^2}\,f_{0i}
+ \frac{\dpar\vBp}{B_0}\lt(\frac{\dd f_{1i}}{\dd\vpar}\rt)_{v}\rt].
\label{eq:B1f3}
\eea
The first term in \eqref{eq:B1f3} exactly cancels when \eqref{eq:B2f2} is added to it. 
Simplifying the double vector product in the second term and 
noticing that the gyroaverage in the third term is equal to 
$\lt(\vperp^2/2\rt)\lt(\vI-\vb_0\vb_0\rt)$, we have
\bea
\nonumber
\lt<I_3\rt> &=& \frac{\vBp}{B_0}\cdot\Biggl\{
\lt(\vpar^2 - \frac{\vperp^2}{2}\rt)
\lt[\frac{2\dpar\vup}{\vthi^2}\,f_{0i}
+ \frac{\dpar\vBp}{B_0}\lt(\frac{\dd f_{1i}}{\dd\vpar}\rt)_{v}\rt]
+ \vpar\lt(\frac{v^2}{\vthi^2}-\frac{5}{2}\rt)\frac{\vdel T_{0i}}{T_{0i}}\,f_{0i}\Biggr\}\\
&=& \frac{2\vpar^2-\vperp^2}{\vthi^2}
\lt[\frac{1}{2}\frac{\did}{\did t}\frac{|\vBp|^2}{B_0^2}\,f_{0i}
+ \frac{\vthi^2}{4}\frac{\dpar|\vBp|^2}{B_0^2}\lt(\frac{\dd f_{1i}}{\dd\vpar}\rt)_{v}\rt]
+ \vpar\lt(\frac{v^2}{\vthi^2}-\frac{5}{2}\rt)
\frac{\vBp}{B_0}\cdot\frac{\vdel T_{0i}}{T_{0i}}\,f_{0i},
\label{eq:term3}
\eea
where we have used the perturbed induction equation \exref{eq:indperp} 
to express $\dpar\vup$ in terms of $\vBp$. 
Finally, the gyroaverage of \eqref{eq:eps3} is 
\bea
\nonumber
\frac{\did f_{0i}}{\did t}
+ \frac{\dd g_{2i}}{\dd t} 
&+& \vpar\lt(\vb_0\cdot\vdel f_{1i} + \dpar g_{3i}\rt)
+ \lt(\vdel\cdot\vb_0\rt)\frac{\vperp^2}{2}\lt(\frac{\dd f_{1i}}{\dd\vpar}\rt)_v\\
\nonumber
&+& \frac{ZT_e}{T_{0i}}\frac{\vpar\dpar n_{3i}}{n_{0i}}\,f_{0i} 
- \frac{ZT_e}{m_i}\frac{\vb_0\cdot\vdel n_{0i}}{n_{0i}}
\lt[\frac{\vpar}{v}\lt(\frac{\dd f_{1i}}{\dd v}\rt)_{\vpar} 
+ \lt(\frac{\dd f_{1i}}{\dd\vpar}\rt)_v\rt]\\
&+& \lt<I_3\rt> 
+ \lt[\frac{2\vpar^2-\vperp^2}{\vthi^2}\,\vb_0\vb_0:\vdel\lt(\vu_{0i}+\vu_{2i}\rt)
+ \frac{\vperp^2}{\vthi^2}\,\vdel\cdot\lt(\vu_{0i}+\vu_{2i}\rt)\rt]f_{0i} 
= \St{g_{2i}} + C[f_{1i},f_{1i}],
\label{eq:eps3_gavg}
\eea
where $\lt<I_3\rt>$ is given by \eqref{eq:term3}.
Note that $g_{2i}$ does not have fast spatial variation (\eqref{eq:dparg2}), 
so, to the order we are keeping, $\did g_{2i}/\did t = \dd g_{2i}/\dd t$. 
The next step is to average 
this equation over small scales: again many terms vanish (in particular, all 
terms where the fast-varying perturbed quantities enter linearly) and we get 
\bea
\nonumber
\frac{\did f_{0i}}{\did t} + \frac{\dd g_{2i}}{\dd t} 
&+& \vpar\vb_0\cdot\vdel f_{1i}
+ \lt(\vdel\cdot\vb_0\rt)\frac{\vperp^2}{2}\lt(\frac{\dd f_{1i}}{\dd\vpar}\rt)_v
- \frac{ZT_e}{m_i}\frac{\vb_0\cdot\vdel n_{0i}}{n_{0i}}
\,\vb_0\cdot\frac{\dd f_{1i}}{\dd\vv}\\
&+& \lt[\frac{2\vpar^2-\vperp^2}{\vthi^2}
\lt(\vb_0\vb_0:\vdel\vu_{0i} 
+ \frac{1}{2}\frac{\dd}{\dd t}\frac{\overline{|\vBp|^2}}{B_0^2}\rt)
+ \frac{\vperp^2}{\vthi^2}\,\vdel\cdot\vu_{0i}\rt]f_{0i} 
= \St{g_{2i}} + C[f_{1i},f_{1i}],
\label{eq:eps3_avg}
\eea
where the overline denotes the small-scale average and the derivatives 
of $f_{1i}$ have been written in a compact form that will prove useful momentarily.
Remarkably, the contribution of the perturbations has survived in \eqref{eq:eps3_avg} 
in the form of a single quadratic term --- in \secref{app:aniso}, we will see that 
it gives rise to precisely the nonlinear feedback on the pressure anisotropy 
that was anticipated qualitatively in \secref{sec:nonlin}. 

By averaging out the gyroangle- and small-scale-dependent parts 
of the third-order kinetic equation \exref{eq:eps3}, we 
have eliminated $f_{4i}$, $g_{3i}$, $\vB_2$, $\vB_3$, and $\vu_{2i}$, 
which are unknown and potentially very cumbersome to calculate.
As we are about to see, in order to calculate the ion pressure tensor to 
the relevant orders, we do not, in fact, need to know any of these quantities, 
so we will neither have to revisit the unaveraged \eqsref{eq:eps3} or \exref{eq:eps3_gavg}
or go to higher orders in the $\eps$ expansion of \eqref{eq:Vlasov_i}. 
\Eqref{eq:eps3_avg} determines $g_{2i}$, which is all that we require 
to calculate $\vP_{2i}$ (\secref{app:aniso}) and $\vP_{3i}$ (\secref{app:fluct}). 
Knowing these tensors will then allow us to close the ion momentum equations 
describing the plasma motion at large (\eqref{eq:momavg})
and small (\eqref{eq:momperp}) scales. 

\subsubsection{Order $\eps^{3}$ continued: transport equations}
\label{app:transp}

Since $g_{2i}$ is gyrotropic, the tensor $\vP_{2i}$ is diagonal:
\beq
\vP_{2i} = \pperpi\lt(\vI - \vb_0\vb_0\rt) + \ppari\vb_0\vb_0
= p_{2i}\vI + (\pperpi-\ppari)\lt(\frac{1}{3}\,\vI - \vb_0\vb_0\rt),
\label{eq:P2diag}
\eeq
where the scalar pressures are
\beq
\pperpi = \int\did^3\vv\,\frac{m_i\vperp^2}{2}\,g_{2i},
\quad
\ppari = \int\did^3\vv\,m_i\vpar^2g_{2i},
\quad
p_{2i} = \frac{2}{3}\,\pperpi + \frac{1}{3}\,\ppari 
= \int\did^3\vv\,\frac{m_i v^2}{3}\,g_{2i}.
\eeq
Then the ion momentum equation \exref{eq:momavg} becomes\footnote{Formally, this 
equation is the result of taking the velocity moment of the kinetic equation 
\exref{eq:Vlasov_i} at the order $\eps^4$. We do not write explicitly 
\eqref{eq:Vlasov_i} at this order because it is not needed for anything 
except the momentum equation, the form of which we already know.}
\beq
m_in_{0i}\frac{\did \vu_{0i}}{\did t} = 
-\vdel\tilde p
+\vdel\cdot\lt[\vb_0\vb_0\lt(\pperpi-\ppari + \frac{B_0^2}{4\pi}\rt)\rt],
\qquad
\tilde p = p_{2i}+\frac{1}{3}\,(\pperpi-\ppari)+\frac{B_0^2}{8\pi},
\label{eq:u0}
\eeq
which is the familiar momentum equation in the long-wavelength limit 
(\eqref{eq:gen_mom}). 
Let us first explain how $p_{2i}$ (or, equivalently, $\tilde p$) is determined 
and then calculate the pressure anisotropy $\pperpi-\ppari$ (\secref{app:aniso}). 

First, let us integrate \eqref{eq:eps3_avg} over velocities. Since 
$\int\did^3\vv\,g_{2i}=0$ and $\int\did^3\vv\,\vpar f_{1i}=0$, we get
\beq
\frac{\did n_{0i}}{\did t} = - n_{0i}\vdel\cdot\vu_{0i},
\label{eq:n0}
\eeq 
an unsurprising result (the continuity equation was already obtained 
in \secref{sec:transp}; see \eqref{eq:contavg}). 
Now multiply \eqref{eq:eps3_avg} by $m_iv^2/3$ and integrate over velocities:
\beq
\frac{\did p_{0i}}{\did t}
+ \frac{\dd p_{2i}}{\dd t}
= - \frac{2}{3}\,\vdel\cdot\bigl(\vb_0 q_{1i}\bigr) - \frac{5}{3}\,p_{0i}\vdel\cdot\vu_{0i},
\label{eq:p0}
\eeq
where $p_{0i}=n_{0i}T_{0i}$
and the parallel collisional heat flux is, 
using \eqref{eq:f1_sln},\footnote{Since we used a very simplified 
collision operator in our calculation of $f_{1i}$ in \secref{app:grads}, 
the numerical prefactor in the expression for the heat flux should not 
be regarded as quantitatively correct. This is not a problem for our 
purposes. Correct numerical coefficients for this and other 
collisional fluxes were calculated by \citet{Braginskii} (see also 
\citealt{Catto1}). The same caveat applies to \eqref{eq:qs} 
and \eqref{eq:gyroq}. \label{fn:Brag}} 
\beq
q_{1i} = \int\did^3\vv\,\frac{m_i v^2\vpar}{2}\, f_{1i}
= - \int\did^3\vv\,\frac{m_i v^2\vpar^2}{2\nuii}
\lt(\frac{v^2}{\vthi^2} - \frac{5}{2}\rt)f_{0i}\,\frac{\vb_0\cdot\vdel T_{0i}}{T_{0i}}
= -\frac{5}{4}\,n_{0i}\,\frac{\vthi^2}{\nuii}\,\vb_0\cdot\vdel T_{0i}.
\label{eq:hflux}
\eeq
The heat flux term in \eqref{eq:p0} arises from the third and fourth 
terms on the left-hand side of \eqref{eq:eps3_avg} (in the fourth term, 
write $\vperp^2=v^2-\vpar^2$ and integrate by parts with respect to $\vpar$
at constant $v$). 

Since formally all terms in \eqref{eq:p0} except the one involving $p_{2i}$ 
have slow time dependence, we may assume that so does $p_{2i}$ and, therefore,  
$\dd p_{2i}/\dd t$ can be dropped from this equation 
(this can be formalized via averaging over short timescales). 
Using now \eqsand{eq:n0}{eq:hflux}, we can rewrite \eqref{eq:p0} as 
an evolution equation for the equilibrium temperature:
\beq
n_{0i}\frac{\did T_{0i}}{\did t} = 
\vdel\cdot\lt(n_{0i}\kappa_i\vb_0\vb_0\cdot\vdel T_{0i}\rt)
- \frac{2}{3}\,n_{0i}T_{0i}\vdel\cdot\vu_{0i},\quad
\kappa_i = \frac{5}{6}\frac{\vthi^2}{\nuii},
\label{eq:T0}
\eeq
where $\kappa_i$ is the ion thermal conductivity. 
The first term on the right-hand side represents collisional heat transport, 
the second compressional heating. 

\Eqsand{eq:n0}{eq:T0} evolve $n_{0i}$ and $T_{0i}$. 
However, the equilibrium density and temperature can only change in such 
a way that pressure balance, \eqref{eq:pr_bal}, is maintained. 
This means that if we know the spatial distribution of $T_{0i}$, 
we also know that of $n_{0i}$, or vice versa.  
Compressive motions will develop to make the density  
and temperature distributions adjust to each other 
and preserve the pressure balance. These 
motions must be consistent with the momentum equation \exref{eq:u0} 
and the isotropic pressure perturbation $p_{2i}$ will adjust to 
make it so. Thus, if we provide the expression for the pressure 
anisotropy $\pperpi-\ppari$, the other 5 equilibrium quantities 
--- $\vu_{0i}$, $n_{0i}$, $T_{0i}$, $p_{2i}$, and $\vB_{0}$ --- 
are determined by the closed set of 5 equations:\footnote{If, 
as discussed in \secref{app:grads}, one 
takes the easy option and assumes $\vdel T_{0i}=0$ and $\vdel n_{0i}=0$, 
then no equations are needed for the constant density and temperature, 
while $p_{2i}$ in \eqref{eq:u0} 
is determined by the incompressibility condition $\vdel\cdot\vu_{0i}=0$
(which follows from \eqref{eq:n0} with $n_{0i}=\const$).} 
momentum \eqref{eq:u0}, 
continuity \eqref{eq:n0}, 
heat conduction \eqref{eq:T0}, 
pressure balance \eqref{eq:pr_bal}, 
and induction \eqref{eq:indavg}. 

\subsubsection{Order $\eps^{3}$ continued: pressure anisotropy}
\label{app:aniso}

In order to calculate the pressure anisotropy $\pperpi-\ppari$, we multiply 
\eqref{eq:eps3_avg} by $m_i(\vperp^2-2\vpar^2)/2$ and integrate over velocities. 
All isotropic terms vanish, as does the term containing $\dd f_{1i}/\dd\vv$, 
which generally cannot contribute to pressure (after integration by parts, 
it is zero because $\int\did^3\vv\,\vv f_{1i}=0$). The result is
\beq
\frac{\dd}{\dd t}\,(\pperpi-\ppari)
= 3p_{0i}\lt(\vb_0\vb_0:\vdel\vu_{0i} - \frac{1}{3}\,\vdel\cdot\vu_{0i}
+ \frac{1}{2}\frac{\dd}{\dd t}\frac{\overline{|\vBp|^2}}{B_0^2}\rt)
- \vdel\cdot\bigl[\vb_0(\qperp-\qpar)\bigr]
- 3\qperp\vdel\cdot\vb_0
- 3\nuii(\pperpi-\ppari),
\label{eq:aniso}
\eeq 
where $\qperp$ and $\qpar$ are parallel fluxes of perpendicular 
and parallel heat, respectively: using \eqref{eq:f1_sln},
\beq
\qperp = \int\did^3\vv\,\frac{m_i\vperp^2\vpar}{2}\,f_{1i} 
= -\frac{1}{2}\,n_{0i}\,\frac{\vthi^2}{\nuii}\,\vb_0\cdot\vdel T_{0i},
\quad
\qpar = \int\did^3\vv\,m_i\vpar^3\,f_{1i},
= -\frac{3}{2}\,n_{0i}\,\frac{\vthi^2}{\nuii}\,\vb_0\cdot\vdel T_{0i},
\label{eq:qs}
\eeq
and we note that $\qperp + \qpar/2=q_{1i}$ (see \eqref{eq:hflux}). 
To work out the collision term in \eqref{eq:aniso}, we have again resorted to 
brutal simplification by using the Lorentz operator (see \eqref{eq:f1_Lorentz}) 
and dropping the nonlinear collision term $C[f_{1i},f_{1i}]$ in \eqref{eq:eps3_avg}
(the consequences of retaining this term, which are mostly small and 
irrelevant for our purposes, have been explored by \citealt{Catto1}). 
The solution of \eqref{eq:aniso} is 
\beq
\Delta(t) \equiv \frac{\pperpi-\ppari}{p_{0i}}
= \Delta_0 + \frac{3}{2}\int_0^t\did t' e^{-3\nuii(t-t')}
\frac{\dd}{\dd t'}\frac{\overline{|\vBp(t')|^2}}{B_0^2},
\label{eq:aniso_sln}
\eeq
where $\Delta_0$ is the part of the anisotropy due to the large-scale dynamics:
\beq
\Delta_0 = \frac{1}{\nuii}\lt\{\vb_0\vb_0:\vdel\vu_{0i} - \frac{1}{3}\,\vdel\cdot\vu_{0i}
- \frac{\vdel\cdot\bigl[\vb_0(\qperp-\qpar)\bigr] + 3\qperp\vdel\cdot\vb_0}{3p_{0i}}\rt\}
\label{eq:D0}
\eeq
and we have assumed $\Delta(0)=\Delta_0$. 
The first two terms in \eqref{eq:D0} are the well known collisional 
contributions to the pressure anisotropy calculated by \citet{Braginskii}. 
The heat-flux terms did not occur in Braginskii's calculation because 
they were small in his assumed sonic-flow ordering ($\vu_{0i}\sim\vthi$). 
They occur here because our ordering is subsonic ($\vu_{0i}\sim\eps\vthi$; 
see \eqref{eq:order_u0}) --- that heat fluxes appear in the pressure 
tensor under such assumptions is also a known fact \citep{Mikh1,Mikh2,Catto1,Catto2}.
The new, nonlinear part of the anisotropy is the second term in \eqref{eq:aniso_sln},
which is due to the firehose fluctuations. This result was predicted 
on heuristic grounds in \secref{sec:nonlin}. 

\Eqref{eq:aniso_sln} completes the set of transport equations derived in 
\secref{app:transp}, but we still need to calculate $\overline{|\vBp|^2}$. 
This is done via \eqsand{eq:indperp}{eq:momperp}. 
The third-order pressure term in the latter equation is be calculated 
in the next section. 

\subsubsection{Pressure tensor for the firehose turbulence}
\label{app:fluct}

In order to close the small-scale momentum equation \exref{eq:momperp}, 
we must calculate the divergence of the third-order ion-pressure tensor 
or, more precisely, the perpendicular part thereof:
since the fast spatial variation is only in the parallel direction, we have, 
from \eqref{eq:f3}
\bea
\nonumber
\lt(\vdel\cdot\vP_{3i}\rt)_\perp &=& \dpar \int\did^3\vv\,m_i\vpar\vvperp f_{3i}
= \dpar\int\did^3\vv\,\frac{m_i\vperp^2}{2}
\Biggl\{\vpar\lt(\frac{\dd g_{2i}}{\dd\vpar}\rt)_v\frac{\vBp}{B_0}
+ \vpar^2\,\frac{\vb_0}{\Omega_i}\times\lt[\frac{2\dpar\vup}{\vthi^2}\,f_{0i} 
+ \frac{\dpar\vBp}{B_0}\lt(\frac{\dd f_{1i}}{\dd\vpar}\rt)_v\rt]\Biggr\}\\
&=& -\dpar\Biggl\{(\pperpi-\ppari)\,\frac{\vBp}{B_0}
+ \frac{1}{\Omega_i}\lt[p_{0i}\dpar\vup
- (2\qperp-\qpar)\,\frac{\dpar\vBp}{B_0}\rt]
\times\vb_0\Biggr\},
\label{eq:P3}
\eea
where the last formula was obtained via integration by parts with respect 
to $\vpar$ (at constant $v$). 
Note that the terms in \eqref{eq:f3} containing $\vB_2$ and $g_{3i}$ 
do not contribute, so, as announced at the end of \secref{app:eps3}, 
we do not need to compute these quantities. 
The pressure anisotropy $\pperpi-\ppari$ is given by \eqref{eq:aniso_sln} 
and the heat fluxes $\qperp$ and $\qpar$ by \eqref{eq:qs}, whence\footnote{As 
explained in footnote \ref{fn:Brag}, the numerical prefactor here should not be taken 
literally because we have used a very simplified collision operator. The correct 
prefactors can be found, e.g., in \citet{Catto1}.}
\beq
\gT \equiv \frac{2\qperp-\qpar}{p_{0i}\vthi} = 
\frac{1}{2}\frac{\vthi}{\nuii}\frac{\vb_0\cdot\vdel T_{0i}}{T_{0i}}. 
\label{eq:gyroq}
\eeq 
The second term in \eqref{eq:P3} is recognizable as the collision-independent 
``gyroviscosity'' \citep{Braginskii} and the third term as 
the collisional heat-flux contribution to it that arises for subsonic 
flows \citep{Mikh1,Mikh2,Catto1}. It is the 
gyroviscous term that will limit the range of wavenumbers susceptible 
to the firehose instability (see \secref{sec:disp_rln}), 
while the heat-flux term will lead to substantial modifications 
of the firehose turbulence and even give rise to an additional 
source of unstable behaviour (the gyrothermal instability; see \secref{sec:GTI}). 

\subsubsection{CGL equations with nonlinear feedback}
\label{app:CGL}

It is perhaps useful to explain how our equations compare to the 
standard ones, due to \citet{CGL}, or CGL. Let us first notice that 
the induction equation \exref{eq:indavg} implies
\beq
\frac{1}{B_0}\frac{\did B_0}{\did t} = 
\vb_0\vb_0:\vdel\vu_{0i} - \vdel\cdot\vu_{0i}.
\eeq
Using \eqref{eq:n0} for $\vdel\cdot\vu_{0i}$, 
we may, therefore, rewrite \eqref{eq:aniso} as follows
\beq
\frac{\dd}{\dd t}\,(\pperpi-\ppari)
= 3p_{0i}\lt(\frac{1}{\Bbar}\frac{\did\Bbar}{\did t} 
- \frac{2}{3}\frac{1}{n_{0i}}\frac{\did n_{0i}}{\did t}\rt)
- \vdel\cdot\bigl[\vb_0(\qperp-\qpar)\bigr]
- 3\qperp\vdel\cdot\vb_0
- 3\nuii(\pperpi-\ppari),
\label{eq:anisoB}
\eeq
where $\Bbar$ includes both the large-scale magnetic field $B_0$
and the averaged firehose fluctuations (see \eqref{eq:Bbar}). 
Now recall that the total perpendicular and parallel pressures 
may be written as follows:
\beq
\piperp = p_{0i} + p_{2i} + \frac{1}{3}\,(\pperpi-\ppari),
\quad
\pipar = p_{0i} + p_{2i} - \frac{2}{3}\,(\pperpi-\ppari),
\eeq
and so, combining \eqsand{eq:p0}{eq:anisoB}, we get
\bea
\label{eq:pperp}
\frac{\did}{\did t}\,\ln\frac{\piperp}{n_{0i}\Bbar} &=& 
- \frac{\vdel\cdot\bigl(\vb_0\qperp\bigr) + \qperp\vdel\cdot\vb_0}{p_{0i}}
- \nuii\,\frac{\piperp-\pipar}{p_{0i}},\\
\frac{\did}{\did t}\,\ln\frac{\pipar\Bbar^2}{n_{0i}^3} &=&
- \frac{\vdel\cdot\bigl(\vb_0\qpar\bigr) - 2\qperp\vdel\cdot\vb_0}{p_{0i}}
- 2\nuii\,\frac{\pipar-\piperp}{p_{0i}}.
\label{eq:ppar}
\eea
These are exactly the CGL equations (without neglecting the heat fluxes; 
see also \citealt{Snyder1,Snyder2,Ramos,Passot}, 
who adapt these equations to a situation in which collisions are weak and the heat 
fluxes are calculated from wave-particle interactions). 
They are also the equations that form the basis for recent 
numerical studies by \citet{Sharma1,Sharma2} 
of the effects of pressure anisotropies in astrophysical plasmas.
As pointed out in \citet{Sharma1}, \eqref{eq:pperp} can be rewritten 
in a form that makes explicit the conservation of the first adiabatic 
invariant (including by the heat-flux terms):
\beq
\frac{\dd}{\dd t}\frac{p_i^\perp}{\Bbar} 
+ \vdel\cdot\lt(\frac{p_i^\perp}{\Bbar}\,\vu_{0i} + \vb_0\,\frac{\qperp}{\Bbar}\rt)
= -\nuii\,\frac{p_i^\perp-p_i^\parallel}{p_{0i}}.
\eeq

The nonlinear feedback in \eqsdash{eq:pperp}{eq:ppar} is provided by the 
small-scale magnetic fluctuations in $\Bbar$. \citet{Sharma1,Sharma2} do not 
have the small-scale fluctuations and model their effect
by introducing strong effective damping terms in \eqsdash{eq:pperp}{eq:ppar} 
that limit the pressure anisotropies to the marginal state of 
the plasma instabilities. The calculation carried out in the present paper 
attempts to derive this feedback from first principles (note the 
discussion in \secref{sec:scatt} regarding the absence in our theory 
of particle scattering by firehose fluctuations). 


\subsubsection{Comparison with the equations of \citet{Shapiro}}
\label{app:ShSh}

Finally, let us make a comparison between our equations and those 
derived in the classic paper by \citet{Shapiro}.\footnote{We thank 
the anonymous referee for suggesting to include this discussion.} 
Their approach is to assume isotropic electrons, an initial 
bi-Maxwellian equilibrium distribution for the ions with 
$\pperp\neq\ppar$, no collisions, a constant uniform 
background magnetic field $B_0$, and to calculate the evolution of 
the ion equilibrium in the quasilinear approximation. 
Recast in our notation, their equations (12-12') for the 
parallel ($\kperp=0$) firehose read
\bea
\label{eq:ShShperp}
\frac{\did\piperp}{\did t} &=& 
\piperp\frac{\did}{\did t}\frac{1}{2}\overline\frac{|\dvBperp|^2}{B_0^2}
-2\lt(\piperp-\pipar\rt)\frac{\did}{\did t}\frac{1}{2}\overline\frac{|\dvBperp|^2}{B_0^2}
- \frac{\did}{\did t}\overline\frac{|\dvBperp|^2}{8\pi},\\
\frac{\did\pipar}{\did t} &=& 
-2\pipar\frac{\did}{\did t}\frac{1}{2}\overline\frac{|\dvBperp|^2}{B_0^2}
-2\lt(\pipar-\piperp\rt)\frac{\did}{\did t}\frac{1}{2}\overline\frac{|\dvBperp|^2}{B_0^2}.
\label{eq:ShShpar}
\eea
The first terms on the right-hand side of these equations 
are the CGL terms and by transferring them to the left-hand side, 
we recover \eqsand{eq:pperp}{eq:ppar} with zeros on the right-hand side.  
Indeed, since \citet{Shapiro} assume $\nuii=0$, $B_0=\const$, and 
bi-Maxwellian ions, there is no collisional relaxation, 
$\did\ln\Bbar/\did t = (1/2B_0^2)\did\overline{|\dvBperp|^2}/\did t$ 
(see \eqref{eq:Bbar}), and there are no flows or heat fluxes. 
The second terms in \eqsand{eq:ShShperp}{eq:ShShpar} could 
be interpreted as the relaxation of the pressure anisotropy by 
effective scattering of particles off the firehose fluctuations. 
However, these terms, as well as the last term in \eqref{eq:ShShperp}, 
are subdominant in our ordering ($\sim\eps^2$ compared to the CGL terms; 
see \eqref{eq:Delta_order} and the discussion in \secref{sec:scatt}). 
They are, in fact, also negligible 
under the assumptions that \citet{Shapiro} have to make in order 
to guarantee the validity of the quasilinear approximation, 
viz., $\pipar-\piperp\ll\pipar$ and $\beta_i\gg1$ (see discussion 
at the end of their paper). 

Note that, dropping the subdominant terms and so retaining only the CGL terms, 
one can integrate either \eqref{eq:ShShperp} or \eqref{eq:ShShpar} and get 
\eqref{eq:Delta_QL}, which then leads to \eqref{eq:QL_standard} 
if $\Delta(t)$ is set to its marginal value. This is the result 
for the saturated fluctuation amplitude 
expressed by equation (17) of \citet{Shapiro}. As discussed in 
\secref{sec:QL_prev}, such a constant saturation level is only possible 
in a system that is collisionless and where anisotropy is assumed 
to relax from an initial value rather than continuously driven.





\label{lastpage}


\begin{thebibliography}{99}
\bibitem[\protect\citeauthoryear{Bahcall}{2000}]{Bahcall} 
  Bahcall N.A., 2000, Phys.\ Rep., 333, 233
\bibitem[\protect\citeauthoryear{Balbus}{2000}]{Balbus1} 
  Balbus S.A., 2000, 
  ApJ 534, 420
\bibitem[\protect\citeauthoryear{Balbus}{2004}]{Balbus2} 
  Balbus S.A., 2004, 
  ApJ, 616, 857
\bibitem[\protect\citeauthoryear{Bale et al.}{2009}]{Bale2} 
  Bale S.D., Kasper J.C., Howes G.G., Quataert E., Salem C., Sundkvist D., 2009, 
  Phys.\ Rev.\ Lett., 103, 211101
\bibitem[\protect\citeauthoryear{Barnes}{1966}]{Barnes} 
  Barnes A., 1966, 
  Phys.\ Fluids, 9, 1483
\bibitem[\protect\citeauthoryear{Berezin \& Vshivkov}{1976}]{Berezin} 
  Berezin Yu.A., Vshivkov V.A., 1976, 
  J.\ Comput.\ Phys., 20, 81
\bibitem[\protect\citeauthoryear{Binney}{2003}]{Binney} 
  Binney J., 2003,
  in The Riddle of Cooling Flows in Galaxies and Clusters of Galaxies, 
  ed.\ T.H.\ Reiprich, J.C.\ Kempner, N.\ Soker, p. 233
  (http://adsabs.harvard.edu/abs/2004rcfg.proc..233B)
\bibitem[\protect\citeauthoryear{Boltzmann}{1872}]{Boltzmann} 
  Boltzmann L., 1872, 
  Sitsungsber.\ Akad.\ Wiss.\ Wien, 66, 275
\bibitem[\protect\citeauthoryear{Bogdanovi\'c et al.}{2009}]{Bogdanovic} 
  Bogdanovi\'c T., Reynolds C.S., Balbus S.A., Parrish I.J., 2009,
  ApJ, 704, 211
\bibitem[\protect\citeauthoryear{Braginskii}{1965}]{Braginskii} 
  Braginskii S.I., 1965, 
  Rev.\ Plasma Phys., 1, 205
\bibitem[\protect\citeauthoryear{Brandenburg \& Nordlund}{2009}]{Brandenburg} 
  Brandenburg, A., Nordlund, A., 2009, 
  Rep.\ Prog.\ Phys., submitted (arXiv:0912.1340)
\bibitem[\protect\citeauthoryear{Br\"uggen \& Scannapieco}{2009}]{Brueggen} 
  Br\"uggen M., Scannapieco E., 2009, 
  arXiv:0905.4726
\bibitem[\protect\citeauthoryear{Califano et al.}{2008}]{Califano} 
  Califano F., Hellinger P., Kuznetsov E., Passot T., Sulem P.L., 
  Tr\'avni\'cek P.M. 2008,
  J.~Geophys.\ Res., 113, A08219
\bibitem[\protect\citeauthoryear{Carilli \& Taylor}{2002}]{Carilli} 
  Carilli C.L., Taylor G.B., 2002, 
  ARA\&A, 40, 319
\bibitem[\protect\citeauthoryear{Catto \& Simakov}{2004}]{Catto1} 
  Catto P.J., Simakov A.N., 2004,
  Phys.\ Plasmas, 11, 90
\bibitem[\protect\citeauthoryear{Catto \& Simakov}{2005}]{Catto2} 
  Catto P.J., Simakov A.N., 2005,
  Phys.\ Plasmas, 12, 114503
\bibitem[\protect\citeauthoryear{Cavagnolo et al.}{2009}]{Cavagnolo} 
  Cavagnolo K.W., Donahue M., Voit G.M., Sun M., 2009,
  ApJS, 182, 12
\bibitem[\protect\citeauthoryear{Chandran \& Cowley}{1998}]{Chandran1} 
  Chandran B.D.G., Cowley S.C., 1998,
  Phys.\ Rev.\ Lett., 80, 3077
\bibitem[\protect\citeauthoryear{Chandran \& Rasera}{2007}]{Chandran2}
  Chandran B.D.G., Rasera Y., 2007,
  ApJ, 671, 1413
\bibitem[\protect\citeauthoryear{Chandrasekhar, Kaufman \& Watson}{1958}]{Chandra} 
  Chandrasekhar S., Kaufman A.N., Watson K.M., 1958, 
  Proc.\ R.\ Soc.\ London A, 245, 435
\bibitem[\protect\citeauthoryear{Chew, Goldberger \& Low}{1956}]{CGL} 
  Chew C.F., Goldberger M.L., Low F.E., 1956,
  Proc.\ R.\ Soc.\ London A, 236, 112
\bibitem[\protect\citeauthoryear{Cho et al.}{2003}]{Cho} 
  Cho J., Lazarian A., Honein A., Knaepen B., Kassinos S., Moin P., 2003,
  ApJ, 589, L77
\bibitem[\protect\citeauthoryear{Churazov et al.}{2004}]{Churazov3}
  Churazov E., Forman W., Jones C., Sunyaev R., B\"\o hringer H., 2004, 
  MNRAS, 347, 29
\bibitem[\protect\citeauthoryear{Clarke \& En\ss lin}{2006}]{Clarke}
  Clarke T.E., En\ss lin T.A., 2006,
  AJ, 131, 2900
\bibitem[\protect\citeauthoryear{David et al.}{2001}]{David}
  David L.P., Nulsen P.E.J., McNamara B.R., Forman W., Jones C., Ponman T., 
  Robertson B., Wise M., 2001,
  ApJ, 557, 546
\bibitem[\protect\citeauthoryear{Davidson \& V\"olk}{1968}]{Davidson}
  Davidson R.C., V\"olk H.J., 1968,
  Phys.\ Fluids, 11, 2259
\bibitem[\protect\citeauthoryear{Dennis \& Chandran}{2005}]{Dennis}
  Dennis T.J., Chandran B.D.G., 2005, 
  ApJ, 622, 205
\bibitem[\protect\citeauthoryear{Dong \& Stone}{2009}]{Dong} 
  Dong R., Stone J.M., 2009, 
  ApJ, 704, 1309
\bibitem[\protect\citeauthoryear{En\ss lin \& Vogt}{2006}]{Ensslin} 
  En\ss lin T.A., Vogt C., 2006,
  A\&A, 453, 447
\bibitem[\protect\citeauthoryear{Fabian}{1994}]{Fabian0} 
  Fabian A.C., 1994, 
  ARA\&A, 32, 277
\bibitem[\protect\citeauthoryear{Fabian et al.}{2003a}]{Fabian1} 
  Fabian A.C., Sanders J.S., Allen S.W., Crawford C.S., Iwasawa K., 
  Johnstone R.M., Schmidt R.W., Taylor G.B., 2003a, 
  MNRAS, 344, L43
\bibitem[\protect\citeauthoryear{Fabian et al.}{2003b}]{Fabian_turb} 
  Fabian A.C., Sanders J.S., Crawford C.S., Conselice C.J., 
  Gallagher III J.S., Wyse R.F.G., 2003b, 
  MNRAS, 344, L48
\bibitem[\protect\citeauthoryear{Fabian et al.}{2005a}]{Fabian2} 
  Fabian A.C., Sanders J.S., Taylor G.B., Allen S.W., 2005a, 
  MNRAS, 360, L20
\bibitem[\protect\citeauthoryear{Fabian et al.}{2005b}]{Fabian3}
  Fabian A.C., Reynolds C.S., Taylor G.B., Dunn R.J.H., 2005b, 
  MNRAS 363, 891
\bibitem[\protect\citeauthoryear{Fabian et al.}{2006}]{Fabian4} 
  Fabian A.C., Sanders J.S., Taylor G.B., Allen S.W., Crawford C.S.,
  Johnstone R.M., Iwasawa K., 2006, 
  MNRAS, 366, 417
\bibitem[\protect\citeauthoryear{Ferrari et al.}{2008}]{Ferrari} 
  Ferrari C., Govoni F., Schindler S., Bykov A.M., Rephaeli Y., 2008, 
  Space Sci.\ Rev., 134, 93
\bibitem[\protect\citeauthoryear{Forman et al.}{2007}]{Forman}
  Forman W., Jones C., Churazov E., Markevitch M., Nulsen P., Vikhlinin A., Begelman M., 
  B\"oringer H., Eilek J., Heinz S., Kraft R., Owen F., Pahre M., 2007,
  ApJ, 665, 1057  
\bibitem[\protect\citeauthoryear{Furth}{1962}]{Furth}
  Furth H.P., 1962, 
  Nucl.\ Fusion Suppl., 1, 169
\bibitem[\protect\citeauthoryear{Gary \& Feldman}{1978}]{Gary0}
  Gary S.P., Feldman W.C., 1978,
  Phys.\ Fluids, 21, 72
\bibitem[\protect\citeauthoryear{Gary et al.}{1998}]{Gary1}
  Gary S.P., Li H., O'Rourke S., Winske D., 1998,
  J.\ Geophys.\ Res., 103, 14567
\bibitem[\protect\citeauthoryear{Gary et al.}{2001}]{Gary2}
  Gary S.P., Skoug R.M., Steinberg J.T., Smith C.W., 2001,
  Geophys.\ Res.\ Lett., 28, 2759
\bibitem[\protect\citeauthoryear{Govoni \& Feretti}{2004}]{Govoni1} 
  Govoni F., Feretti L., 2004,
  Int.\ J.\ Mod.\ Phys.\ D, 13, 1549
\bibitem[\protect\citeauthoryear{Govoni et al.}{2006}]{Govoni2} 
  Govoni F., Murgia M., Feretti L., Giovannini G., Dolag K., Taylor G.B., 2006,
  A\&A, 460, 425
\bibitem[\protect\citeauthoryear{Graham et al.}{2006}]{Graham} 
  Graham J., Fabian A.C., Sanders J.S., Morris R.G., 2006,
  MNRAS, 368, 1369
\bibitem[\protect\citeauthoryear{Guidetti et al.}{2008}]{Guidetti} 
  Guidetti D, Murgia M., Govoni F., Parma P., Gregorini L., de Ruiter H.R., 
  Cameron R.A., Fanti R., 2008,
  A\&A, 483, 699
\bibitem[\protect\citeauthoryear{Guo, Oh \& Ruszkowski}{2008}]{Guo}
  Guo F., Oh S.P., Ruszkowski M., 2008,
  ApJ, 688, 859
\bibitem[\protect\citeauthoryear{Hall}{1981}]{Hall2} 
  Hall A.N., 1981, 
  MNRAS, 195, 685
\bibitem[\protect\citeauthoryear{Hall \& Sciama}{1979}]{Hall1}
  Hall A.N., Sciama D.W., 1979, 
  ApJ, 228, L15
\bibitem[\protect\citeauthoryear{Hasegawa}{1969}]{Hasegawa} 
  Hasegawa A., 1969, 
  Phys.\ Fluids, 12, 2642
\bibitem[\protect\citeauthoryear{Helander \& Sigmar}{2002}]{Helander}
  Helander P., Sigmar D.J., 2002, Collisional Transport in Magnetized
  Plasmas, Cambridge: Cambridge University Press
\bibitem[\protect\citeauthoryear{Hellinger}{2007}]{Hellinger2} 
  Hellinger P., 2007, 
  Phys.\ Plasmas, 14, 082105
\bibitem[\protect\citeauthoryear{Hellinger \& Matsumoto}{2000}]{Hellinger3} 
  Hellinger P., Matsumoto H., 2000, 
  J.\ Geophys.\ Res., 105, 10519
\bibitem[\protect\citeauthoryear{Hellinger \& Matsumoto}{2001}]{Hellinger4} 
  Hellinger P., Matsumoto H., 2001, 
  J.\ Geophys.\ Res., 106, 13215
\bibitem[\protect\citeauthoryear{Hellinger et al.}{2006}]{Hellinger1}
  Hellinger P., Tr\'avn\'i\^cek P., Kasper J.C., Lazarus A.J., 2006,
  Geophys.\ Res.\ Lett., 33, L09101
\bibitem[\protect\citeauthoryear{Horton, Xu \& Wong}{2004}]{Horton1}
  Horton W., Xu B.-Y., Wong H.V., 2004,
  Geophys.\ Res.\ Lett., 31, L06807
\bibitem[\protect\citeauthoryear{Horton et al.}{2004}]{Horton2}
  Horton W., Xu B.-Y., Wong H.V., Van Dam J.W., 2004,
  J.\ Geophys.\ Res., 109, A09216
\bibitem[\protect\citeauthoryear{Islam \& Balbus}{2005}]{Islam}
  Islam T., Balbus S., 2005,
  ApJ, 633, 328
\bibitem[\protect\citeauthoryear{Istomin, Pokhotelov \& Balikhin}{2009}]{Istomin}
  Istomin Ya.N., Pokhotelov O.A., Balikhin M.A. 2009, 
  Phys.\ Plasmas, 16, 062905
\bibitem[\protect\citeauthoryear{Kaiser \& Binney}{2003}]{Kaiser}
  Kaiser C.R., Binney, J., 2003,
  MNRAS, 338, 837
\bibitem[\protect\citeauthoryear{Kasper, Lazarus \& Gary}{2002}]{Kasper}
  Kasper J.C., Lazarus A.J., Gary S.P., 2002,
  Geophys.\ Res.\ Lett., 29, 1839
\bibitem[\protect\citeauthoryear{Kennel \& Sagdeev}{1967}]{Kennel}
  Kennel C.F., Sagdeev R.Z., 1967,
  J.\ Geophys.\ Res., 72, 3303
\bibitem[\protect\citeauthoryear{Kuchar \& En\ss lin}{2009}]{Kuchar} 
  Kuchar P., En\ss lin T., 2009,
  arXiv:0912.3930
\bibitem[\protect\citeauthoryear{Kulsrud}{1983}]{Kulsrud} 
  Kulsrud R.M., 1983, in Galeev A.A., Sudan R.N., eds., Handbook of Plasma Physics, 
  Amsterdam: North-Holland, Vol.~1, p.~115
\bibitem[\protect\citeauthoryear{Kunz et al.}{2011}]{Kunz}
  Kunz M.W., Schekochihin A.A., Cowley S.C., Binney J.J., Sanders J.S. 2011, 
  MNRAS, 410, 2446
\bibitem[\protect\citeauthoryear{Lagan\'a, Andrade-Santos \& Lima Neto}{2010}]{Lagana}
  Lagan\'a T.F., Andrade-Santos F., Lima Neto G.B., 2010,
  A\&A, 511, A15
\bibitem[\protect\citeauthoryear{Leccardi \& Molendi}{2008}]{Leccardi}
  Leccardi A., Molendi S., 2008, 
  A\&A, 486, 359
\bibitem[\protect\citeauthoryear{Lithwick \& Goldreich}{2001}]{Lithwick}
  Lithwick Y., Goldreich P., 2001, 
  ApJ, 562, 279
\bibitem[\protect\citeauthoryear{Longmire}{1963}]{Longmire}
  Longmire C.L., 1963, 
  Elementary Plasma Physics, New York: Interscience
\bibitem[\protect\citeauthoryear{Lyutikov}{2007}]{Lyutikov}
  Lyutikov M., 2007, 
  ApJ, 668, L1 
\bibitem[\protect\citeauthoryear{Malyshkin}{2001}]{Malyshkin} 
  Malyshkin L., 2001,
  ApJ, 554, 561
\bibitem[\protect\citeauthoryear{Markevitch \& Vikhlnin}{2007}]{Markevitch1}
  Markevitch M., Vikhlinin A., 2007, 
  Phys.\ Rep., 443, 1
\bibitem[\protect\citeauthoryear{Markevitch et al.}{2003}]{Markevitch2}
  Markevitch M., Mazzotta P., Vikhlinin A., Burke D., Butt Y., David L, 
  Donnelly H., Forman W.R., Harris D., Kim D.-W., Virani S., Vrtilek J. 2003,
  ApJ, 586, L19
\bibitem[\protect\citeauthoryear{Marsch, Ao \& Tu}{2004}]{Marsch}
  Marsch E., Ao X.-Z., Tu C.-Y., 2004, 
  J.~Geophys.\ Res., 109, A04102
\bibitem[\protect\citeauthoryear{Matteini et al.}{2006}]{Matteini1}
  Matteini L., Landi S., Hellinger P., Velli M., 2006,
  J.\ Geophys.\ Res., 111, A10101
\bibitem[\protect\citeauthoryear{Matteini et al.}{2007}]{Matteini2}
  Matteini L., Landi S., Hellinger P., Pantellini F., Maksimovic M., 
  Velli M., Goldstein B.E., Marsch E., 2007,
  Geophys.\ Res.\ Lett., 34, L20105
\bibitem[\protect\citeauthoryear{Mikhailovskii \& Tsypin}{1971}]{Mikh1} 
  Mikhailovskii A.B., Tsypin V.S., 1971,
  Plasma.\ Phys., 13, 785
\bibitem[\protect\citeauthoryear{Mikhailovskii \& Tsypin}{1984}]{Mikh2} 
  Mikhailovskii A.B., Tsypin V.S., 1984,
  Beitr.\ Plasmaphys., 24, 335
\bibitem[\protect\citeauthoryear{Million \& Allen}{2009}]{Million}
  Million, E.T., Allen S.W., 2009,
  MNRAS, 399, 1307
\bibitem[\protect\citeauthoryear{Narayan \& Medvedev}{2001}]{Narayan} 
  Narayan R., Medvedev M.V., 2001,
  ApJ, 562, L129
\bibitem[\protect\citeauthoryear{Omma \& Binney}{2004}]{Omma2} 
  Omma H., Binney J., 2004, 
  MNRAS, 350, L13
\bibitem[\protect\citeauthoryear{Omma et al.}{2004}]{Omma1} 
  Omma H., Binney J., Bryan G., Slyz A., 2004, 
  MNRAS, 348, 1105
\bibitem[\protect\citeauthoryear{Ogrean et al.}{2010}]{Ogrean} 
  Ogrean G.A., Hatch N.A., Simionescu A., B\"oringer H., Br\"uggen M., 
  Fabian A.C., Werner N. 2010, 
  MNRAS, 406, 354
\bibitem[\protect\citeauthoryear{Parker}{1958}]{Parker} 
  Parker E.N. 1958, 
  Phys.\ Rev., 109, 1874
\bibitem[\protect\citeauthoryear{Parrish, Quataert \& Sharma}{2009}]{Parrish2}
  Parrish I.J., Quataert E., Sharma P., 2009,
  ApJ, 703, 96 
\bibitem[\protect\citeauthoryear{Parrish, Quataert \& Sharma}{2010}]{Parrish3}
  Parrish I.J., Quataert E., Sharma P., 2010,
  ApJ, 712, L194
\bibitem[\protect\citeauthoryear{Parrish, Stone \& Lemaster}{2008}]{Parrish1} 
  Parrish I.J., Stone J.M., Lemaster N., 2008,
  ApJ, 688, 905
\bibitem[\protect\citeauthoryear{Passot \& Sulem}{2007}]{Passot}
  Passot T., Sulem P.L., 2007,
  Phys.\ Plasmas, 14, 082502
\bibitem[\protect\citeauthoryear{Peterson \& Fabian}{2006}]{Peterson}
  Peterson J.R., Fabian A.C., 2006, 
  Phys.\ Rep., 427, 1
\bibitem[\protect\citeauthoryear{Piffaretti et al.}{2005}]{Piffaretti}
  Piffaretti R., Jetzer P., Kaastra J.S., Tamura T., 2005,
  A\&A, 433, 101
\bibitem[\protect\citeauthoryear{Podesta}{2009}]{Podesta}
  Podesta J.J., 2009,
  ApJ, 698, 986
\bibitem[\protect\citeauthoryear{Quataert}{2008}]{Quataert2}
  Quataert E., 2008, 
  ApJ, 673, 758
\bibitem[\protect\citeauthoryear{Quataert, Dorland \& Hammett}{2002}]{Quataert1}
  Quataert E., Dorland W., Hammett G.W., 2002, 
  ApJ, 577, 524
\bibitem[\protect\citeauthoryear{Quest \& Shapiro}{1996}]{Quest} 
  Quest K.B., Shapiro V.D., 1996, 
  J.\ Geophys.\ Res., 101, 24457
\bibitem[\protect\citeauthoryear{Ramos}{2005}]{Ramos}
  Ramos J.J., 2005,
  Phys. Plasmas, 12, 052102
\bibitem[\protect\citeauthoryear{Rebusco et al.}{2005}]{Rebusco1}
  Rebusco P., Churazov E., B\"oringer H., Forman W., 2005,
  MNRAS, 359, 1041
\bibitem[\protect\citeauthoryear{Rebusco et al.}{2006}]{Rebusco2}
  Rebusco P., Churazov E., B\"oringer H., Forman W., 2006,
  MNRAS, 372, 1840
\bibitem[\protect\citeauthoryear{Rebusco et al.}{2008}]{Rebusco3}
  Rebusco P., Churazov E., Sunyaev R,  B\"oringer H., Forman W., 2008,
  MNRAS, 384, 1511
\bibitem[\protect\citeauthoryear{Rincon, Schekochihin \& Cowley}{2010}]{Rincon}
  Rincon F., Schekochihin A.A., Cowley S.C., 2010, 
  MNRAS, in preparation 
\bibitem[\protect\citeauthoryear{Rosenbluth}{1956}]{Rosenbluth}
  Rosenbluth M.N., 1956, 
  Los Alamos Sci.\ Lab.\ Rep.\ LA-2030
\bibitem[\protect\citeauthoryear{Ruszkowski \& Oh}{2010}]{Ruszkowski2}
  Ruszkowski M., Oh S.P., 2010, 
  ApJ, 713, 1332
\bibitem[\protect\citeauthoryear{Ruszkowski et al.}{2007}]{Ruszkowski1} 
  Ruszkowski M., En\ss lin T.A., Br\"uggen M., Heinz, S., Pfrommer C., 2007,
  MNRAS, 378, 662
\bibitem[\protect\citeauthoryear{Sanders \& Fabian}{2006}]{Sanders1} 
  Sanders J.S., Fabian A.C., 2006,
  MNRAS, 371, L65
\bibitem[\protect\citeauthoryear{Sanders \& Fabian}{2008}]{Sanders2} 
  Sanders J.S., Fabian A.C., 2008,
  MNRAS, 390, L93
\bibitem[\protect\citeauthoryear{Sanders et al.}{2010a}]{Sanders3} 
  Sanders J.S., Fabian A.C., Frank K.A., Peterson J.R., Russell H.R., 2010a,
  MNRAS, 402, 127
\bibitem[\protect\citeauthoryear{Sanders et al.}{2010b}]{Sanders4} 
  Sanders J.S., Fabian A.C., Smith R.K., Peterson J.R., 2010b,
  MNRAS, 402, L11
\bibitem[\protect\citeauthoryear{Sanders et al.}{2011}]{Sanders5} 
  Sanders J.S., Fabian A.C., Smith R.K., 2011,
  MNRAS, 410, 1797
\bibitem[\protect\citeauthoryear{Sarazin}{2003}]{Sarazin} 
  Sarazin C.L., 2003, 
  Phys.\ Plasmas, 10, 1992
\bibitem[\protect\citeauthoryear{Schekochihin \& Cowley}{2006}]{Schek2} 
  Schekochihin A.A., Cowley S.C., 2006, 
  Phys.\ Plasmas, 13, 056501
\bibitem[\protect\citeauthoryear{Schekochihin et al.}{2005}]{Schek3}
  Schekochihin A.A., Cowley S.C., Kulsrud R.M., Hammett G.W., Sharma P., 2005, 
  ApJ, 629, 139
\bibitem[\protect\citeauthoryear{Schekochihin et al.}{2008}]{Schek4}
  Schekochihin A.A., Cowley S.C., Kulsrud R.M., Rosin M.S., Heinemann T., 2008, 
  Phys.\ Rev.\ Lett., 100, 081301
\bibitem[\protect\citeauthoryear{Schekochihin et al.}{2009}]{Schek5}
  Schekochihin A.A., Cowley S.C., Dorland W., Hammett G.W., Howes G.G., Quataert E., Tatsuno T., 2009,
  ApJS, 182, 310
\bibitem[\protect\citeauthoryear{Schekochihin et al.}{2010}]{GTI}
  Schekochihin A.A., Cowley S.C., Rincon F., Rosin M.S., 2010,
  MNRAS, 405, 291
\bibitem[\protect\citeauthoryear{Schuecker et al.}{2004}]{Schuecker}
  Schuecker P., Finoguenov A., Miniati F., B\"ohringer H., Briel U.G., 2004, 
  A\&A, 426, 387
\bibitem[\protect\citeauthoryear{Shakura \& Sunyaev}{1973}]{Shakura}
  Shakura N.I., Sunyaev R.A., 1973,
  A\&A, 24, 337
\bibitem[\protect\citeauthoryear{Shapiro \& Shevchenko}{1964}]{Shapiro}
  Shapiro V.D., Shevchenko V.I., 1964, 
  Sov.\ Phys. --- JETP, 18, 1109
\bibitem[\protect\citeauthoryear{Sharma, Hammett \& Quataert}{2003}]{Sharma0}
  Sharma P., Hammett G.W., Quataert E., 2003,
  ApJ, 596, 1121
\bibitem[\protect\citeauthoryear{Sharma, Quataert \& Stone}{2008}]{Sharma3}
  Sharma P., Quataert E., Stone J.M., 2008,
  MNRAS, 389, 1815
\bibitem[\protect\citeauthoryear{Sharma et al.}{2009}]{Sharma4}
  Sharma P., Chandran B.D.G., Quataert E., Parrish I.J., 2009,
  ApJ, 699, 348
\bibitem[\protect\citeauthoryear{Sharma et al.}{2006}]{Sharma1}
  Sharma P., Hammett G.W., Quataert E., Stone J.M., 2006,
  ApJ, 637, 952
\bibitem[\protect\citeauthoryear{Sharma et al.}{2007}]{Sharma2}
  Sharma P., Quataert E., Hammett G.W., Stone J.M., 2007,
  ApJ, 667, 714
\bibitem[\protect\citeauthoryear{Simionescu et al.}{2001}]{Simionescu} 
  Simionescu A., B\"oringer H., Br\"uggen M., Finoguenov A., 2001,
  A\&A, 465, 749
\bibitem[\protect\citeauthoryear{Snyder \& Hammett}{2001}]{Snyder2}
  Snyder P.B., Hammett G.W., 2001, 
  Phys.\ Plasmas, 8, 3199
\bibitem[\protect\citeauthoryear{Snyder, Hammett \& Dorland}{1997}]{Snyder1}
  Snyder P.B., Hammett G.W., Dorland W., 1997, 
  Phys.\ Plasmas, 4, 3974
\bibitem[\protect\citeauthoryear{Southwood \& Kivelson}{1993}]{Southwood} 
  Southwood D.J., Kivelson M.G. 1993,
  J.\ Geophys.\ Res.\ 98, 9181
\bibitem[\protect\citeauthoryear{Subramanian, Shukurov \& Haugen}{2006}]{Subramanian} 
  Subramanian K., Shukurov A., Haugen N.E.L., 2006, 
  MNRAS, 366, 1437
\bibitem[\protect\citeauthoryear{Tajiri}{1967}]{Tajiri}
  Tajiri M., 1967, 
  J.\ Phys.\ Soc.\ Japan, 22, 1482
\bibitem[\protect\citeauthoryear{Teyssier et al.}{2010}]{Teyssier}
  Teyssier R., Moore B., Martizzi D., Dubois Y., Mayer L., 2010,
  MNRAS, submitted (arXiv:1003.4744)
\bibitem[\protect\citeauthoryear{Vedenov \& Sagdeev}{1958}]{Vedenov1}
  Vedenov A.A., Sagdeev R.V., 1958, 
  Sov.\ Phys. --- Dokl., 3, 278
\bibitem[\protect\citeauthoryear{Vedenov, Velikhov \& Sagdeev}{1961}]{Vedenov2}
  Vedenov A.A., Velikhov E.P., Sagdeev R.Z., 1961,
  Nucl.\ Fusion, 1, 83
\bibitem[\protect\citeauthoryear{Vikhlinin et al.}{2005}]{Vikhlinin} 
  Vikhlinin A., Markevitch M., Murray S.S., Jones C., Forman W., Van Speybroeck L., 2005,
  ApJ, 628, 655
\bibitem[\protect\citeauthoryear{Vogt \& En\ss lin }{2005}]{Vogt} 
  Vogt C., En\ss lin T.A., 2005, 
  A\&A, 434, 67
\bibitem[\protect\citeauthoryear{Voigt \& Fabian}{2004}]{Voigt}
  Voigt L.M., Fabian A.C., 2004, 
  MNRAS, 347, 1130
\bibitem[\protect\citeauthoryear{Wicks et al.}{2010}]{Wicks}
  Wicks R.T., Horbury T.S., Chen C.H.K., Schekochihin A.A., 2010, 
  MNRAS, 407, L31
\bibitem[\protect\citeauthoryear{Xu et al.}{2009}]{Xu}
  Xu H., Li H., Collins D.C., Li S., Norman M.L., 2009, 
  ApJ, 698, L14
\bibitem[\protect\citeauthoryear{Yoon, Wu \& de Assis}{1993}]{Yoon}
  Yoon P.H., Wu C.S., de Assis A.S., 1993,
  Phys.\ Fluids B, 5, 1971
\bibitem[\protect\citeauthoryear{Zakamska \& Narayan}{2003}]{Zakamska}
  Zakamska N.L., Narayan R., 2003,
  ApJ, 582, 162
\end{thebibliography}
\end{document}